\documentclass[twocolumn,dvipsnames]{aastex63}

\usepackage{graphicx}
\usepackage{graphics}
\usepackage{amsmath}
\usepackage{amssymb}
\usepackage{url}
\usepackage{hyperref}
\usepackage{xcolor}

\submitjournal{\apj}
\shorttitle{FOGGIE VI: CGM Supported in Emergent, Non-Hydrostatic Equilibrium}
\shortauthors{Lochhaas et al.}

\newcommand{\rvir}{R_{200}}

\newcommand{\red}[1]{\textcolor{Red}{#1}}
\newcommand{\green}[1]{\textcolor{ForestGreen}{#1}}
\newcommand{\magenta}[1]{\textcolor{Fuchsia}{#1}}
\newcommand{\blue}[1]{\textcolor{blue}{#1}}
\newcommand{\gold}[1]{\textcolor{Dandelion}{#1}}
\newcommand{\new}[1]{\textcolor{Bittersweet}{\textbf{#1}}}

\begin{document}

\title{Figuring Out Gas \& Galaxies In Enzo (FOGGIE) VI:\\ The Circumgalactic Medium of $L^*$ Galaxies is Supported in an Emergent, Non-Hydrostatic Equilibrium}
\correspondingauthor{Cassandra Lochhaas}
\email{clochhaas@stsci.edu}

\author[0000-0003-1785-8022]{Cassandra Lochhaas}
\affiliation{Space Telescope Science Institute, 3700 San Martin Dr., Baltimore, MD 21218}

\author[0000-0002-7982-412X]{Jason Tumlinson}
\affiliation{Space Telescope Science Institute, 3700 San Martin Dr., Baltimore, MD 21218}
\affiliation{Department of Physics \& Astronomy, Johns Hopkins University, 3400 N.\ Charles Street, Baltimore, MD 21218}

\author[0000-0003-1455-8788]{Molly S.\ Peeples}
\affiliation{Space Telescope Science Institute, 3700 San Martin Dr., Baltimore, MD 21218}
\affiliation{Department of Physics \& Astronomy, Johns Hopkins University, 3400 N.\ Charles Street, Baltimore, MD 21218}

\author[0000-0002-2786-0348]{Brian W.\ O'Shea}
\affiliation{Department of Computational Mathematics, Science, and Engineering, 
Department of Physics and Astronomy, 
National Superconducting Cyclotron Laboratory,  
Michigan State University}

\author[0000-0002-0355-0134]{Jessica K.\ Werk}
\affiliation{Department of Astronomy, University of Washington, Seattle, WA 98195}

\author[0000-0002-6386-7299]{Raymond C.\ Simons}
\affiliation{Space Telescope Science Institute, 3700 San Martin Dr., Baltimore, MD 21218}
\affiliation{Department of Physics, University of Connecticut, 196A Auditorium Road, Storrs, CT 06269}

\author[0000-0001-6835-273X]{James Juno}
\affiliation{Princeton Plasma Physics Laboratory, 100 Stellarator Road, Princeton, NJ 08540}

\author[0000-0001-5158-1966]{Claire Kopenhafer}
\affiliation{Department of Physics and Astronomy,
Department of Computational Mathematics, Science, and Engineering,
Michigan State University, 567 Wilson Road
East Lansing, MI 48824}

\author[0000-0001-7472-3824]{Ramona Augustin}
\affiliation{Space Telescope Science Institute, 3700 San Martin Dr., Baltimore, MD 21218}

\author[0000-0002-1685-5818]{Anna C.\ Wright}
\affiliation{Department of Physics \& Astronomy, Johns Hopkins University, 3400 N.\ Charles Street, Baltimore, MD 21218}

\author[0000-0003-4804-7142]{Ayan Acharyya}
\affiliation{Department of Physics \& Astronomy, Johns Hopkins University, 3400 N.\ Charles Street, Baltimore, MD 21218}

\author[0000-0002-6804-630X]{Britton D. Smith}
\affiliation{Institute for Astronomy, University of Edinburgh, Royal Observatory, EH9 3HJ, UK}

\begin{abstract}

The circumgalactic medium (CGM) is often assumed to exist in or near hydrostatic equilibrium with the regulation of accretion and the effects of feedback treated as perturbations to a stable balance between gravity and thermal pressure. We investigate global hydrostatic equilibrium in the CGM using four highly-resolved $L^*$ galaxies from the Figuring Out Gas \& Galaxies In Enzo (FOGGIE) project. The FOGGIE simulations were specifically targeted at fine spatial and mass resolution in the CGM ($\Delta x \lesssim 1$ kpc $h^{-1}$ and $M \simeq 200M_\odot$). We develop a new analysis framework that calculates the forces provided by thermal pressure gradients, turbulent pressure gradients,  ram pressure gradients of large-scale radial bulk flows, centrifugal rotation, and gravity acting on the gas in the CGM. Thermal and turbulent pressure gradients vary strongly on scales of $\lesssim5$ kpc throughout the CGM. Thermal pressure gradients provide the main supporting force only beyond $\sim 0.25\rvir$, or $\sim50$ kpc at $z=0$. Within $\sim0.25\rvir$, turbulent pressure gradients and rotational support provide stronger forces than thermal pressure. More generally, we find that global equilibrium models are neither appropriate nor predictive for the small scales probed by absorption line observations of the CGM. Local conditions generally cannot be derived by assuming a global equilibrium, but an \emph{emergent} global equilibrium balancing radially inward and outward forces is obtained when averaging over the non-equilibrium local conditions on large scales in space and time. Approximate hydrostatic equilibrium holds only at large distances from galaxies even when averaging out small-scale variations.

\end{abstract}


\section{Introduction}
\label{sec:intro}

Galaxies grow by accreting gas from their surroundings and turning it into stars. A galaxy's immediate surroundings, the circumgalactic medium (CGM), thus play an important role in the galaxy's evolution \citep*[see review by][]{Tumlinson2017}. As the intermediary of galactic accretion and feedback, the CGM likely plays a significant role in setting the well-known evolutionary patterns of galaxy populations, such as the stellar mass/halo mass fractions \citep{Behroozi2010}, star formation rates versus mass \citep{Noeske2007,Wuyts2011}, and the mass-metallicity relation \citep{Tremonti2004}. Isolating the effects of the CGM on these global patterns requires understanding its mass, energetics, kinematics, and evolution, and relating these physically to galaxy properties. 

For galaxies near the mass of the Milky Way, fundamental galaxy formation theory suggests that gas accreting from the cosmic web passes through a virial shock as it falls onto a galaxy's halo, filling the CGM with a hot (``virialized") gas near $T\sim10^6-10^7$ K \citep{Rees1977,Silk1977,White1978}. This hot halo is theorized to either develop a cooling flow in the inner regions to feed the central galaxy (at low redshift), or has pristine cold filaments piercing it that can feed the central galaxy without being shock-heated to a high temperature \citep[at high redshift,][]{Birnboim2003,Keres2005,Dekel2006,Dekel2009,Nelson2013}. Hot, pressurized gas in a gravitational potential well is generally expected to arrange itself into approximate hydrostatic equilibrium, where the pull of gravity is opposed by the thermal pressure of the gas, before cooling and flowing inward. This hot, volume-filling halo is a generic prediction of both classic and modern galaxy evolution theories. 

However, observational surveys of the CGM have found a significant mass of metal-enriched cool ($T\sim10^4$ K) gas located in the halos of $L^*$ galaxies that is difficult to explain in the simplest version of the hydrostatic hot halo paradigm \citep{Tumlinson2011,Werk2014,Keeney2017,Chen2018}, even when large-scale accretion filaments are included. Cold gas persists even in the halos of red galaxies that are quenched \citep{Gauthier2009,Thom2012,Zhu2014,Berg2019,Zahedy2019}, and which (presumably) have developed fully virialized hot halos.

Analytic works find many different ways to fit cold gas into a hot halo: perhaps the cold gas is located in small clouds in pressure equilibrium with the hot gas \citep{Maller2004,Stern2016}, or is the signature of the hot gas cooling through thermal instability \citep{Voit2015,McQuinn2018,Stern2019}. \citet{Voit2017} and \citet{Faerman2020} present analytic models for the CGM that assume the hot halo is in hydrostatic equilibrium, and show that these models can describe the observations of the CGM quite well \citep{Voit2019a,Voit2019b}. However, recent simulation work that investigates the degree of hydrostatic equilibrium obtained by the CGM of Milky Way mass galaxies finds that there is a significant amount of non-thermal pressure support, such that a purely hydrostatic equilibrium model is not a good description of the halo gas \citep{Oppenheimer2018,Lochhaas2020}. \citet{Faerman2022} applied the analytic model of \citet{Faerman2020} to a semi-analytic model for galaxy formation and also found that non-thermal support was necessary to fit the observations. These recent works suggest that perhaps the hydrostatic hot halo model needs to be modified in order to describe the complexities of the observed and simulated CGM. 

A common and broadly successful class of models for galaxy evolution assumes that galaxies can be described in a slowly evolving pseudo-equilibrium between star formation and the gas flows entering and leaving the galaxy \citep[e.g.,][]{Finlator2008,Dave2012,Lilly2013,Dekel2014,Peng2014}. Such a model describes the formation of observed scaling relations in galaxy populations, such as the mass-metallicity relation or the star formation main sequence \citep{Bouche2010,Mitra2015,Somerville2015a}. Successful semi-analytic models of galaxy evolution are also built on the assumption of equilibrium within the gas flows \citep[e.g.,][]{Somerville2015b,Lacey2016}. It is natural to extend the assumption of equilibrium to the properties of the CGM through hydrostatic equilibrium, but this assumption should be explicitly tested. In particular, if equilibrium is found to hold in the CGM, the bigger question of \emph{why} equilibrium is obtained is then important to answer.

\citet{Oppenheimer2018} and \citet{Lochhaas2020} both found that a certain fraction of non-thermal pressure support was necessary to bring the CGM of different simulated galaxies into equilibrium against the pull of gravity. However, \citet{Lochhaas2020} used idealized, isolated CGM simulations that were missing the important physics of cosmological structure, and while \citet{Oppenheimer2018} did use cosmological simulations, more recent work has pointed out the need for enhanced resolution focused in the CGM to identify small-scale structure \citep{Hummels2019,Peeples2019,vandeVoort2019}. It is as yet unclear how enhanced CGM resolution affects the hot hydrostatic halo model, and it is this question we address in this study.

The Figuring Out Gas \& Galaxies in Enzo (FOGGIE) simulations are cosmological zoom-in simulations that focus on roughly Milky Way mass galaxies \citep{Peeples2019}. They resolve the CGM surrounding these galaxies to a high spatial resolution, allowing detailed characterization of the kinematics of CGM gas, which is significantly important to measuring the degree of non-thermal pressure that may be supporting the CGM. The cosmological nature of the simulations means that no equilibrium laws, such as hydrostatic equilibrium, are imposed \emph{a priori}, so investigating the degree to which such equilibrium may emerge will determine the relevancy of applying such laws to CGM observations or to the galaxy-CGM connection in galaxy evolution theories.

In this paper, we use the FOGGIE simulations of four $L^\star$ galaxies to investigate the validity of hydrostatic equilibrium and thermal pressure profile models in describing the gas in the CGM. We characterize the radially inward and outward forces exerted on CGM gas from thermal pressure gradients, turbulent pressure gradients, ram pressure gradients, rotation, and gravity, and determine how these forces vary on large and small scales, and in particular which forces are dominant in supporting CGM gas against gravitational infall as functions of time and space.

In Section~\ref{sec:methods}, we give a brief overview of the FOGGIE simulations and describe our methods for calculating the various forces acting on CGM gas. Our main results are split between Section~\ref{sec:local_results}, which describes the small-scale variation of the CGM forces, and Section~\ref{sec:global_results}, which describes the trends of the various CGM forces when averaged over large scales in space and time. We discuss the implications of our results for CGM modeling and compare to other works in Section~\ref{sec:discussion} and summarize and conclude in Section~\ref{sec:summary}.

\section{Methods}
\label{sec:methods}
\subsection{Figuring Out Gas \& Galaxies In Enzo}
\label{subsec:FOGGIE}

We give a brief overview of the FOGGIE simulations and the four galaxies and their halos used in this work, but refer the reader to previous papers FOGGIE I through V \citep{Peeples2019,Corlies2020,Zheng2020,Simons2020,Lochhaas2021} for a more detailed description of the simulations.

FOGGIE is a cosmological zoom-in suite of simulations run using the adaptive mesh refinement code Enzo \citep{Bryan2014,BrummelSmith2019}.\footnote{https://enzo-project.org} Six roughly Milky Way-mass (at $z=0$) galaxies are chosen to be ``zoomed in" and simulated with a spatial refinement of $\leq 1.10$ comoving kpc within a ``forced refinement" box that is 287.77 comoving kpc ($200/h$ comoving for $h = 0.695$) on a side and centered on the galaxy and tracking it as it moves through the domain. Within this forced refinement region, cells can further refine wherever the product of the gas cooling time and sound speed is smaller than the cell size, down to a minimum cell size of 274.44 comoving pc (``cooling refinement"). This adaptive refinement scheme tends to lead to smaller cell sizes near the center of the halo, within and close to the central galaxy, and larger cell sizes further from the galaxy, up to $1.10$ comoving kpc within the forced refinement region. The cooling refinement based on the gas cooling time automatically allows for higher resolution in rapidly cooling parts of the CGM, while the forced refinement maintains a high spatial resolution even within the warm, diffuse gas \citep{Simons2020}. Because it defines resolution in terms of physical cell size, this refinement scheme can result in very fine mass elements in warm diffuse gas; the median cell mass in the FOGGIE CGM regions is $100-400 M_{\odot}$.  

Of the six FOGGIE halos, four halos -- named Tempest, Squall, Maelstrom, and Blizzard -- have reached $z=0$ at the time of analysis, and we focus on these four in this paper. These galaxies had their last major merger (mass ratio $<10:1$) before $z=2$, and \citet{Lochhaas2021} showed that Tempest, Squall, and Maelstrom had approached a constant, low-level of star formation on the order of a $\sim$few $M_\odot$/yr and rough virial equilibrium in the halo gas by $z\sim0.1-0$ (it was not analyzed in that paper, but Blizzard follows the same trends). Table~\ref{tab:masses} lists the dark matter, stellar, gaseous, and total mass within the virial radius of these galaxies at $z=0$ \citep[see][for the time evolution of these quantities over $z=2\rightarrow 0$]{Lochhaas2021}.

For those properties that we compute over cosmic time, we use 194 snapshots in time between $z=2$ and $z=0$ at a constant $\sim54$ Myr cadence. For those properties that we compute in bins of radial distance from the central galaxy, we use 100 bins linearly spread between 0 and $1.5R_{200}$ such that each has a width of $0.015R_{200}$, equivalent to $\sim2.5-3.3$ kpc at $z=0$ for these galaxies. $R_{200}$ is the the virial radius, calculated as the radius enclosing an overdensity $200\times$ the redshift-evolving critical density of the universe \citep{Bryan1998}. The virial radius is not fully enclosed within the forced refinement region of the FOGGIE simulations at low redshifts, so some of the simulation cells within $R_{200}$ are at a much lower resolution than the refinement box requires. Throughout the paper, we do not restrict our calculations to only the forced refinement region so as to show results out to a consistent radius that does not depend on the refine box size for all halos. However, we verified that including lower resolution cells in the large-scale averaging of Section 4 does not affect our results, by re-calculating using only the highest-resolution cells within the forced refinement region and finding no difference.

\begin{table}
    \centering
    \begin{tabular}{l r r r r}
        Property & Tempest & Squall & Maelstrom & Blizzard \\ \hline
        $^\mathrm{a}R_{200}$ & 168.30 & 195.93 & 211.87 & 220.33 \\
        $^\mathrm{b}M_{200}$& 5.04 & 8.02 & 10.12 & 11.41 \\
        $^\mathrm{c}M_\mathrm{DM,200}$ & 4.26 & 6.56 & 8.45 & 9.38 \\
        $^\mathrm{d}M_{\star,200}$ & 5.44 & 12.34 & 11.55 & 14.59 \\
        $^\mathrm{e}M_\mathrm{gas,200}$ & 2.33 & 2.28 & 5.15 & 5.68 \\
    \end{tabular}
    \caption{Properties of the four FOGGIE halos studied in this paper at $z=0$.\\$^\mathrm{a}$ Radius enclosing an average density of $200\times$ the critical density of the universe at $z=0$ in kpc.\\$^\mathrm{b}$ Total mass enclosed within $\rvir$ in units of $10^{11}M_\odot$.\\$^\mathrm{c}$ Dark matter mass enclosed within $\rvir$ in units of $10^{11}M_\odot$.\\$^\mathrm{d}$ Stellar mass enclosed within $\rvir$ in units of $10^{10}M_\odot$. Includes satellites.\\$^\mathrm{e}$ Gas mass enclosed within $\rvir$ in units of $10^{10}M_\odot$. Includes ISM of central and satellites.}
    \label{tab:masses}
\end{table}

\subsection{Segmenting the CGM}
\label{subsec:segmenting}

\begin{figure*}
    \centering
    \includegraphics[width=0.32\linewidth]{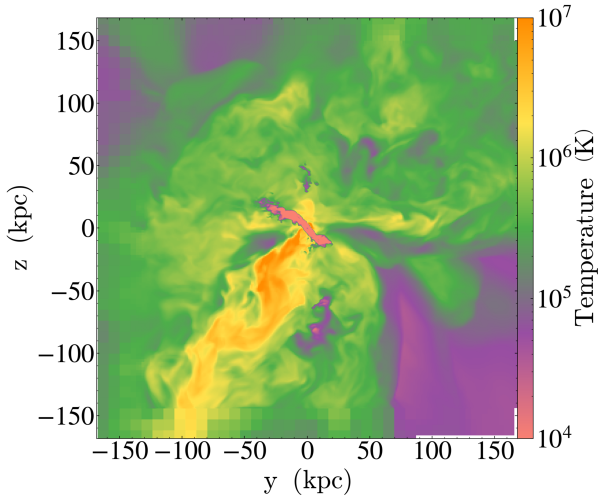}
    \includegraphics[width=0.32\linewidth]{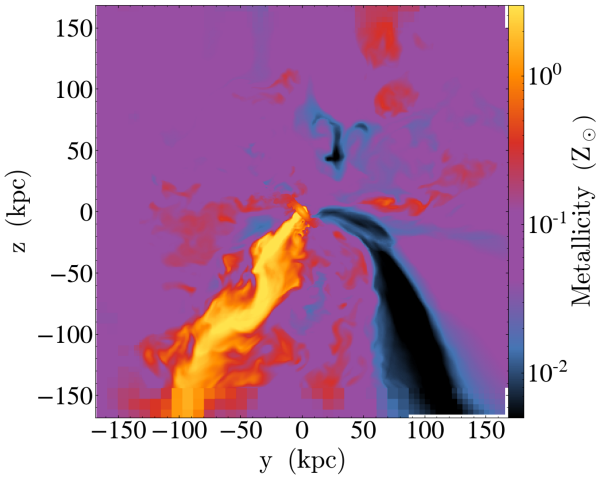}
    \includegraphics[width=0.32\linewidth]{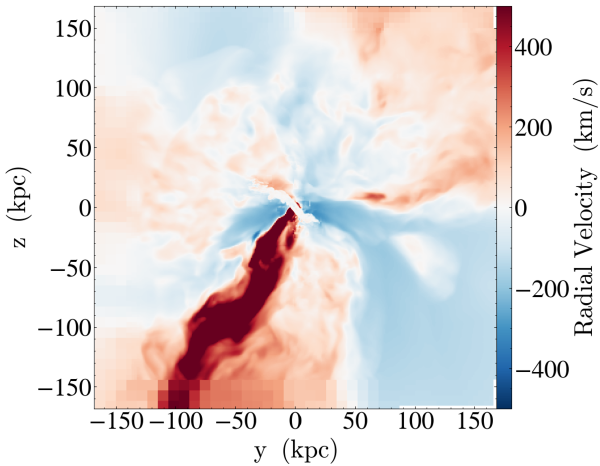}
    \caption{Density-weighted, 10-kpc-thick projections of the temperature (\textbf{\emph{left}}), metallicity (\textbf{\emph{center}}), and radial velocity (\textbf{\emph{right}}) of the Tempest halo at $z=0$, centered on the host dark matter halo center.}
    \label{fig:CGM_segments}
\end{figure*}

It is common for models incorporating hydrostatic equilibrium to consider the CGM to be one dimensional, with temperature, density, and/or other properties depending only on the radial distance from the central galaxy. However, full-hydro simulations like FOGGIE make it clear that there are specific geometric regions of the volume surrounding the galaxy that contain drastically different gas properties from one another and are related to different physical processes. We call these ``segments'' of the CGM. The most obvious CGM segments in FOGGIE are (1) hot ($T>2\times10^6$ K), metal-enriched, and fast ($v_\mathrm{out}\gg v_\mathrm{circ}$) outflows, (2) cool ($T<10^5$ K), metal-poor inflows falling near the free-fall velocity of the halo, and (3) the warm ($10^5\ \mathrm{K}\ <T<2\times10^6$ K), intermediate-metallicity volume-filling medium that is not participating in any bulk inflows or outflows but is turbulent and/or rotating. It is this third segment that is most analogous to the ``hot halo", since its bulk radial velocity is close to zero and its temperature is close to the virial temperature of the halo. However, \citet{Lochhaas2021} shows that in the FOGGIE simulations, the ``hot halo" is actually a factor of two cooler than the standard virial temperature, so here we refer to it as ``warm".

\begin{figure*}
    \centering
    \includegraphics[width=0.8\linewidth]{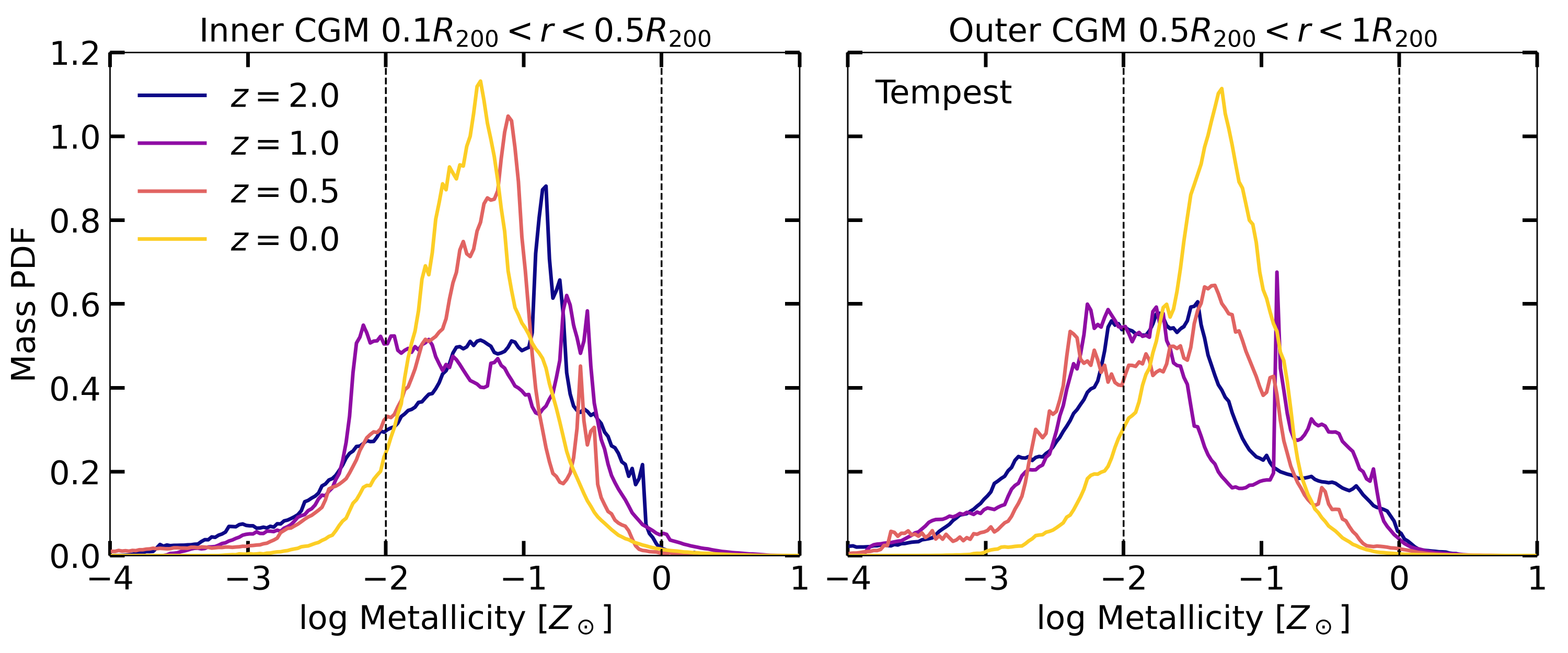}
    \caption{Mass-weighted distributions of the metallicity of gas in the inner CGM ($0.1R_{200}<r<0.5R_{200}$, \textbf{\emph{left}}) and the outer CGM ($0.5R_{200}<r<R_{200}$, \textbf{\emph{right}}), at four different times in the evolution of the Tempest halo. Vertical dashed lines show the choices for the cuts in metallicity that separate the low-metallicity inflows, the high-metallicity outflows, and the intermediate-metallicity ambient material. The vast majority of the gas mass falls into the intermediate metallicity regime.}
    \label{fig:metallicity_pdf}
\end{figure*}

Figure~\ref{fig:CGM_segments} shows density-weighted projections of the temperature (left), metallicity (center), and radial velocity (right) of the Tempest halo at $z=0$ (the projections are 10 kpc thick along the axis of projection). The gas density of the CGM for this snapshot ranges from $10^{-7}-10^{-3}$ cm$^{-3}$, with a median gas density of $\sim10^{-5}$ cm$^{-3}$. A one-sided conical outflow made of hot, metal-enriched material is outflowing at several hundred km s$^{-1}$, extending down and to the left from the roughly edge-on galaxy disk at the center of the image. A cool, metal-poor accretion filament falls at the free fall velocity onto the galaxy ($\sim100$ km/s beyond 100 kpc from the galaxy, increasing to $200-300$ km/s within $3-20$ kpc from the center of the halo, for Tempest at $z=0$), entering the image from the bottom right corner and curving up toward the galaxy at the center. Outside of the outflow cone and inflow filament, there is a warm ($\sim10^{5.5}$ K), intermediate-metallicity ($\sim0.1Z_\odot$) volume-filling medium with radial velocities either inward or outward at speeds slower than either the outflow cone or the inflow filament, which we term the ``no strong flow" segment.

These segments are easy to identify in temperature, metallicity, or radial velocity, and these three gas properties are clearly correlated within the segments. The numeric values of each segment's temperature, metallicity, and radial velocity are dependent on the specific physics and feedback implementation within the FOGGIE simulations, but the general trend of hot, metal-enriched galactic winds launched by stellar feedback and cool, metal-poor filamentary inflows is generally expected from galaxy formation theory and seen in other cosmological simulations \citep[e.g.,][]{Keres2005,Brooks2009}.

Given the distinct properties of the gas in these CGM segments, it seems clear that they are not in equilibrium with one another, and that equilibrium may not hold within them either. Therefore, in what follows we explore not only the globally-averaged support of the CGM gas against gravity, but also this support within the different segments. For simplicity, we choose metallicity as the defining property between different segments, although we verify our results are not dependent on this choice over temperature or radial velocity, nor on the exact value of the cuts that define the segments. Whereas temperature and radial velocity can be affected by heating, cooling, and dynamical processes, metallicity is nearly a tracer fluid. Fresh accretion onto the halo is nearly pristine, and metals are injected by feedback, so metallicity is a natural choice for roughly classifying a segment of gas by its category in the baryon cycle. Because our results depend sensitively on gas dynamic properties (see Section~\ref{subsec:forces}), we avoid using instantaneous radial velocities to assign parcels of gas to different CGM segments. We use $Z>1Z_\odot$ to identify wind/outflow material, $Z<0.01Z_\odot$ for filament/inflow material, and $0.01Z_\odot < Z < 1Z_\odot$ for the volume-filling no strong flow medium, and do not vary the value of these metallicity cuts over redshift. These cuts are designed to be exclusive rather than inclusive: perhaps some outflow or inflow material is not captured by the cuts, but we can be confident in saying that only outflow or inflow material falls into the segments we define as outflow or inflow. For example, we find that recycled, inflowing gas is not metal-enriched enough to be classified in the outflow segment, and there is essentially no pristine outflow, so the inflow and outflow segments accurately select only that material that originated in the cosmic web or in star formation feedback, respectively. Figure~\ref{fig:metallicity_pdf} shows mass-weighted probability distribution functions of the metallicity of gas in the Tempest halo at four times as it evolves. There is little mass in the low-metallicity inflow or high-metallicity outflow segments at all times, despite drastically different thermodynamical properties of the gas in these segments, as well as drastically different mechanisms of support of the gas against gravity within these segments as we will show below.

\subsection{Calculating forces}
\label{subsec:forces}

In order to determine what supports the CGM, we must calculate the forces that oppose gravity. The most commonly understood force acting against gravity is that exerted by the thermal gas pressure gradient. Hydrostatic equilibrium is defined such that the thermal gas pressure gradient is equal and opposite to the force of gravity:
\begin{equation}
    \frac{1}{\rho}\frac{\mathrm{d}P}{\mathrm{d}r} = -\frac{GM_\mathrm{enc}(r)}{r^2} \label{eq:hydrostatic}
\end{equation}
where $\rho$ is the gas density, $P$ is the gas pressure, $r$ is the radial distance from the center of the galaxy halo, $M_\mathrm{enc}(r)$ is the mass enclosed within radius $r$, and $G$ is the gravitational constant. The left side of Equation~(\ref{eq:hydrostatic}) is the outward supporting force exerted by the gas pressure while the right side is the opposite inward force of gravity. When we balance thermal pressure against gravity in this way, with no other forces included, we are assuming the system to be in hydrostatic equilibrium.

However, there are other forces acting on the gas in the CGM other than thermal pressure and gravity. Turbulence, random motions, or bulk flows can exert forces in addition to the thermal pressure. Large-scale rotation also provides the pseudo-force of centrifugal support. In addition, these inward and outward forces need not balance one another. They can, in principle, support the system against gravity by summing up to be exactly in balance with gravity, but in a real system it is much more likely that these many forces will put the system out of strict equilibrium, perhaps far from it. Investigating this balance of forces over time is our main goal. As we are interested in the balance of other forces against gravity, we focus here purely on \emph{radial} directions of all forces. Pressures, like thermal and turbulent pressure, are isotropic, but we consider only the radial direction of the forces exerted by these pressures, calculated by the radial direction of their gradients.

Importantly, we cast the support of the CGM in this work in terms of \emph{forces} rather than \emph{pressures} (or pressure gradients). We choose the analytic framework in this way because not all of the forces working to oppose gravity in the CGM have associated pressures, e.g., rotational support. In addition, the values of the various thermal and non-thermal pressures in the CGM have no bearing on their ability to support the CGM against gravity -- it is their gradients that matter rather than their values. More specifically, we normalize the results by the mass of gas the forces act upon, such that we are really discussing accelerations. We normalize in this way to distinguish variations in the acting forces from variations in the gas density. The CGM in the FOGGIE simulations has significant density variations (e.g., see Figure~\ref{fig:den_cut}) that are not our main focus in this work, so we normalize them out.

We therefore recast Equation~(\ref{eq:hydrostatic}) to include these other forces and do away with the assumption of equilibrium:
\begin{equation}
    \frac{F_\mathrm{net}}{m_\mathrm{gas}} = -\red{\frac{1}{\rho}\frac{\mathrm{d}P_\mathrm{th}}{\mathrm{d}r}} - \green{\frac{1}{\rho}\frac{\mathrm{d}P_\mathrm{turb}}{\mathrm{d}r}} - \magenta{\frac{1}{\rho}\frac{\mathrm{d}P_\mathrm{ram}}{\mathrm{d}r}} + \blue{\frac{v_\mathrm{rot}^2}{r}} - \gold{\frac{GM_\mathrm{enc}(r)}{r^2}} \label{eq:Fnet}
\end{equation}
In Equation~(\ref{eq:Fnet}), each force is color-coded as they will appear in all figures throughout the paper. The force exerted by the thermal pressure gradient is given by the first term on the right-hand side (\red{red}), where we introduce the subscript `th' for `thermal' to differentiate between different types of pressure. The second term (\green{green}) gives the force exerted by the turbulent pressure gradient, the third term (\magenta{magenta}) gives the force exerted by the ram pressure gradient, the fourth term (\blue{blue}) gives the centrifugal force exerted by gas moving with a rotational velocity $v_\mathrm{rot}$, and the last term on the right-hand side (\gold{gold}) gives the force of gravity as in Equation~(\ref{eq:hydrostatic}).

Recasting the standard hydrostatic equilibrium (Eq.~\ref{eq:hydrostatic}) as Equation~(\ref{eq:Fnet}) allows us to answer the two main questions of interest for this study:
\begin{enumerate}
    \item Is there a balance of radial forces in the CGM?
    \item What forces contribute the most to supporting the CGM gas against gravity?
\end{enumerate}
If $F_\mathrm{net}$ in Equation~(\ref{eq:Fnet}) is zero for a given parcel of gas with mass $m_\mathrm{gas}$ in the CGM or for the CGM as a whole, then there is a radial force balance, answering question 1. The value of each of the terms on the right-hand side of Equation~(\ref{eq:Fnet}) relative to each other quantitatively answers question 2. Note that Equation~(\ref{eq:Fnet}) reduces to Equation~(\ref{eq:hydrostatic}) if both $F_\mathrm{net}=0$ and all terms other than the first (thermal pressure gradient) and last (gravity) on the right-hand side of Equation~(\ref{eq:Fnet}) are zero. We simplify the gravitational force to be one-dimensional in radius only, so when evaluating force balance against gravity, we consider only the radial direction of each force, even though any of these forces could be acting in any direction in principle. We define the center of each galaxy halo as the peak of the dark matter density distribution.

To compute each of these forces in the FOGGIE simulations, we start by translating the simulation data onto a uniform-resolution grid at the level of the forced refinement (see \S\ref{subsec:FOGGIE}). Uniform resolution is necessary to ensure all gradients are calculated on the same cell-to-cell scale, and the resolution level was chosen to capture as much small-scale variation as possible, while keeping the calculation of gradients from being too computationally expensive to run in reasonable time. Higher-resolution cells (those that are cooling refined) are combined down to this resolution while lower-resolution cells are over-sampled into higher resolution. We compute all pressure gradients cell-by-cell in the three Cartesian directions of the uniform-resolution grid, then convert to radial gradients as $\frac{\mathrm{d}P}{\mathrm{d}r}=\frac{\mathrm{d}P}{\mathrm{d}x}\hat{x} + \frac{\mathrm{d}P}{\mathrm{d}y}\hat{y} + \frac{\mathrm{d}P}{\mathrm{d}{z}}\hat{z}$. We calculate each type of force in Equation~(\ref{eq:Fnet}) as follows:

\textbf{\red{Thermal pressure force:}} The thermal pressure is a quantity calculated cell-by-cell within the simulation as the internal energy per unit mass converted into pressure assuming an ideal equation of state. We take the gradient in the radial direction of the thermal pressure to obtain the radial force exerted by the thermal pressure.

\begin{figure*}
    \centering
    \includegraphics[width=\linewidth]{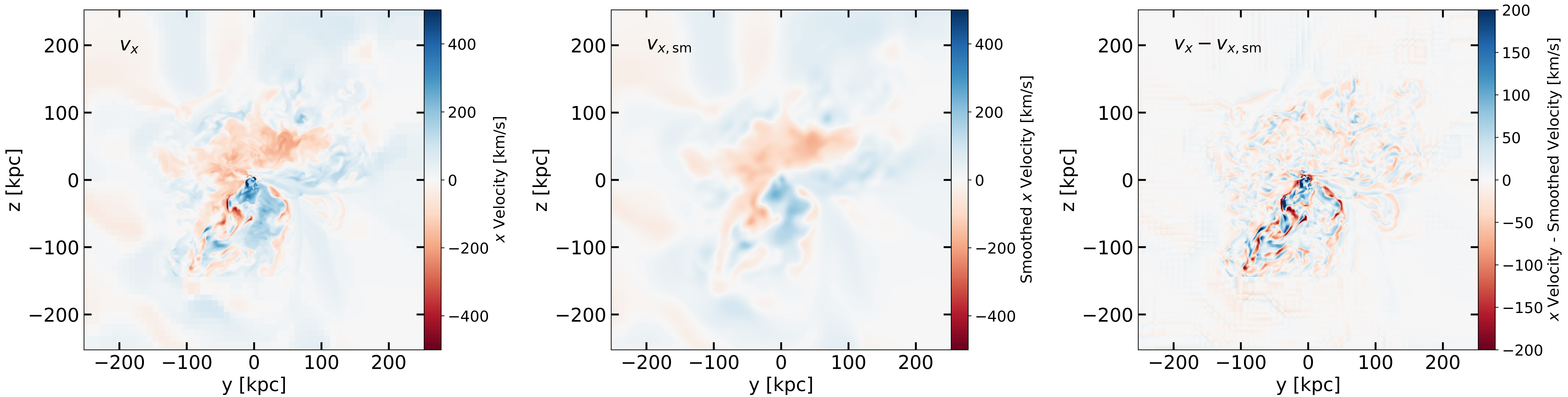}
    \caption{\textbf{\emph{Left:}} A slice in the $y$-$z$ coordinate plane showing the $x$-velocity field within the halo of the Tempest galaxy at $z=0$. \textbf{\emph{Middle:}} The smoothed $x$-velocity field of the same slice, smoothed over a scale of 25 kpc. \textbf{\emph{Right:}} The difference between the $x$-velocity field and the smoothed $x$-velocity field. The right panel shows the deviation of each cell's $v_x$ relative to the average $v_x$ within a 12.5 kpc radius from each cell.}
    \label{fig:velocities}
\end{figure*}

\textbf{\green{Turbulent pressure force:}} Turbulent pressure is a statistical quantity provided by turbulent motions on small scales. We start by computing first order velocity structure functions (VSFs) of the gas in the CGM for all analysis snapshots between $z=2$ and $z=0$ for all four halos as
\begin{equation}
    \langle \delta v \rangle = \langle| \vec{v}(\vec{x} + \vec{s}) - \vec{v}(\vec{x})|\rangle
\end{equation}
where $\vec{v}(\vec{x})$ is the velocity of the gas at a point $\vec{x}$ and $\vec{v}(\vec{x}+\vec{s})$ is the velocity of the gas at another point that is separated from the first by a distance $\vec{s}$. The average denoted by $\langle \rangle$ is done over a random selection of points that fall into bins of $|\vec{s}|$ in the simulation, such that the VSF is the average difference in velocity between two points in the CGM as a function of the separation distance $s$ between those two points. In idealized, Kolmogorov turbulence \citep{Kolmogorov1941}, the VSF shows the turbulent cascade at small separation scales and a turnover at some larger scale that is the driving scale of turbulence \citep{Frisch1995,Davidson2015}. For all four halos and all times between $z=2$ and $z=0$, we find the turbulence driving scale is $\sim25$ physical kpc (not comoving) when calculating the VSF for radii within $0.5R_{200}$. It is possible the driving scale of turbulence is set by the characteristic scale of feedback in the FOGGIE simulations, as the VSF at radii $>0.5R_{200}$, further from the galaxy where feedback operates, shows a turnover at $\sim200$ kpc. We defer to future work for an in-depth study of the turbulent cascade in the CGM of the FOGGIE simulations and what sets the driving scale, but here use the driving scale of turbulence in the inner halo as the physical scale on which the turbulent pressure acts. We use this scale to smooth the velocity fields in each Cartesian direction by convolving with a spherical Gaussian filter that has a standard deviation of $25/6$. The factor of $1/6$ in the standard deviation ensures the full width of the Gaussian, $\pm\sim3$ standard deviations from its center, corresponds to the driving scale of turbulence. We then assign a turbulent velocity dispersion to each cell defined by
\begin{equation}
    v_\mathrm{turb}^2 = \frac{1}{3}\left[(v_x-v_{x,sm})^2 + (v_y-v_{y,sm})^2 + (v_z-v_{z,sm})^2\right] \label{eq:vdisp}
\end{equation}
where $v_x$, $v_y$, and $v_z$ are the velocity fields in the Cartesian directions, $v_{x,sm}$, $v_{y,sm}$, and $v_{z,sm}$ are the smoothed velocity fields in these directions, and the factor of $1/3$ averages the velocity dispersion over the three spatial dimensions. The $v_x$, $v_{x,sm}$, and $(v_x-v_{x,sm})$ fields for one example snapshot of one example halo are shown in Figure~\ref{fig:velocities}. The turbulent pressure for each cell is then given by:
\begin{equation}
    P_\mathrm{turb}=\rho_{sm}v_\mathrm{turb}^2 \label{eq:turb_pressure}
\end{equation}
where $\rho_{sm}$ is the density field smoothed over the same scale (25 physical kpc). We then take the radial gradient of $P_\mathrm{turb}$ to find the radial force exerted by the turbulent pressure. Note that, even though a velocity dispersion is a statistical quantity, the smoothing is done centered on each simulation cell such that each cell is assigned its own velocity dispersion defined as its velocity difference from the smoothed region surrounding it. This allows us to define cell-by-cell turbulent pressures and their gradients, although the cell-to-cell variation is low due to overlapping smoothing regions for adjacent cells.

\textbf{\magenta{Ram pressure force:}} Turbulent pressure is essentially a ram pressure due to motion on scales smaller than the smoothing scale. Therefore, in order to avoid double counting, we are interested only in scales larger than smoothing scale for bulk ram pressure. In addition, ram pressure acts only when a flow is colliding with another flow. We are only interested in ram pressure forces in the radial directions, so to compute $P_\mathrm{ram}$ we start by calculating the spherical coordinates of the velocity fields in $r,\theta,\phi$ rather than $x,y,z$. We then smooth the radial velocity $v_r$ over the same 25 kpc smoothing scale. To capture the effects of the ram pressure of a flow interacting with another flow, we calculate the radial gradient of the smoothed radial velocity to determine where the flow is slowing down by ramming into other gas. Then, the ram pressure is given by: 
\begin{equation}
    P_\mathrm{ram}=\rho_{sm}\left(\frac{\mathrm{d}v_{r,sm}}{\mathrm{d}r}\Delta x\right)^2
\end{equation}
where $\frac{\mathrm{d}v_{r,sm}}{\mathrm{d}r}$ is the radial gradient of the smoothed radial velocity and $\Delta x$ is the cell size in the uniform-resolution grid ($\sim1$ kpc at $z=0$). To turn this ram pressure into a force, we take its radial gradient.

\textbf{\blue{Rotational force:}} The rotational directions of motion are given by the velocities in the angular, non-radial directions, so we convert the Cartesian velocity fields into spherical coordinates and focus on the $v_\theta$ and $v_\phi$ directions for the rotational force. We again wish to avoid double-counting turbulent motions as rotational motions, so we smooth the $v_\theta$ and $v_\phi$ fields on the same smoothing scale of 25 kpc and use the smoothed fields to calculate the rotational force. The rotational force is computed as:
\begin{equation}
    \frac{v_\mathrm{rot}^2}{r} = \frac{v_{\theta,sm}^2+v_{\phi,sm}^2}{r} \label{eq:rot_force}
\end{equation}
where $v_{\theta,sm}$ and $v_{\phi,sm}$ are the smoothed angular velocity fields. Note that we do \emph{not} take the radial gradient of Equation~(\ref{eq:rot_force}), as it is not a pressure that needs to be converted into a force.

\textbf{\gold{Gravitational force:}} The only part of the gravitational force that requires computation is the mass enclosed within a radius $r$. We calculate this as the total mass, including all dark matter, stars, and gas.

\medskip

\begin{figure*}
    \centering
    \includegraphics[width=0.345\linewidth]{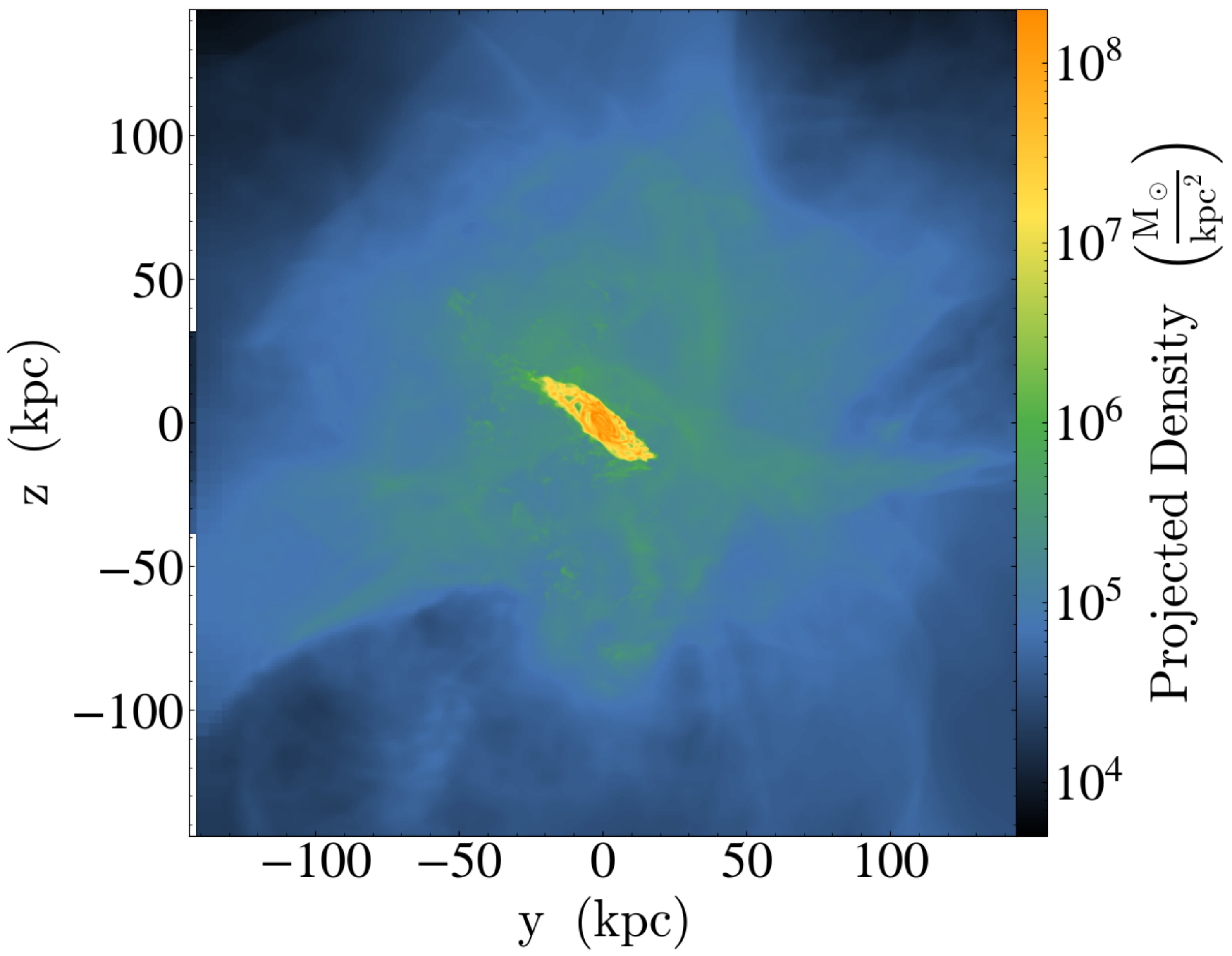}
    \includegraphics[width=0.345\linewidth]{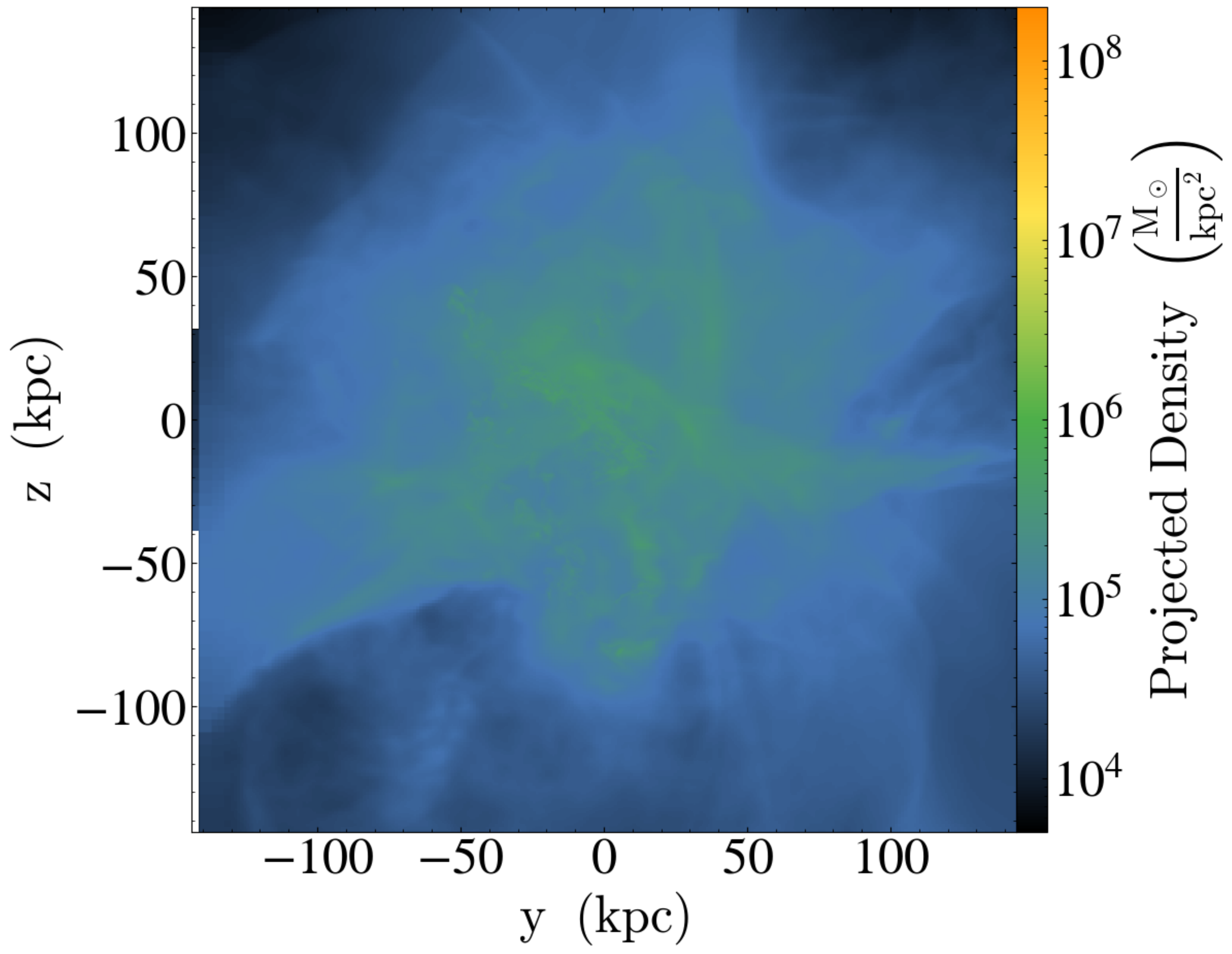}
    \includegraphics[width=0.29\linewidth]{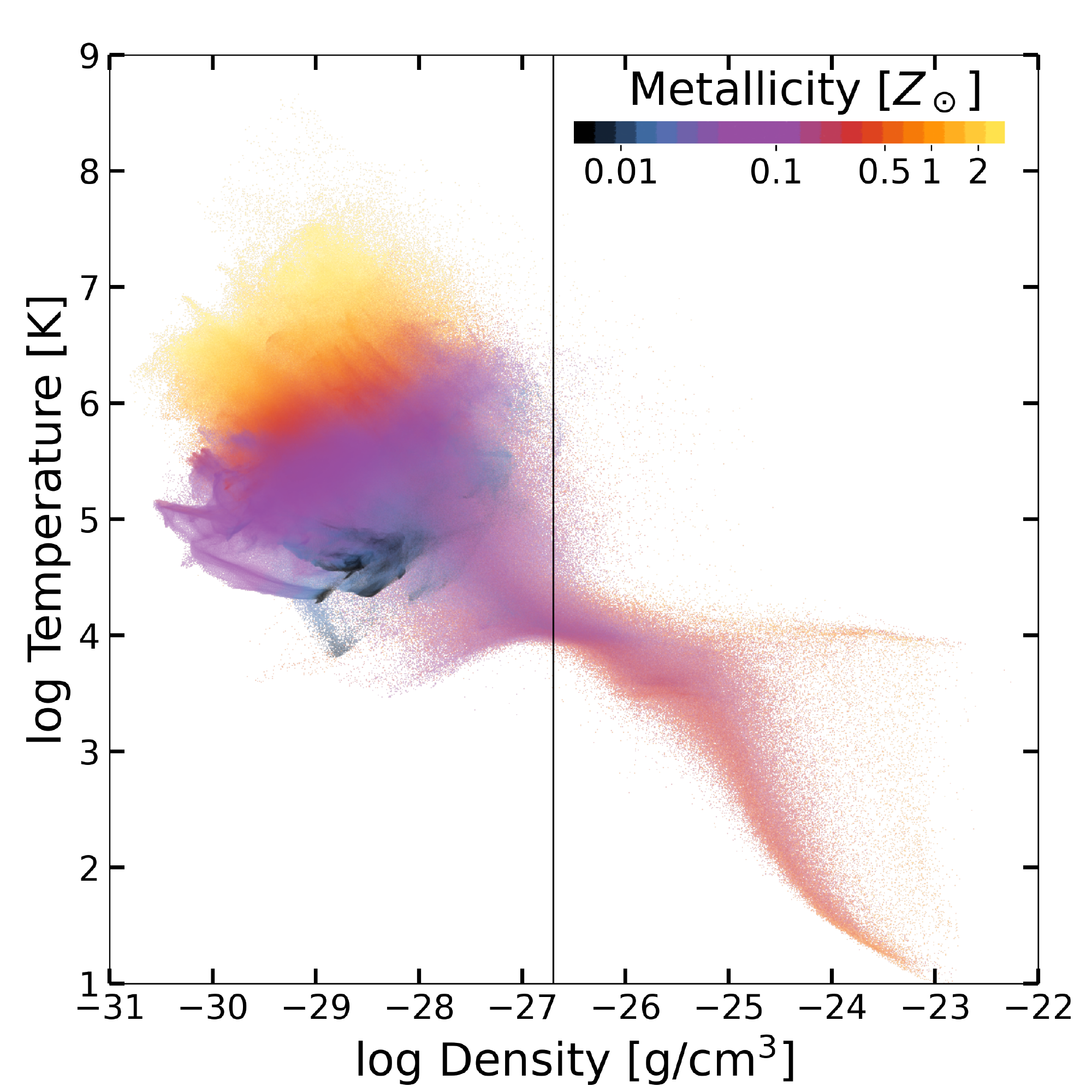}
    \caption{Density projections of the Tempest halo at $z=0$ illustrating the density cut used to separate the ISM and the CGM. \emph{\textbf{Left}}: Density projection before making the cut. \emph{\textbf{Center:}} Density projection after using the chosen cut to remove the ISM. \emph{\textbf{Right}}: Density-temperature phase diagram, color-coded by gas metallicity, where the density cut removes all cells to the right of the vertical black line.}
    \label{fig:den_cut}
\end{figure*}

We are most interested in the forces operating within the CGM, so we define a cut to separate the CGM from the gas disk of each central galaxy, which extends beyond the stellar disk. The distinction between CGM and extended gas disk is somewhat arbitrary, but we find a density cut that seems to separate the two fairly well by eye. CGM is identified as any cell with $\rho<\rho_\mathrm{cut}$ and the disk is identified as any cell with $\rho>\rho_\mathrm{cut}$. At redshifts $z>0.5$, a density cut of $\rho_\mathrm{cut}=2\times10^{-26}$ g~cm$^{-3}$ ($2.96\times10^5\ M_\odot$/kpc$^3$) separates the disk and the CGM well. At redshifts $z<0.25$, $\rho_\mathrm{cut}=2\times10^{-27}$ g~cm$^{-3}$ ($2.96\times10^4\ M_\odot$/kpc$^3$), and at redshifts $0.5>z>0.25$, $\rho_\mathrm{cut}$ varies linearly in time (not linearly in redshift) from $2\times10^{-26}$ g~cm$^{-3}$ to $2\times10^{-27}$ g~cm$^{-3}$. This density cut is arbitrary and determined by-eye to seemingly remove the disk in a projection plot without removing too much of the denser clumps of the CGM, as shown in Figure~\ref{fig:den_cut}. The exact value of the cut likely depends on the physics implementations of the FOGGIE simulations, but it seems to separate the CGM from the extended gas disk surrounding the central galaxy fairly cleanly. In the density-temperature phase diagram in the right panel of Figure~\ref{fig:den_cut}, the chosen dividing line cleanly separates the thin relation for the disk at high densities and low temperatures from the cloud of diffuse and warmer cells that make up the CGM. We verified that small changes to the value of the density cut of a factor of two larger or smaller do not affect our results in Sections~\ref{sec:local_results} and~\ref{sec:global_results} by more than 10\%, and indeed only affect the very inner parts of the CGM within 15 kpc. Note that while we use this cut to isolate the CGM for the purposes of calculating the forces acting strictly on the CGM gas, we do still include the gas and stellar mass of the disk material when calculating the gravitational force that the CGM gas feels.

\section{Local Variation in CGM Forces}
\label{sec:local_results}

\begin{figure*}
    \centering
    \includegraphics[width=0.32\linewidth]{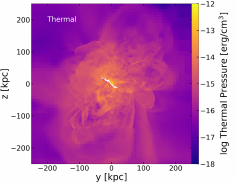}
    \includegraphics[width=0.32\linewidth]{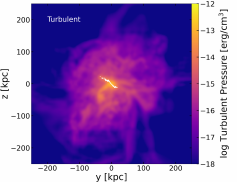}
    \includegraphics[width=0.32\linewidth]{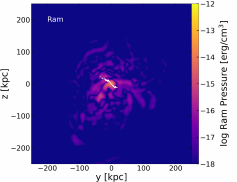}
    \caption{Slices of the thermal (\textbf{\emph{left}}), turbulent (\textbf{\emph{middle}}), and ram (\textbf{\emph{right}}) pressures through the center of the Tempest halo at $z=0$, as calculated in Section~\ref{subsec:forces}. Note that the removed disk appears as white.}
    \label{fig:pressure_slices}
\end{figure*}

\begin{figure*}
    \begin{interactive}{animation}{thermal-turbulent-ram-rotation_force_slice_x.mp4}
    \centering
    \includegraphics[width=0.49\linewidth]{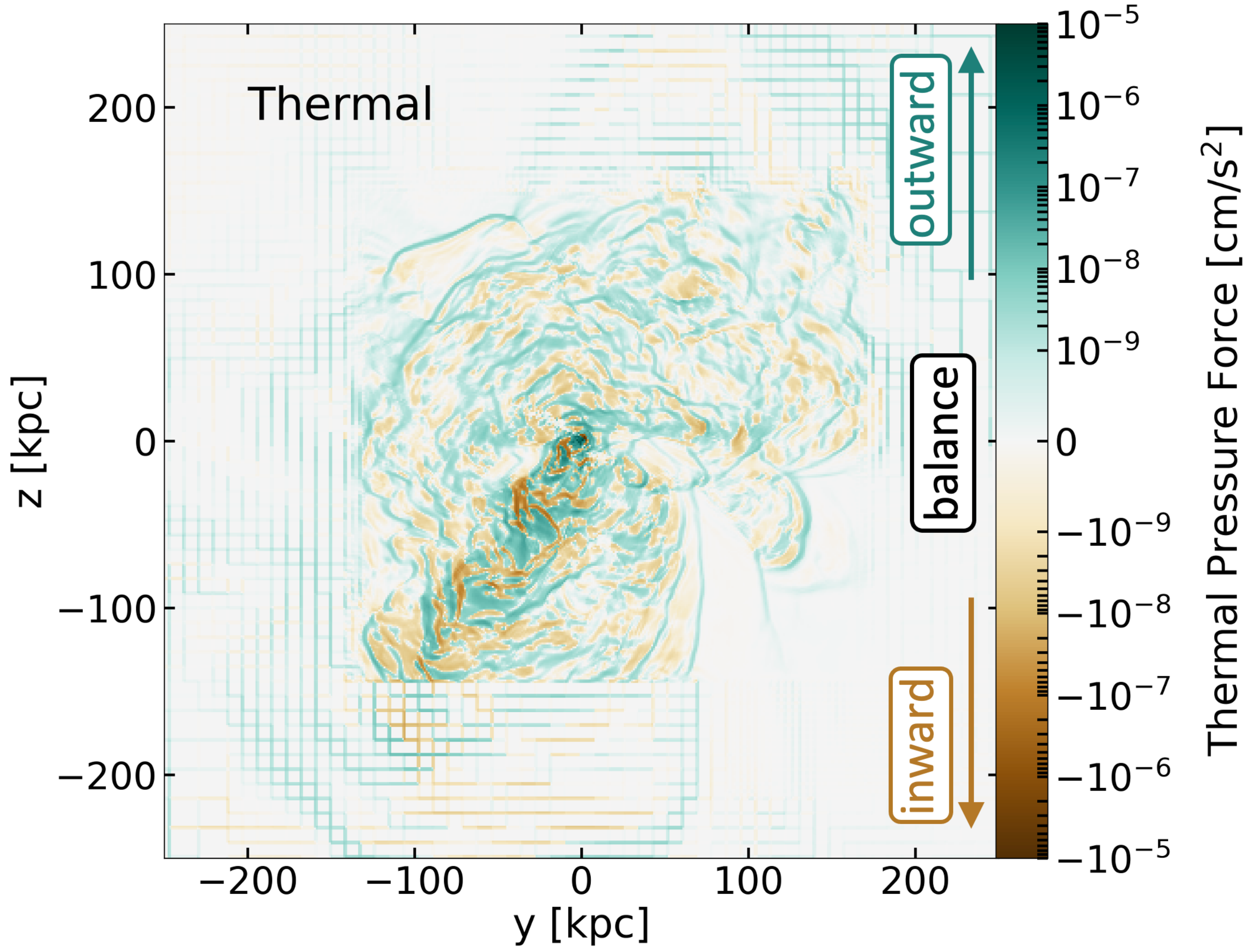}
    \includegraphics[width=0.49\linewidth]{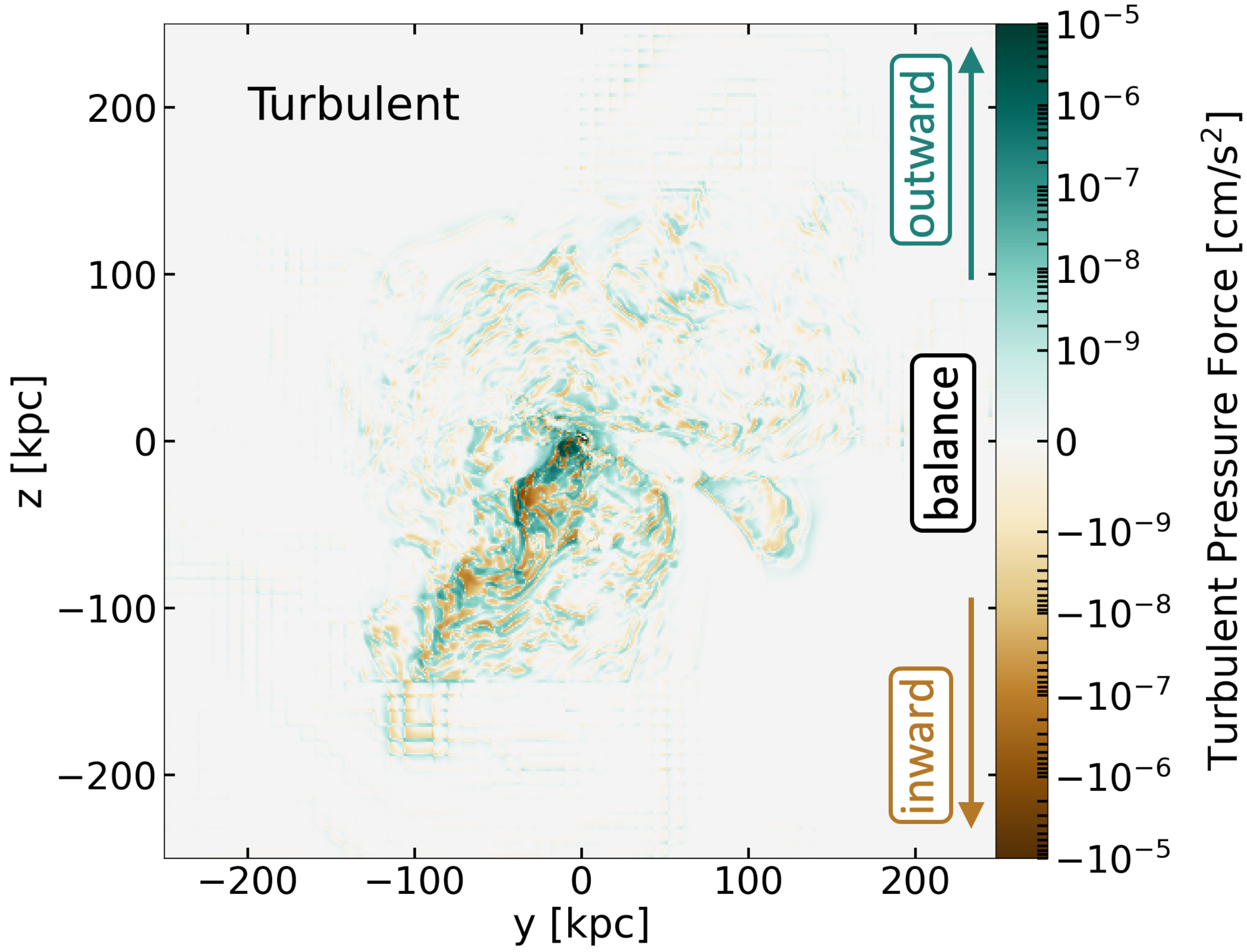}
    \includegraphics[width=0.49\linewidth]{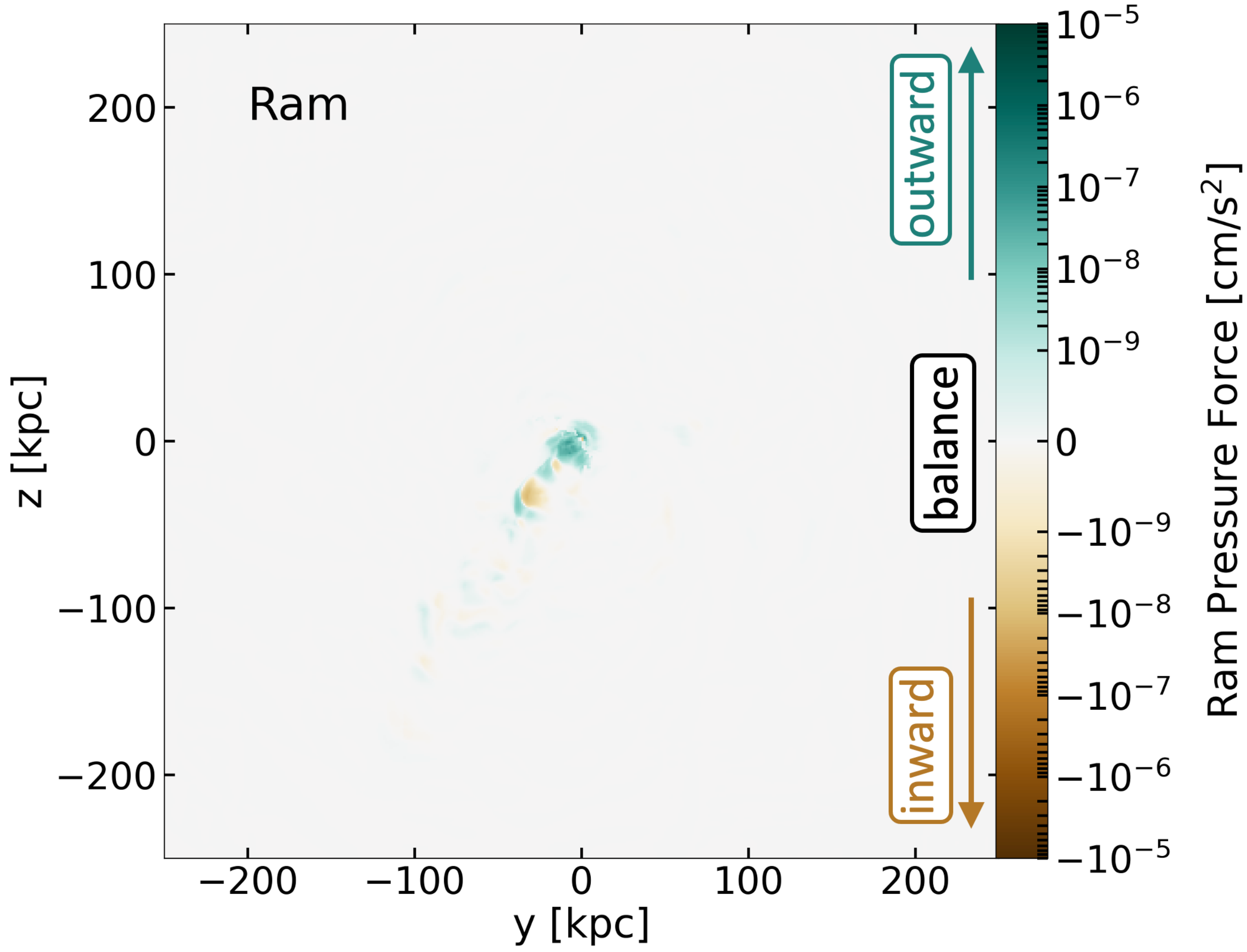}
    \includegraphics[width=0.49\linewidth]{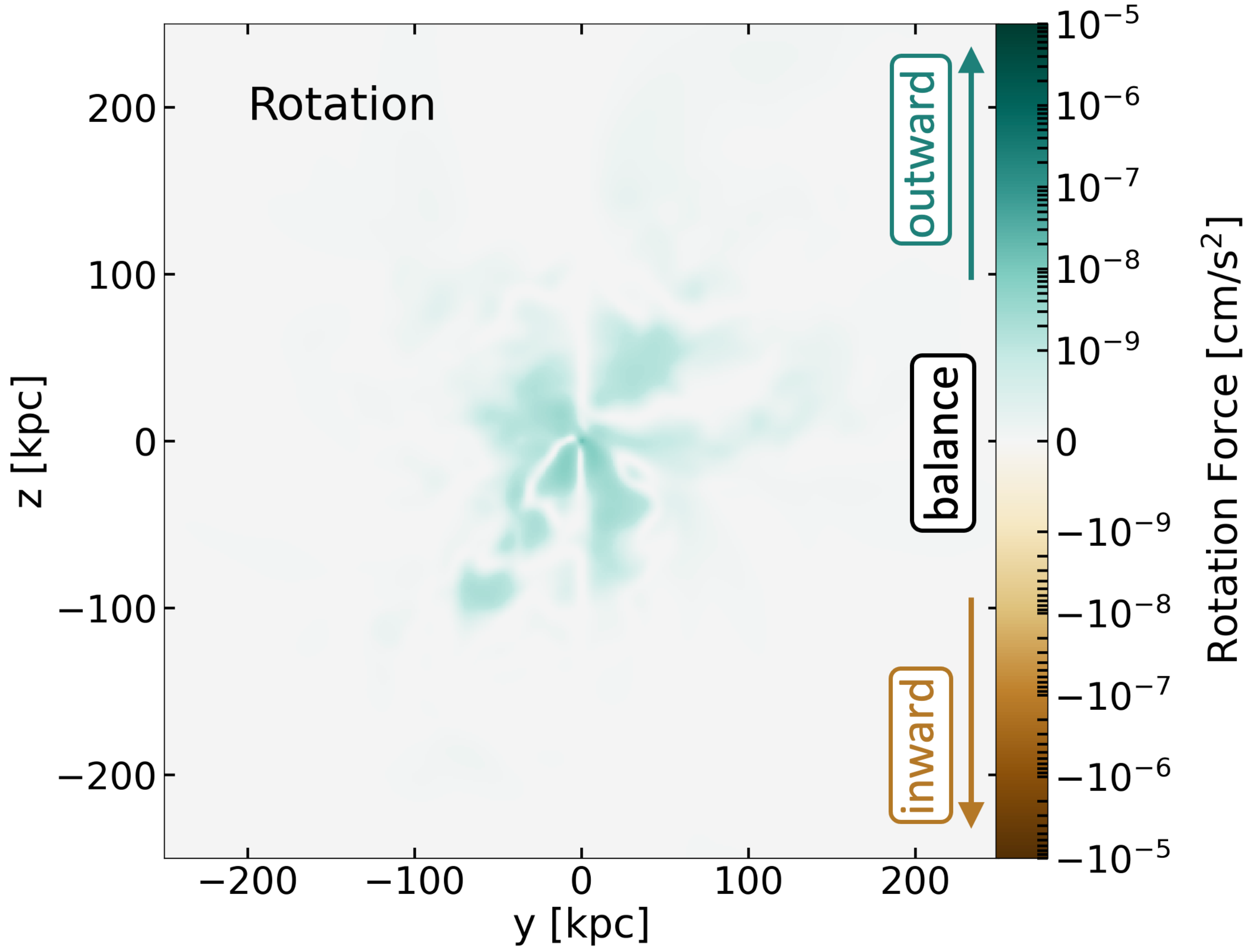}
    \end{interactive}
    \caption{Slices of the thermal (\textbf{\emph{top left}}), turbulent (\textbf{\emph{top right}}), ram (\textbf{\emph{bottom left}}), and rotation (\textbf{\emph{bottom right}}) radial-direction forces through the middle of the Tempest halo at $z=0$. Blue-green colors indicate radially outward forces and brown colors indicate radially inward forces. An animated version of this figure showing the evolution of these slices as Tempest evolves from $z=2$ to $z=0$ is available in the online journal.}
    \label{fig:force_slices}
\end{figure*}

We calculate the cell-by-cell pressures and forces acting on CGM gas using the methods described in Section~\ref{subsec:forces}. Figure~\ref{fig:pressure_slices} shows the thermal, turbulent, and ram pressure measured in a slice through the middle of Tempest's CGM at $z=0$. The morphological similarities between the thermal, turbulent, and ram pressures are directly related to the CGM segments discussed in Section~\ref{subsec:segmenting}: the hot outflows have significant thermal pressure associated with them, as well as driving strong turbulence and exerting significant ram pressure on their surroundings. Figure~\ref{fig:force_slices} shows the forces exerted by the thermal gas pressure, the turbulent pressure, the ram pressure, and the centrifugal rotation force in this same slice, all normalized by the gas mass they are acting on (such that what is plotted is really accelerations, see Section~\ref{subsec:forces}). In Figure~\ref{fig:force_slices}, only the radial direction of the forces are shown, where a positive number (teal) indicates a force directed radially outward and a negative number (brown) indicates a force directed radially inward. Note that the inward- or outward-directed forces shown in Figure~\ref{fig:force_slices} are \emph{not} necessarily aligned with inflowing or outflowing gas: the force vector is an instantaneous acceleration, not a velocity. The forces acting on gas in the inflowing and outflowing CGM segments will be discussed in Section~\ref{subsec:support_segments}.
 
The thermal pressure (left panel of Figure~\ref{fig:pressure_slices}) is measured at a higher resolution than the turbulent and ram pressures (middle and right panels of Figure~\ref{fig:pressure_slices}). Thermal pressure is calculated in the simulation in every cell, but the turbulent and ram pressures require smoothing of the velocity fields to calculate the statistical properties of the gas velocity on small (for turbulent pressure) and large (for ram pressure) scales. This makes the pressure panels appear to have lower resolution, but they are computed at the same resolution as all other fields so that all summation and division operations on the fields are performed at the same scale. 

The thermal pressure is clearly more extended than the turbulent and ram pressure for Tempest at $z=0$. All three types of pressure are strongest in the center of the halo near the (excised) galaxy disk and decline with distance from the galaxy. There appears to be a rough outer edge to the turbulent pressure, outside of which the turbulent pressure is nearly zero. This edge roughly corresponds with the edge of the forced refinement region in FOGGIE, and indicates the transition from the high-resolution to the low-resolution region of the simulation. Because the turbulent pressure requires resolving velocities on small scales, a high resolution is needed to capture the turbulent cascade, and thus there is essentially no turbulent motion outside of the forced refinement region (see Section~\ref{subsec:turb_res} for discussion on this point). In addition, numerical viscosity dampens turbulence at low resolution. When measuring the driving scale of turbulence (Section~\ref{subsec:forces}), we use only the high-resolution forced refinement box and focus on scales close to the galaxy.

The ram pressure is the weakest of the three types of pressure. Ram pressure acts only where a flow impacts another flow, so it is most visible in Figure~\ref{fig:pressure_slices} at the edges of inflow or outflow regions. Note that only the radial velocity is used in calculating the ram pressure as presented in this study; we do not account for ram pressure due to motions in non-radial flows (see Section~\ref{subsec:forces}).

\begin{figure*}
    \begin{interactive}{animation}{total_force_slice_x_updated.mp4}
    \centering
    \includegraphics[width=0.6\linewidth]{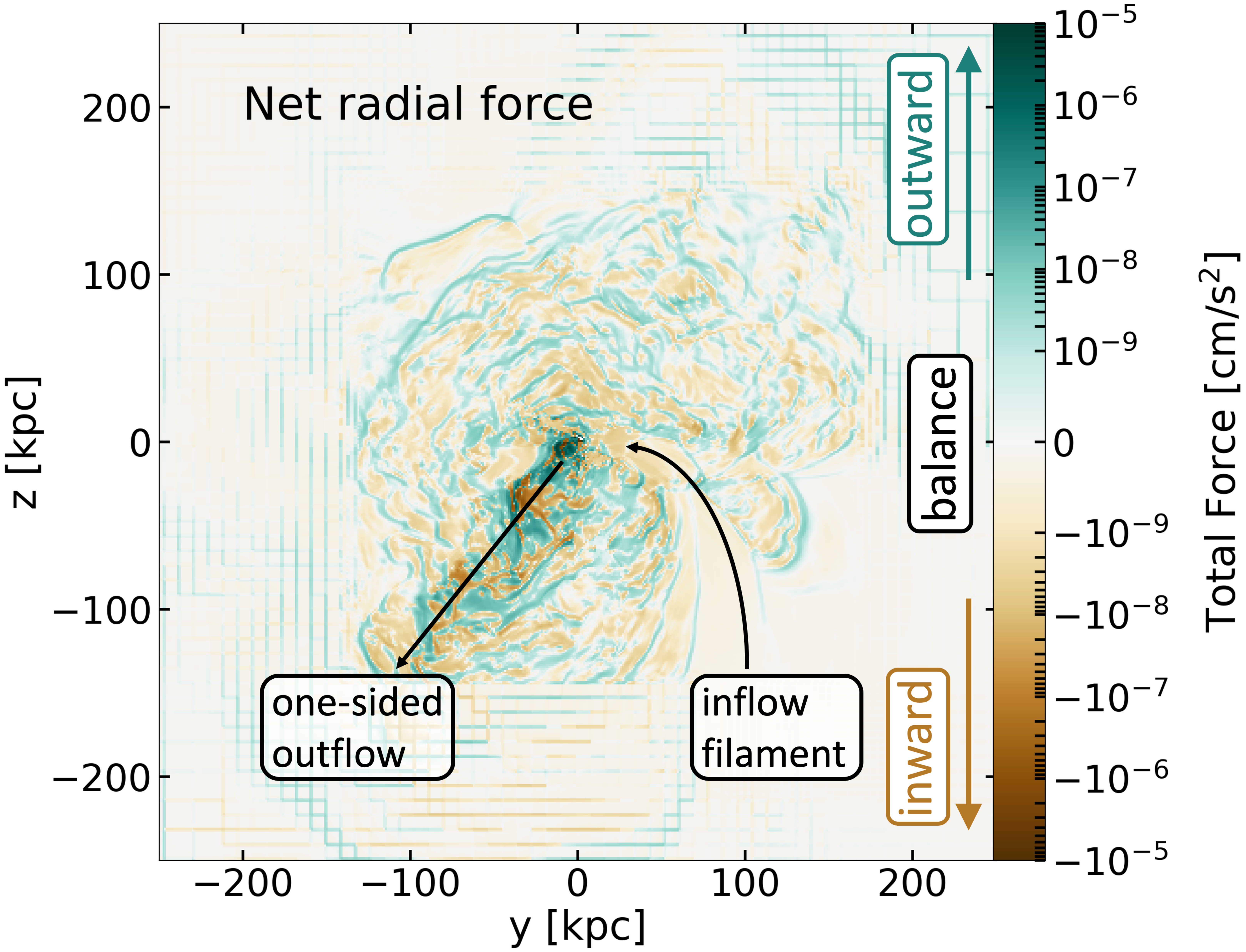}
    \end{interactive}
    \caption{A slice through the center of the Tempest halo at $z=0$ showing the sum total of all investigated forces acting on the CGM gas, including the spherically symmetric and inward gravity. An animated version of this figure showing the evolution of this slice as Tempest evolves from $z=2$ to $z=0$ is available in the online journal.}
    \label{fig:total_force_slice}
\end{figure*}


The most striking feature of the force slices in Figure~\ref{fig:force_slices} is the high degree of variation across small spatial scales, in some cases changing from strong outward forces (teal) to strong inward forces (brown) in adjacent simulation cells ($\sim1$ kpc). The adjacent regions of inward and outward forces may still be unresolved even at this high spatial resolution, as the size of these regions are in some cases just a few cells. The importance of simulation resolution is clearly seen in the thermal and turbulent force panels, where the detailed structure in the forces disappears near the edges of the panels and instead there is a ``hatching" that is an artifact of over-sampling the low-resolution cells and then computing cell-to-cell gradients. The edges of each panel show the forces calculated on the gas outside of FOGGIE's forced refinement region, where the spatial resolution becomes worse by a factor of 10 (cell sizes become $\sim10$ kpc). Contrary to the inherent assumption in hydrostatic equilibrium models, the force exerted by the thermal pressure is not directed outward, such that it is opposing gravity, everywhere. There are places within the halo where the thermal pressure gradient causes an inward-directed force. This strong variation between inward and outward forces is also present in the turbulent pressure force and the ram pressure force, and is indicative of the high degree of spatial variation of these pressures. In short, because the pressures are not smoothly varying, neither are their gradients. 

The fact that the thermal pressure is dominant throughout much of the halo does not necessarily mean that the force exerted by the thermal pressure (top left panel of Figure~\ref{fig:force_slices}) is strongest throughout much of the halo. The force does not depend on the magnitude of the pressure, but rather its gradient. Indeed, the turbulent pressure exhibits a stronger gradient within the inner 15 kpc than the thermal pressure, leading to a stronger force exerted by the turbulent pressure despite a stronger thermal pressure magnitude in the same region (see Section~\ref{sec:global_results} for more discussion on this point).

There are a handful of intriguing features evident in Figure~\ref{fig:force_slices}. This particular galaxy at this particular snapshot in time has a collimated, hot outflow that appears one-sided in the plane of the plotting slice, originating at the central galaxy and extending down and to the left. This outflow has stronger accelerations associated with it than most of the rest of the CGM at this time, although it is also diffuse with little mass, so the overall force associated with the outflow is not necessarily stronger than anywhere else in the CGM that may have a higher density. Interestingly, the accelerations within the collimated outflows are not purely in the outward direction; there are very strong inward-directed accelerations as well. The ram pressure force is weak and negligible everywhere other than within this fast flow. In addition to the collimated hot outflow exhibiting strong accelerations of all types, there is also an inflowing cool filament coming from the bottom right of the slice and flowing toward the right side of the galaxy disk. This filament does not have strong forces associated with it; it is instead identified by a region where there are no forces acting (white in Figure~\ref{fig:force_slices}), other than some mild rotation. The lack of forces acting within the filament indicates it is made up of cool gas without strong thermal pressure gradients, has little turbulence within it, and is not ramming into other gas flows in a way that greatly affects its radial velocity.

The bottom right panel of Figure~\ref{fig:force_slices} shows the centrifugal rotation force, which is only present in the radially outward direction by definition. The rotation is strong only in the inner halo near the galaxy, both in a structure that is roughly aligned with the galaxy disk and within the collimated outflow.

Figure~\ref{fig:total_force_slice} shows the same slice through Tempest at $z=0$, now showing the net force $F_\mathrm{net}/m_\mathrm{gas}$ (equation~\ref{eq:Fnet}) as the sum total of the three pressure forces, rotation, and gravity. The enhanced accelerations, both inward and outward, within the collimated outflow are still present. With the inclusion of gravity, the filament that had no forces associated with it is now colored a faint brown, indicating the primary force acting on it is gravity and the gas in this filament is essentially in free-fall. For brevity, we show these slices of the forces for just one of the four FOGGIE galaxies at just one snapshot in time, but the general trends described here are applicable to other times and galaxies. Videos of slices of each type of force (Figures~\ref{fig:force_slices} and~\ref{fig:total_force_slice}) for all time snapshots between $z=2$ and $z=0$ of all four halos are available on the FOGGIE website.\footnote{\hyperlink{https://foggie.science/cgm_forces.html}{https://foggie.science/cgm\_forces.html}}

In addition to strong spatial variation in the forces acting on the CGM, there is also strong temporal variation. From one snapshot to the next 54 Myr later (one frame to the next in the videos of the force slices), a region of the CGM with an outward-directed force can swap to an inward-directed force. The median sound speed for the Tempest halo at $z=0$ is $\sim80$ km/s, which corresponds to a travel distance of 4.5 kpc in the time between analysis snapshots, roughly the same size as the spatial variations. Thus, a particular small region of the CGM could oscillate between a positive and negative pressure gradient due to a sound wave passing over it between analysis snapshots. \citet{Fabian2003} identified structures in the Perseus cluster that are consistent with being pressure (sound) waves traveling through the intracluster medium. These structures appear visually similar to our thermal force slice (top left panel of Figure~\ref{fig:force_slices}), and so the thermal force structure we find may very well be due to sound waves expanding outward from the central galaxy.

The FOGGIE simulations have outputs every $\sim5$ Myr, a time resolution a factor of 10 better than what we use here. While it would be computationally expensive to calculate the force slices for the full $z=2$ to $z=0$ evolution of each halo at this higher time resolution, we do this calculation for a small length of time for the Tempest halo, from $z=0.2$ to $z=0$. At this higher time resolution, we still see structures in the total force appearing and disappearing from one snapshot to the next, indicating that they are even shorter-lived than $\sim5$ Myr. We would not expect that such short-lived force variations would have a significant impact on the velocity structure of the gas. However, some of the force variations are much longer-lived: the ``fronts" of outward-directed forces in the outer regions of the CGM persist for times $>50$ Myr (see the animated version of Figure~\ref{fig:total_force_slice}). These structures appear to be shock fronts at the edge of outflows from the central galaxy. The outward-directed force corresponds to inflows slowing down as they impact the shock front, while the inward-directed forces just interior to the fronts corresponds to outflows slowing down at the shock.

Clearly, one of the two assumptions behind hydrostatic equilibrium, that the outward and inward forces on the gas balance each other, does not hold everywhere within the CGM. If $F_\mathrm{net}=0$ for a given simulation cell, that cell appears as white within Figure~\ref{fig:total_force_slice}, so the general lack of white cells within the forced refinement region indicates a lack of force balance on the resolution of the simulation grid. The other assumption behind hydrostatic equilibrium, that the only force opposing gravity is exerted by the thermal gas pressure, also clearly does not hold because there are significant forces being exerted by other pressures, as well as rotation, within the CGM.

\begin{figure*}
    \centering
    \includegraphics[width=\linewidth]{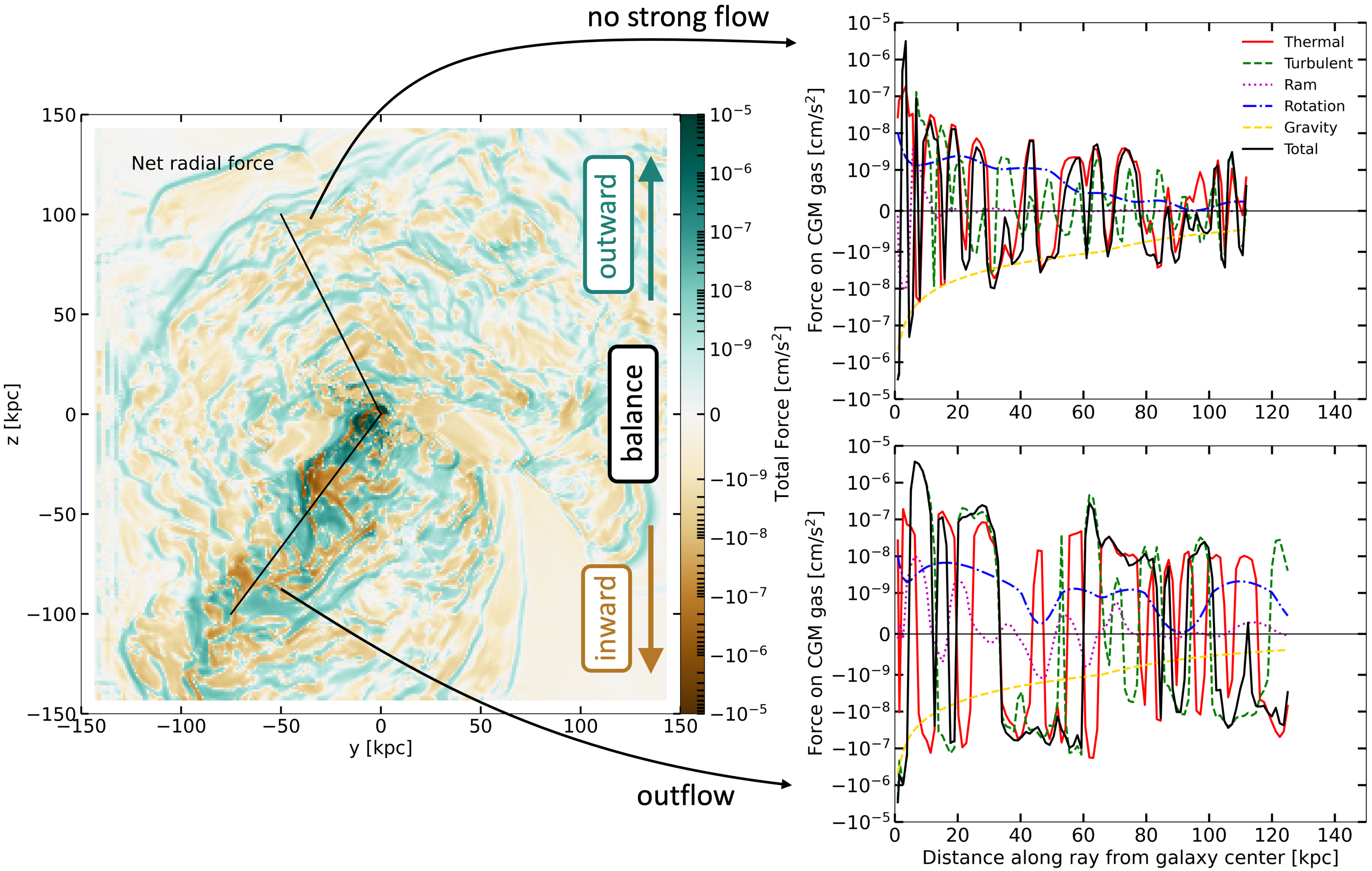}
    \caption{\textbf{\emph{Left:}} A slice through the middle of the Tempest galaxy at $z=0$ showing the total force $F_\mathrm{net}$ (equation~\ref{eq:Fnet}), with two rays overplotted. \textbf{\emph{Right:}} Each normalized force term (curve colors and styles as in legend in figure) is plotted as a function of distance along the rays shown on the left, on a symlog scale to allow the forces to change direction through zero. The top right panel corresponds to the ray on the left that is extending upwards through the halo, and the bottom right panel corresponds to the ray on the left that is extending downwards.}
    \label{fig:force_rays}
\end{figure*}

To better visualize the drastic variation in forces across small spatial scales, Figure~\ref{fig:force_rays} shows each type of force along two selected one-dimensional rays extending from the center of the galaxy outward. Nearly all of the forces are characterized by strong variation in both the strength and direction of the force on scales of $\sim5$ kpc. The bottom right panel in Figure~\ref{fig:force_rays} plots the accelerations along a ray placed through the middle of the strong collimated outflow, where it is clear that the thermal, turbulent, and ram accelerations are stronger in both outward (positive) and inward (negative) directions than in the other ray (top right panel), which is not placed in an outflow region. However, the degree of variation in the two rays is similar: they both show swings from positive to negative values over spatial scales of $\sim5$ kpc. The only forces that do not alternate between outward and inward are the centrifugal rotation force (purely outward) and the force of gravity (purely inward), both of which are defined to operate in only one radial direction. Because the gravitational force is defined spherically, it does not show the variation that the other forces do. The centrifugal rotation force also appears somewhat smoother, but it does have some variation. Note that because Figure~\ref{fig:force_rays} plots normalized forces, i.e. accelerations, it is not possible to ``sum by eye" to determine the total force acting along the rays from the figure. However, plotting the accelerations rather than the forces makes it clear that the strong variations are due to the physics of the driving forces acting on the CGM, rather than on gas density variations.

The cell size resolution in FOGGIE is specified in comoving coordinates, which means it varies in physical coordinates over cosmic time as the universe expands. At higher redshift, the resolution is better (smaller cell sizes) in physical space. To determine if the $\sim5$ kpc scale of the smallest-scale variations in the accelerations (Figure~\ref{fig:force_rays}) at $z=0$ is different when the resolution is improved, we carried out the same analysis as in Figure~\ref{fig:force_rays} on a $z\approx1.5$ output of the Tempest halo. At this redshift, the cell size in the forced refinement region is $\approx0.55$ kpc, about half of that at $z=0$. We find that the variations in the accelerations also occur on scales about half, or $\sim2.5$ kpc, of the variations at $z=0$ (not shown). We sample the $z\approx1.5$ output at the same (physical) resolution as the $z=0$ output and find that the scale of the acceleration variations increases to $\sim5$ kpc. This means the smallest scale of force variation is $\sim5\times$ the simulation resolution, regardless of the simulation resolution. Thus, it appears that the smallest variation scale is dependent on simulation resolution rather than being set by the physics of CGM gas. It is possible that in reality, the acceleration of CGM gas, and thus forces acting on it, are varying on smaller scales than the FOGGIE simulation resolution. 

In summary, the forces exerted by thermal gas pressure, turbulent pressure, ram pressure, centrifugal rotation, and gravity, as well as the net sum of all these forces, vary significantly on spatial scales of $\lesssim5$ kpc throughout the CGM. This suggests that any kind of global force balance model, like that of hydrostatic equilibrium, fails at describing the properties of the gas on $\lesssim5$ kpc scales within the CGM.

\section{Global Trends of CGM Forces}
\label{sec:global_results}

\begin{figure*}
    \centering
    \includegraphics[width=0.32\linewidth]{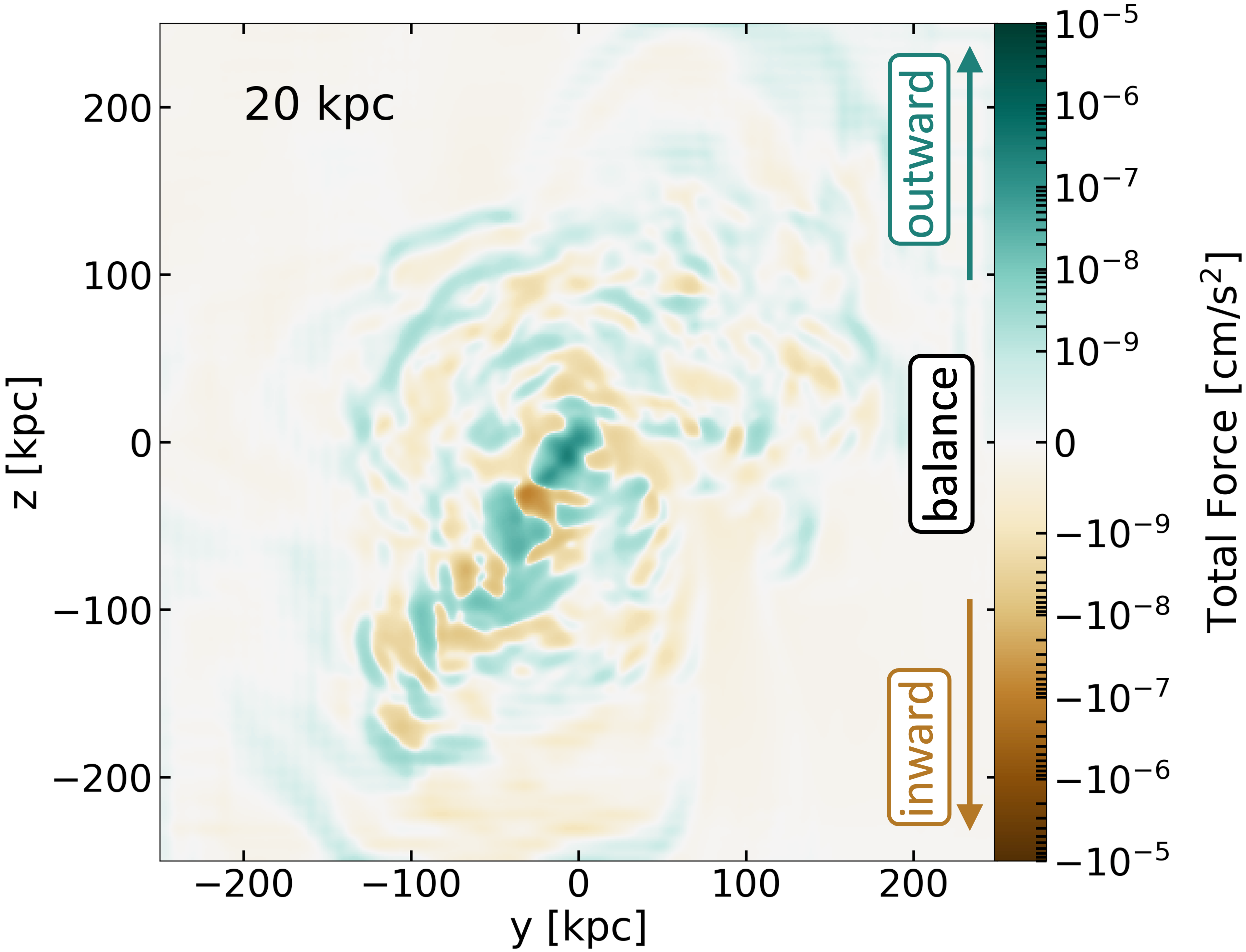}
    \includegraphics[width=0.32\linewidth]{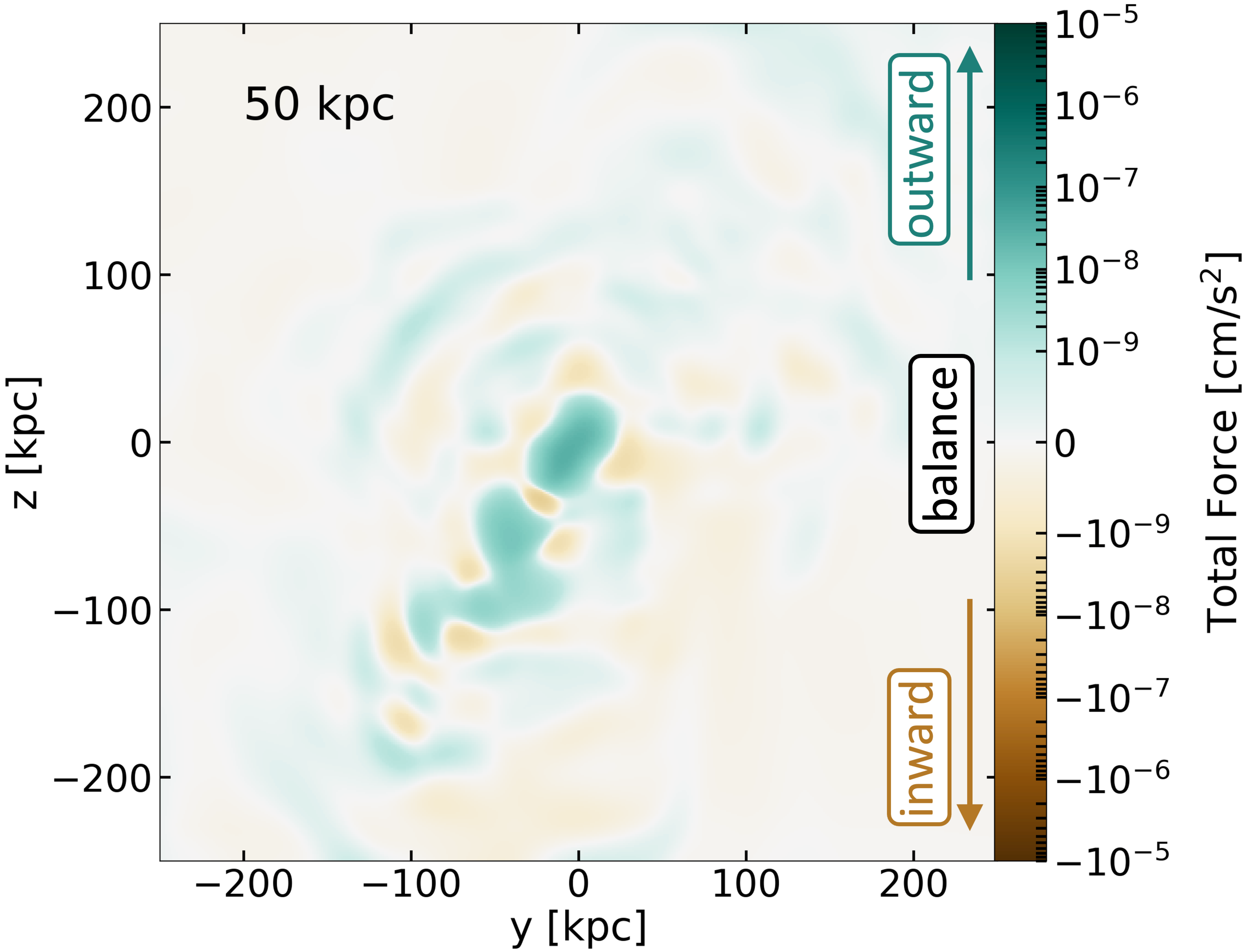}
    \includegraphics[width=0.32\linewidth]{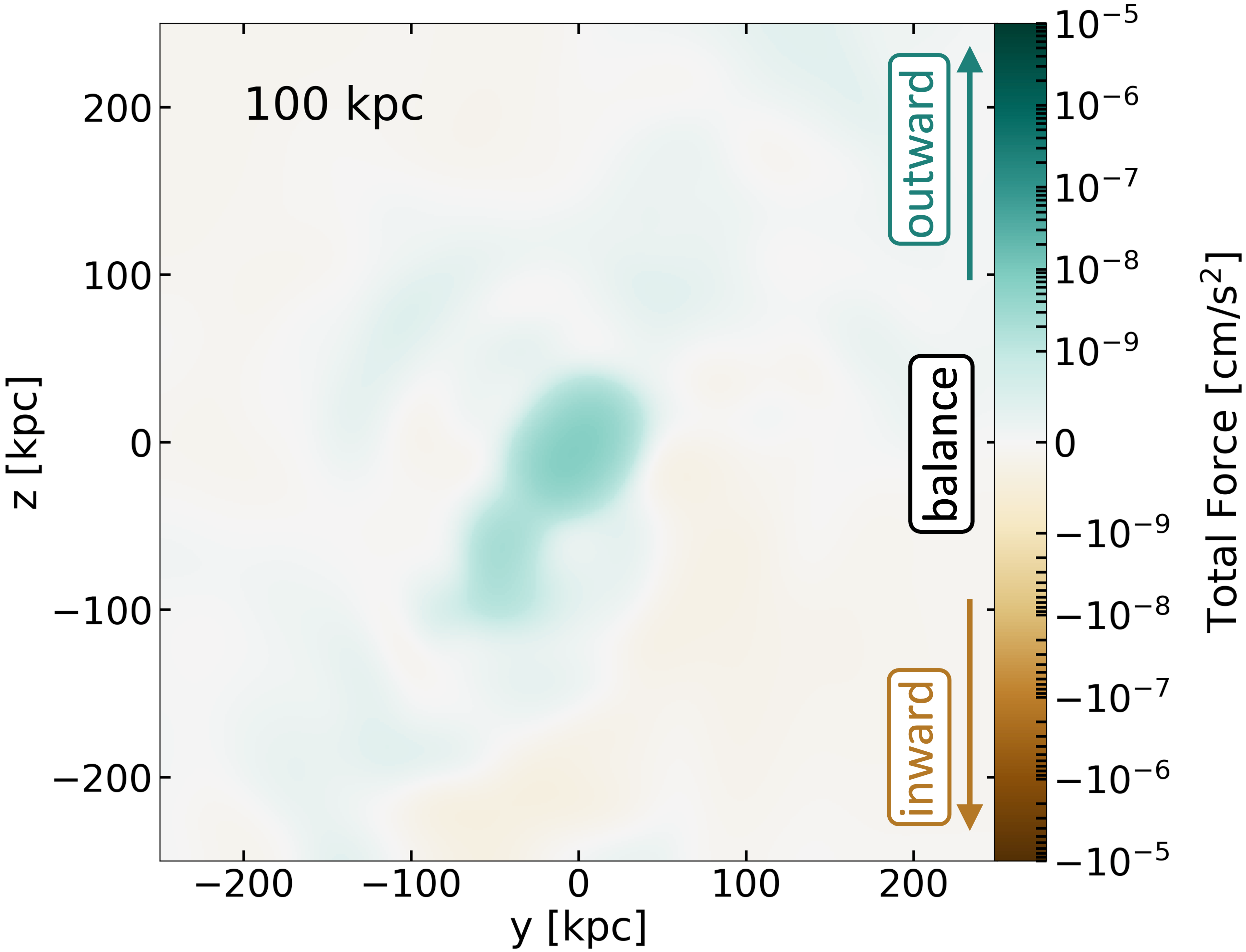}
    \caption{The same slice showing the $F_\mathrm{net}$, the sum of all forces (including gravity), as in Figure~\ref{fig:total_force_slice}, with different sizes of spherical Gaussian smoothing filters applied: 20 kpc, 50 kpc, and 100 kpc, from \textbf{\emph{left}} to \textbf{\emph{right}}. The small-scale variation averages out to near equilibrium (white colors) in the balance of forces directed radially inward and outward, with a weak outward force in the center of the halo near the galaxy.}
    \label{fig:smoothing}
\end{figure*}

Despite the small-scale variation in forces acting on the CGM gas, it appears as though a general force balance equilibrium may hold on larger scales: the inward and outward variation of forces seem as though they could cancel each other out when smoothed over larger scales. Figure~\ref{fig:smoothing} shows what happens to the force variations when they are smoothed on successively larger and larger scales. The small-scale variation (Section~\ref{sec:local_results}) is averaged out and what remains is much closer to an equilibrium in the radial force balance, with a slight positive (outward) force in the center of the halo near the galaxy. We will see in Figure~\ref{fig:support_vs_r_Tempest} that the outward force in the center of the halo is driven by turbulence very close to the galaxy. Figure~\ref{fig:smoothing} encouragingly shows that a balance of the radially inward and outward forces could be recovered by averaging over large scales. In this section, we quantify the degree of this global balance and determine which forces contribute to it most strongly, in the whole CGM (Section~\ref{subsec:support_all}), within CGM sections (Section~\ref{subsec:support_segments}), and in simulations run with different feedback strengths (Section~\ref{subsec:support_feedback}).

\subsection{Support against gravity in whole CGM}
\label{subsec:support_all}

\begin{figure}
    \centering
    \includegraphics[width=\linewidth]{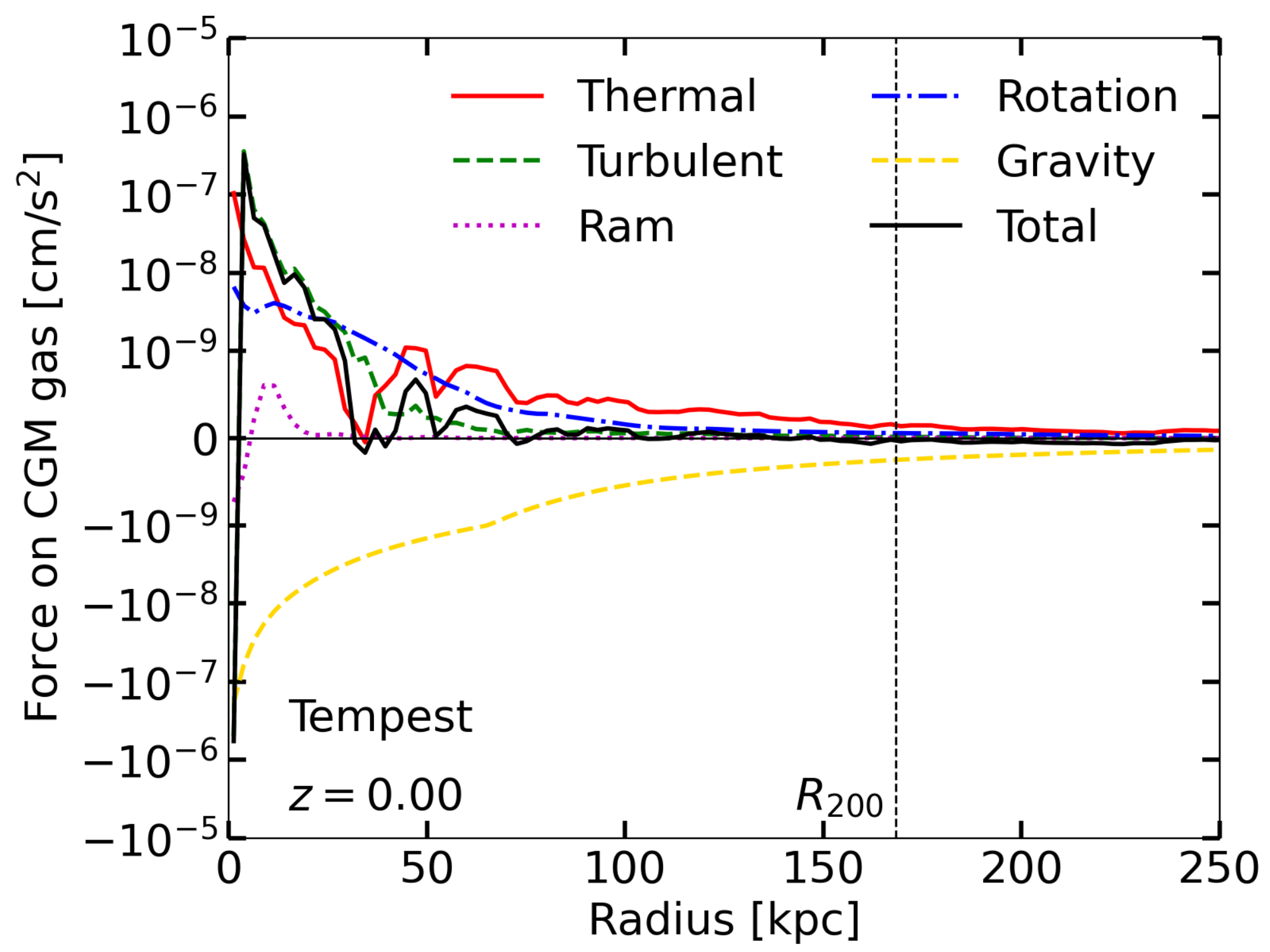}
    \caption{Each type of force (components in Equation~\ref{eq:Fnet}) summed in radial shells (of width $0.015R_{200}$, or $2.5$ kpc at $z=0$ for Tempest) as a function of distance from the galaxy, for the Tempest halo at $z=0$.}
    \label{fig:force_vs_r}
\end{figure}

The simplest and most commonly-done form of averaging is in radial shells. Figure~\ref{fig:force_vs_r} shows each type of force as in Equation~(\ref{eq:Fnet}) as a function of radius within the Tempest halo at $z=0$, the same halo and snapshot shown in all previous figures. The force is calculated by first summing the force (not the acceleration) acting on each cell within each radial shell, then normalizing by the mass of gas within each radial shell (of width $0.15R_{200}$). Most of the different force terms are positive or close to zero, with the exception of gravity (yellow dashed curve), which is negative. Despite large local variation in each force term and in $F_\mathrm{net}$, averaging over radial shells allows $F_\mathrm{net}$ to approach zero in the outer CGM, beyond $\sim60$ kpc. Thus, inward and outward force balance in equilibrium holds in the outer CGM. Within $\sim30$ kpc of the galaxy, $F_\mathrm{net}>0$, indicating outward-directed forces on the whole. The outward-directed $F_\mathrm{net}$ in this inner part of the CGM is primarily driven by strong outward-directed turbulent pressure forces and the outward-directed thermal pressure forces. Close to the FOGGIE galaxies, there are significant kinematics driven by strong shocks as hot outflows escape the galaxies. We defer a detailed investigation of the driving and properties of the turbulence in the FOGGIE CGM to future work.

Because the numerical values of the forces are difficult to interpret in context, and because what we want to investigate is how these forces compare to gravity, we introduce a new term that we call ``support" and that we will use throughout the rest of the paper. It is defined as the ratio of each type of force (each component in Equation~\ref{eq:Fnet}) to the absolute value of the force of gravity. Since gravity is directed inward, its force is negative in our formalism, so we use the absolute value of gravitational force to ensure the support ratio is positive for outward-directed forces and negative for inward-directed forces, for easier understanding. When the support ratio equals one for a particular force component, that indicates that component is exactly balancing against gravity. When the support ratio is between zero and one, that type of force is contributing partially to the support against gravity. If the support ratio is negative, that force is acting in the same direction as gravity (inward), rather than supporting against it. When the support ratio of all non-gravity forces (black curve on all following plots) is close to one, that indicates the CGM is roughly in an equilibrium where the sum of all outward radial forces are balancing gravity. If the support ratio of all non-gravity forces is greater or less than one, that indicates an over- or under-supported CGM, respectively.

\begin{figure*}
    \centering
    \includegraphics[width=0.7\linewidth]{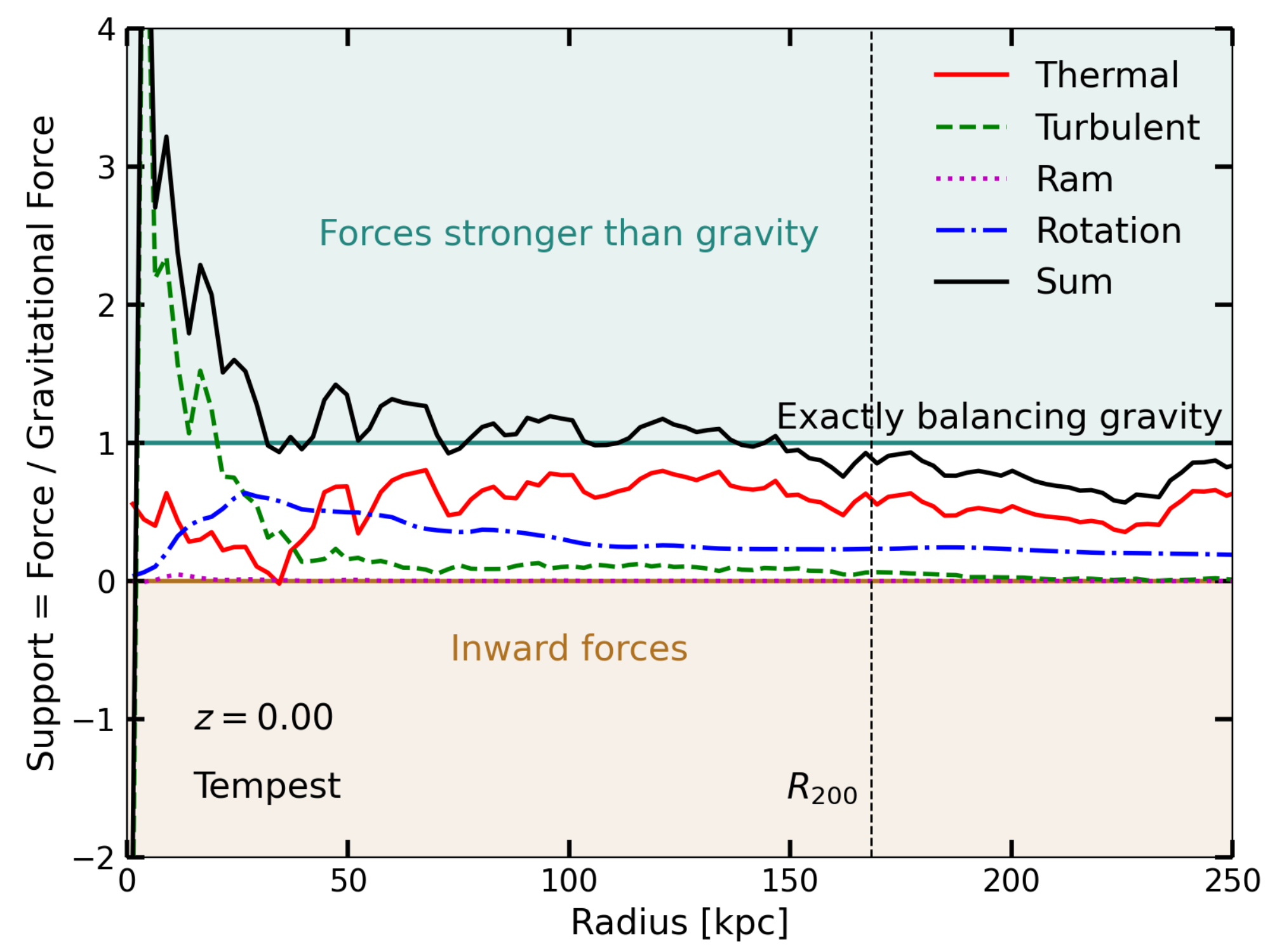}
    \caption{The ratio of each type of force (components in Equation~\ref{eq:Fnet}) to gravity as a function of the distance from the galaxy, for the Tempest halo at $z=0$. The colored curves show the different force types while the black curve shows the sum of all non-gravity forces. The teal horizontal line at a support value of exactly 1 indicates where a force is exactly balancing gravity, and the teal shaded region at support values $>1$ indicates outward-directed forces stronger than the inward pull of gravity. The brown horizontal line at a support value of exactly 0 indicates where a radial force is zero and does not contribute to support against gravity at all. The brown shaded region for support values $<0$ indicates inward-directed forces. The fact that the black curve falls close to a support value of 1 for $40\ \mathrm{kpc}<r<R_{200}$ indicates the halo gas at these radii is in rough radial force balance equilibrium. If the halo gas were in hydrostatic equilibrium, the thermal force (red curve) would lie along a support value of 1. This is not the case, and so the halo gas is not in hydrostatic equilibrium even though approximate force balance equilibrium is achieved. Instead, turbulent and rotation forces also contribute to the overall force balance in this halo.}
    \label{fig:support_vs_r_Tempest}
\end{figure*}

Figure~\ref{fig:support_vs_r_Tempest} shows an annotated example of how each type of force contributes to support, for the Tempest halo at $z=0$. Figure~\ref{fig:support_vs_r} in Appendix~\ref{sec:other_halos} shows the support ratio for each type of force for the other three FOGGIE halos at $z=0$. In the outer CGM, beyond $\sim0.5R_{200}$, there is an overall force balance in $F_\mathrm{net}$ (black solid line), as indicated by the value of total support being approximately one. For most of the FOGGIE galaxies, the main contributor to the support of gas against gravity in the outer CGM is the force exerted by the thermal gas pressure, with some additional support from large-scale coherent rotation. In the inner CGM, within $\sim0.25R_{200}$, the strong turbulent forces drive $F_\mathrm{net}$ to large values and there is an over-abundance of outward forces compared to gravity. For most of the FOGGIE galaxies, there is a point in the inner CGM, at $\sim20$ kpc from the galaxy, where the contributions of thermal, turbulent, and rotation forces contribute roughly equally to the support against gravity. This is also the smallest radius at which rough inward and outward force balance is present. Note that we compute the forces and support of only that gas that survives after performing a cut in density to separate the CGM from the central galaxy's ISM (and any satellite ISM), as described in Section~\ref{sec:methods}, so a net outward force in the CGM gas does not necessarily indicate the galaxies are constantly exploding without ever accreting gas (although note that \emph{all} gas mass is included in the $M_\mathrm{enc}$ term of equation~\ref{eq:Fnet}). We verified that when we calculate forces for all gas without performing this density cut (not shown), we find predominantly negative (inward) forces, mostly driven by the gravitational forces on the large mass of disk gas, indicating the central galaxy is, in fact, accreting material to fuel its growth. We continue with the density cut throughout the rest of the paper so as to focus on the properties of the CGM rather than the central galaxy's ISM, but note that beyond $\sim0.15R_{200}$, the results with and without the density cut are roughly equivalent.

\begin{figure*}
    \centering
    \includegraphics[width=\linewidth]{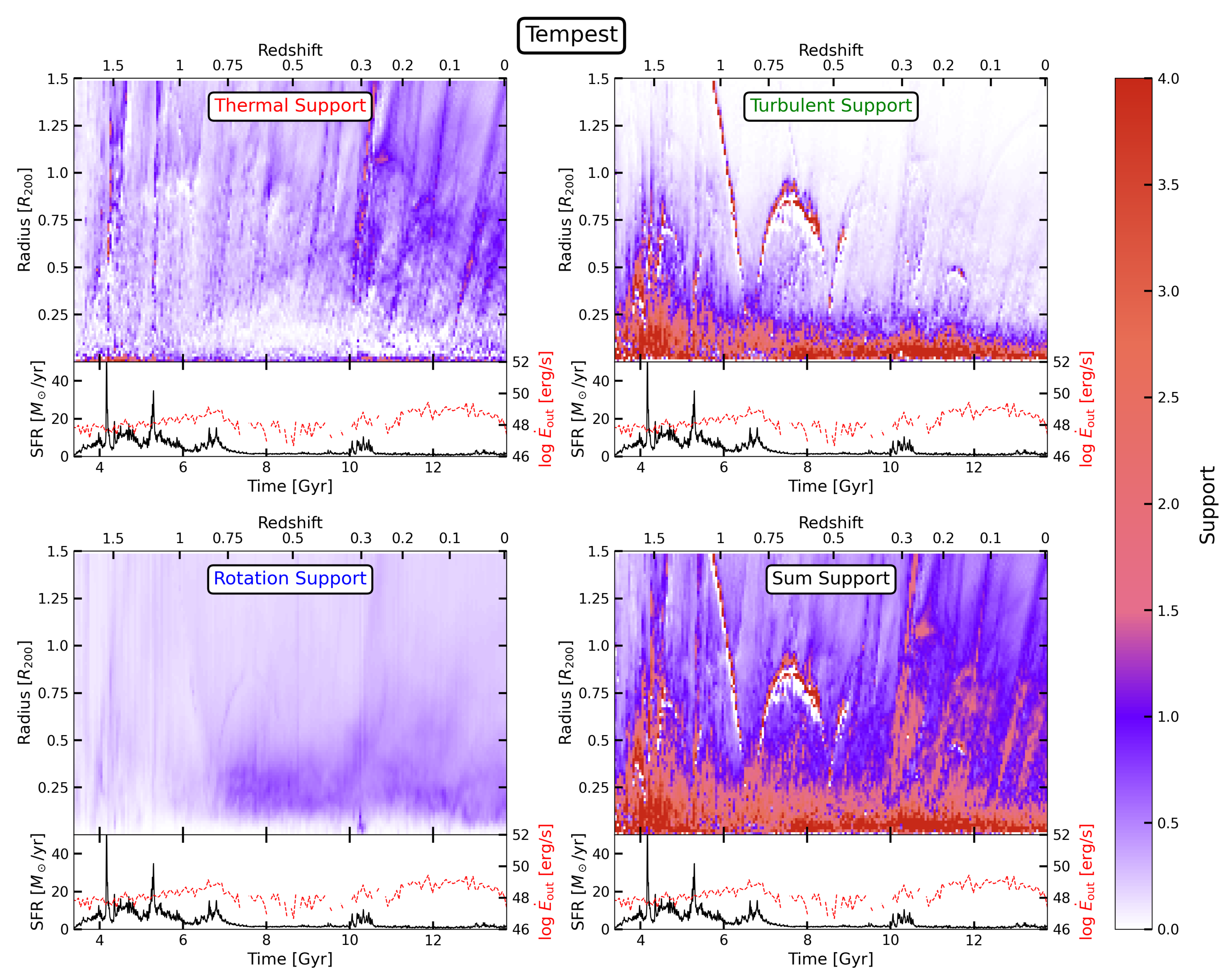}
    \caption{Each 2D panel shows the support against gravity as functions of both time, from $z=2$ to $z=0$, and radius, in the Tempest halo. Each panel shows the support provided by a different type of force as in Equation~(\ref{eq:Fnet}): thermal (\textbf{\emph{top left}}), turbulent (\textbf{\emph{top right}}), and rotational (\textbf{\emph{bottom left}}) support and the sum support of all non-gravity forces (\textbf{\emph{bottom right}}). The color scale is chosen to emphasize fractional contributions to perfect support against gravity in the white-purple colors and over-support provided by outward forces stronger than gravity in the salmon-orange colors. Inward-directed forces that work with gravity are also colored white because they do not contribute to support. Underneath each panel is the SFR of the central galaxy over time as the black curve and the thermal + kinetic energy flowing through a $0.1R_{200}$ sphere centered on the central galaxy as the red dashed curve. Gaps in the $\dot{E}_\mathrm{out}$ curve are where the net energy flux is inward onto the galaxy, and therefore negative and not visible on the log scale.}
    \label{fig:support_vs_t_Tempest}
\end{figure*}

To investigate the time dependence of the forces acting on CGM gas, Figure~\ref{fig:support_vs_t_Tempest} shows 2D plots of the support ratios for each type of force as functions of both time and radius as the Tempest galaxy evolves, from $z=2$ to $z=0$. Each panel shows the support against gravity contributed by a different type of force: thermal force in the top left, turbulent in the top right, rotation in the bottom left, and the sum of all non-gravity forces in the bottom right. The colorbar is chosen to highlight two general regions of support values: support $<1$, corresponding to a fractional contribution to the overall support against gravity, is indicated by white-purple colors, while support $>1$, corresponding to forces stronger than the force of gravity, is indicated by purple-salmon-orange colors. Inward-directed forces that work with gravity, rather than opposing it, are colored white. The star formation rate of the central galaxy, calculated as the mass of new star particles formed within 20 kpc of the center of the halo over the previous 10 Myr, is shown underneath each panel as the black curve. The bottom panels also show $\dot{E}_\mathrm{out}$, the thermal + kinetic energy flux passing through a $0.1R_{200}$ radius sphere centered on the central galaxy as the red dashed curve. The energy is on a log scale, so times when the energy flux is inward, onto the galaxy, are represented by gaps in the curve because the values of the energy flux are negative for these times. In each panel, there are structures that look like a bouncing ball being dropped: coming in from large radii and then ``bouncing" several times before disappearing, most noticeable in the turbulent support (top right) and sum support (bottom right) panels. These are artifacts due to incomplete satellite removal --- the gas surrounding satellites as they spiral in toward the central (which looks like ``bouncing" in this time-radius plot) has disturbed kinematics that appear as strong turbulent support at the location of the satellite. There are also diagonal streaks in each panel that appear to begin mostly following bursts in the star formation rate -- these indicate expanding outflow fronts launched by outflows from the central galaxy.

The inner CGM ($r\lesssim0.5R_{200}$) spends more time out of radial force balance, as indicated by the support of all non-gravity forces (bottom right panel) being further from a value of one for more of the time evolution from $z=2$ to $z=0$ than in the outer CGM ($r\gtrsim0.5R_{200}$). The inner CGM is also more time-variable than the outer CGM. In both the inner and outer CGM, the thermal (top left panel) and turbulent (top right panel) support, as well as the sum support of all non-gravity forces (bottom right panel), increase following strong bursts of star formation in the central galaxy, or when $\dot{E}_\mathrm{out}$ is particularly high. This response to the central galaxy's feedback is stronger in the inner CGM, but it is clear that sufficiently strong bursts can affect the forces acting on the outer CGM gas as well. In general, it is only at lower redshifts ($z\lesssim0.3$) when the star formation history becomes smoother with fewer large bursts, that the inward and outward forces roughly balance in either the inner or outer CGM, as indicated by the sum support of all non-gravity forces (bottom right panel) approaching a value of one with fewer large deviations. In the inner CGM, turbulent support is strong at high redshift and slowly decreases over time, while thermal support starts low and slowly increases over time. Rotational support stays fairly constant or perhaps increases slightly over time, and is somewhat stronger in the inner CGM. In the outer CGM, it is only at high redshift that turbulent support is non-negligible and roughly equal to thermal and rotational support. Over time, the turbulent support in the outer CGM drops off while the thermal support continues to build. The other three FOGGIE halos follow similar trends, and plots analogous to Figure~\ref{fig:support_vs_t_Tempest} for them can be found in Appendix~\ref{sec:other_halos}.

To reduce the impact of time-dependent events on the interpretation of force balance within the CGM of the FOGGIE galaxies, Figures~\ref{fig:support_t_avg_Tempest} and~\ref{fig:support_t_avg} (in Appendix~\ref{sec:other_halos}) show the median support against gravity across all time snapshots for the four FOGGIE galaxies, as functions of distance from the central galaxy. The radius coordinate on the horizontal axis of these plots is normalized by each galaxy's virial radius, $R_{200}$, so as not to confuse any trends with the growing size of the halo over time. Figure~\ref{fig:support_t_avg_Tempest} shows each force type in its own panel, with annotations as in Figure~\ref{fig:support_vs_r_Tempest}, to help orient the reader. Figure~\ref{fig:support_t_avg} in Appendix~\ref{sec:other_halos} combines each force type into a single panel for the other three FOGGIE halos, where we see the same qualitative trends as in the Tempest halo. We computed the radial functions of forces and ratios of forces to gravity for every time snapshot between $z=2$ and $z=0$, separated by $\sim50$ Myr each, and found the median with equal weighting to each time snapshot. The colored shading around each curve indicates the interquartile range in values between 25\% and 75\% of the time variation.

\begin{figure*}
    \centering
    \includegraphics[width=0.8\linewidth]{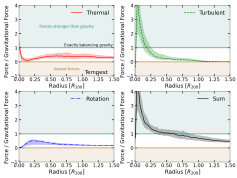}
    \caption{The median of all time snapshots between $z=2$ and $z=0$ of the ratio of each force (components in Equation~\ref{eq:Fnet}) to gravity as a function of the distance from the galaxy normalized by virial radius, for the Tempest halo. Each force type is shown in its own panel, and we have neglected the ram pressure force because its value is nearly zero everywhere. Curves indicate median values while shading around curves indicates the $25\%-75\%$ interquartile range of the time variation. The teal horizontal line at support values of exactly 1 indicates where forces would exactly balance the inward pull of gravity, and the teal shaded region at support values $>1$ indicates where outward-directed forces are stronger than the force of gravity. The brown horizontal line at support values of exactly 0 indicates where forces do not contribute to the support against gravity. The brown shaded region at support values $<0$ indicates where forces are inward-directed.}
    \label{fig:support_t_avg_Tempest}
\end{figure*}

The trends of Figure~\ref{fig:support_vs_r_Tempest} become clearer when considering the median of all time snapshots: the outer CGM, with $0.5R_{200} \lesssim r\lesssim R_{200}$, is roughly in force balance between radially inward and outward forces, while the inner CGM with $r\lesssim0.5R_{200}$ is dominated by outward-directed forces mostly driven by large turbulent forces close to the central galaxy. The shading indicates more time variance in the turbulent and thermal forces than in the rotation force. The spread in time in the turbulent and thermal forces leads to a spread in $F_\mathrm{net}$ and in the combined total support, which, in turn, indicates there are some times when the inward and outward forces in a galaxy's CGM are significantly unbalanced.

The results of this section put forward a picture in which the CGM of high-redshift galaxies is chaotic, dominated by small-scale velocity motions, and not necessarily in any sort of equilibrium balance with the force of gravity. Over time, as star formation becomes smoother and less bursty and the central galaxies grow, the outer CGM approaches an equilibrium force balance and shifts to being primarily supported by thermal forces, although large-scale rotation provides $25-50\%$ of the support against gravity at any given time. The inner CGM remains dominated by dynamic, non-thermal forces even at low redshift. This picture is consistent with the one put forward by \citet{Lochhaas2021}, wherein the outer CGM of these simulated galaxies approached virial equilibrium only at low redshift, and even then, dynamic gas motions still contribute to the equilibrium in addition to the thermal properties of the gas.

\subsection{Support against gravity in CGM segments}
\label{subsec:support_segments}

The drastic local variation and hints that different CGM segments are home to different forces in Section~\ref{sec:local_results} suggests that the radial averaging of forces and support in Section~\ref{subsec:support_all} is not telling the full story of what supports the CGM gas against gravity. In this section, we examine the support in the outflow, inflow, and no strong flow CGM segments (see Section~\ref{subsec:segmenting}) to determine if different processes are responsible for force balance (or non-balance) in these regions.

\begin{figure*}
    \centering
    \includegraphics[width=0.32\linewidth]{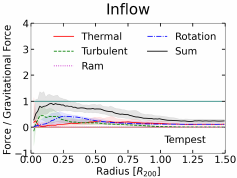}
    \includegraphics[width=0.32\linewidth]{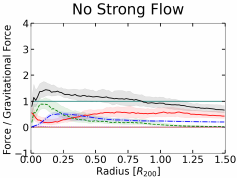}
    \includegraphics[width=0.32\linewidth]{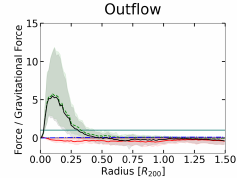}
    \caption{The median of all time snapshots from $z=2$ to $z=0$ of each type of force as in Equation~(\ref{eq:Fnet}) as a function of distance from the galaxy, similar to Fig.~\ref{fig:support_t_avg_Tempest}, for the Tempest halo, within the three CGM segments as defined by cuts in gas metallicity: inflow (\textbf{\emph{left}}), no strong flow (\textbf{\emph{center}}), and outflow (\textbf{\emph{right}}). Note that the outflow panel (right) has a different $y$-axis scale than the other two panels.}
    \label{fig:support_segments_Tempest}
\end{figure*}

Figure~\ref{fig:support_segments_Tempest} shows the median support of CGM gas against gravity across all time snapshots between $z=2$ and $z=0$, as functions of halo radius relative to the time-evolving virial radius, within the inflow (left), no strong flow (center), and outflow (right) CGM segments, for the Tempest halo. This figure is similar to Figure~\ref{fig:support_t_avg_Tempest}, but with all forces shown in each panel and the teal and brown shading removed for clarity. The shading around each curve shows the $25\%-75\%$ interquartile range of the time variation. The inflow, outflow, and no strong flow segments are defined using a metallicity cut of $Z<0.01Z_\odot$ for inflow, $Z>1Z_\odot$ for outflow, and $0.01Z_\odot<Z<1Z_\odot$ for no strong flow (see Section~\ref{subsec:segmenting}). Figure~\ref{fig:support_segments} in Appendix~\ref{sec:other_halos} shows the support in the CGM segments for the other three FOGGIE halos, which again show the same qualitative trends as in the Tempest halo discussed below.

The forces acting on the gas in the inflow segment (left panel) are generally weaker than the forces acting on the other segments, as shown by each curve in Figure~\ref{fig:support_segments_Tempest} falling at or below a support value of one (and a support value $<1$ is expected for gas that is primarily inflowing). The inflow exhibits roughly equal force contributions from thermal, turbulent, and rotational forces, or perhaps slightly less important turbulent forces (especially in the other three halos; see Figure~\ref{fig:support_segments} in Appendix~\ref{sec:other_halos}). The outflow segments show a drastically different story: the total cumulative support of all forces against gravity is significantly above a value of one for $r<0.5R_{200}$, so much so that the right panel in Figure~\ref{fig:support_segments_Tempest} is on a different scale than the other panels. The dominant force is turbulence, with very little contribution from rotation or (somewhat surprisingly) ram pressure, and a negative force (indicating inward, in the same direction as gravity) provided by the thermal pressure gradient. A support value $<1$ for the inflow segment and $>1$ for the outflow segment are expected: a perfect force balance between inward and outward forces is expected to hold gas in place, not drive large-scale inflows or outflows. In addition, the time variation (indicated by the shading around the curves) is much larger in the outflow segment than the other two segments. Above, we saw variation in the value of support appears to be driven by bursts in the SFR (see Figure~\ref{fig:support_vs_t_Tempest}), so it is not surprising that the segment that most closely probes the direct outflowing material driven by star formation feedback has a large variation in time.

The thermal force within the outflowing material varies strongly between inward and outward (Figure~\ref{fig:force_rays}), but when summing the force within the outflow segment and then finding the median over significant spans of time, the thermal force is predominantly inward (Figure~\ref{fig:support_segments_Tempest}). This is somewhat counter-intuitive, as we might expect the hot outflows are driven outward (not inward) by their thermal pressure. There are significant shocks within the outflow region, with Mach numbers ranging from $\sim1$ to $\sim10$, and the thermal pressure rises significantly across shocks within the flow. This leads to a positive thermal pressure gradient across shocks and therefore a negative (inward) force exerted by the thermal pressure. While the large-scale thermal pressure gradient within the outflow does weakly decline with distance from the galaxy as expected (e.g., Figure~\ref{fig:pressure_slices}), the handful of strong shocks within this segment cause the inward force to win out when summed over space and time. This is expected from classical models of shocked wind bubbles \citep[e.g.][]{Weaver1977}: while the forward shock of the wind on the surrounding material would produce a negative pressure gradient, we have selected the outflow region to contain only wind material, so this forward shock is not captured in this CGM segment. Instead, the shocks we identify within the outflow segment may be reverse shocks, across which the pressure gradient is strongly positive, thus producing an inward-directed force.

Finally, the support of the gas in the no strong flow segment, which is not directly participating in coherent inflows or outflows, indicates a nearly perfect force balance within the virial radius for the four FOGGIE galaxies. Values of the sum total support (black curve) slightly greater than one in the inner CGM may be indicating that the metallicity cut for the outflows is slightly too aggressive, such that it is missing some outflowing material that instead falls into the no strong flow segment. Overall, the no strong flow segment seems to be capturing the material in the CGM that is closest to a force balance, with thermal support dominating slightly in most of the halos followed by roughly equal amounts of turbulent and rotational support. The no strong flow segment dominates both the mass and volume of the gas in the CGM (see Figure~\ref{fig:metallicity_pdf}), so this segment describes the force balance of the bulk of the CGM. In the very inner regions of the CGM ($r<0.2R_{200}$), turbulent support dominates or is roughly equal to thermal and rotational support. Beyond $\sim0.8-0.9R_{200}$, the no strong flow segment is under-supported, perhaps indicating inflow onto the halo on average. This inflow must be somewhat metal-enriched, as all gas in this segment has $Z>0.01Z_\odot$, so this may be a signature of enriched gas recycling back onto the outskirts of the halo, or of fresh inflow mixing up to a higher metallicity even in the very outskirts of the halo.

In summary, evaluating the support against gravity of gas in the different CGM segments reveals that these cuts are correctly selecting under-supported inflow, over-supported outflow, and nearly perfectly-balanced intermediate, volume-filling material with no strong bulk flows. What little support the inflow segment does have is provided nearly equally by thermal, turbulent, and rotational forces, with centrifugal rotational support slightly dominating, while the outflow segment is driven mostly by turbulence. The no strong flow segment has a nearly equal contribution from thermal, turbulent, and rotational forces in the inner CGM and slightly dominating thermal force in the outskirts of the halo.

\subsection{How feedback affects the support against gravity}
\label{subsec:support_feedback}

\begin{figure*}
    \centering
    \includegraphics[width=\linewidth]{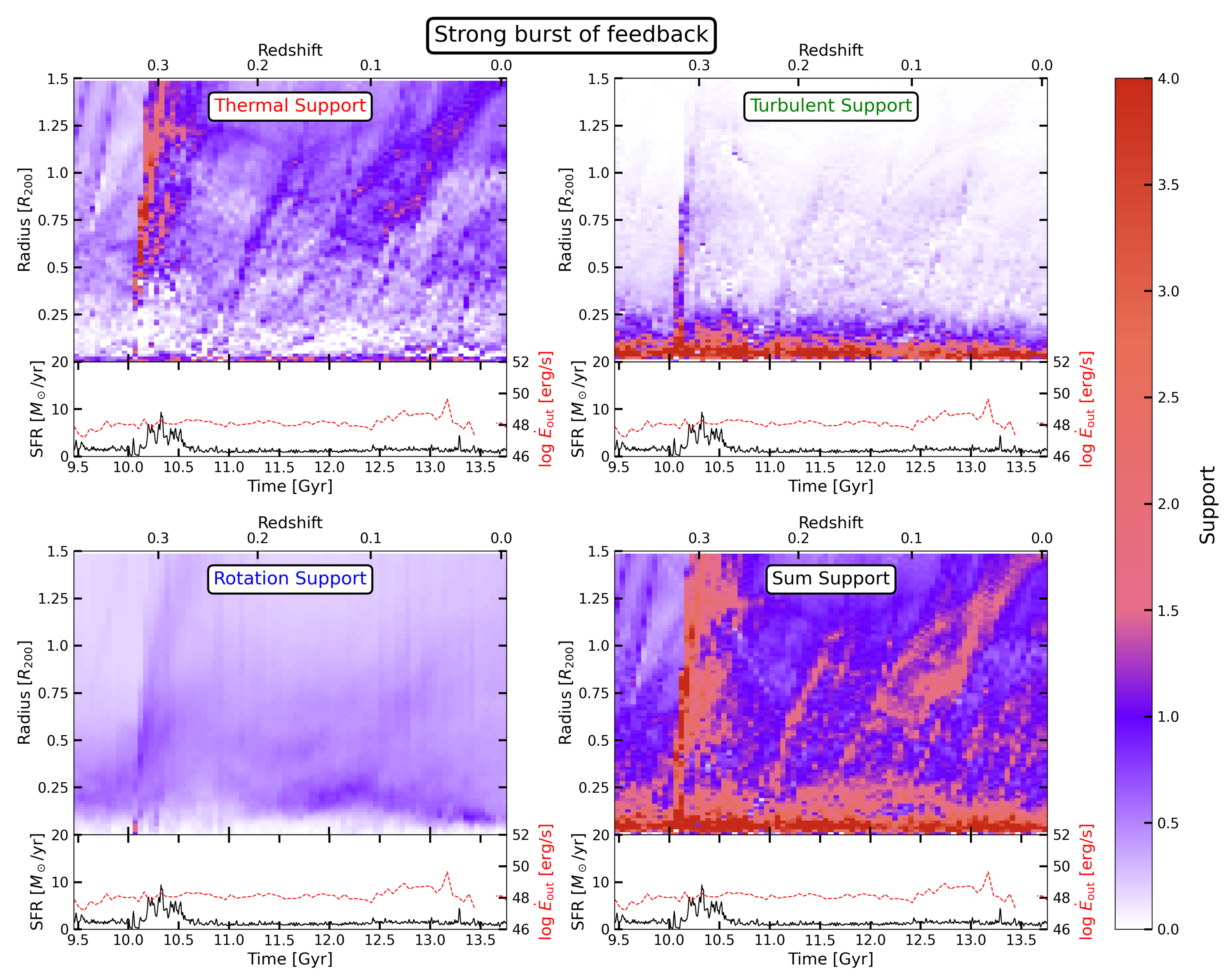}
    \caption{As in Figure~\ref{fig:support_vs_t_Tempest}, each 2D panel shows the support against gravity as functions of both time, from $z=0.4$ to $z=0$, and radius, in the strong burst re-run of Tempest (see text). Each panel shows the support provided by a different type of force as in Equation~(\ref{eq:Fnet}): thermal (\textbf{\emph{top left}}), turbulent (\textbf{\emph{top right}}), and rotational (\textbf{\emph{bottom left}}) support and the sum support of all non-gravity forces (\textbf{\emph{bottom right}}). The color scale is chosen to emphasize fractional contributions to perfect support against gravity in the white-purple colors and over-support provided by outward forces stronger than gravity in the salmon-orange colors. Underneath each panel is the SFR of the central galaxy over time. The increased feedback strength occurs in a 50 Myr burst starting at 10 Gyr.}
    \label{fig:support_feedback_strong}
\end{figure*}

To explore the effect of feedback strength on the support of CGM gas against gravity, we re-started the Tempest halo at low redshift with different values of the Enzo simulation parameter that controls the efficiency of converting supernova energy into feedback energy. This parameter sets how much of the stellar mass formed in a time step is converted into thermal feedback energy. For a stronger feedback run, we identified a burst in the star formation at $z\sim0.3$ in the fiducial Tempest simulation, changed the feedback parameter to $10^{-4}$, and restarted Tempest from just before this feedback burst. After 100 Myr, when the burst has subsided, we turned the feedback parameter back down to the fiducial $10^{-5}$ and continue running to $z=0$. For a low feedback run, we changed the feedback parameter from its fiducial value of $10^{-5}$ to $10^{-6}$, and ran Tempest with this lower feedback strength from $z=0.4$ to $z=0$ (starting at a higher redshift well before the burst in the fiducial run to avoid confusion with the burst). The strong feedback re-run of Tempest immediately blows out a significant amount of hot material, which drastically increases the computational run time, so setting the feedback strength parameter back down to its fiducial value after the burst of star formation is necessary to reduce computational expense.

\begin{figure*}
    \centering
    \includegraphics[width=\linewidth]{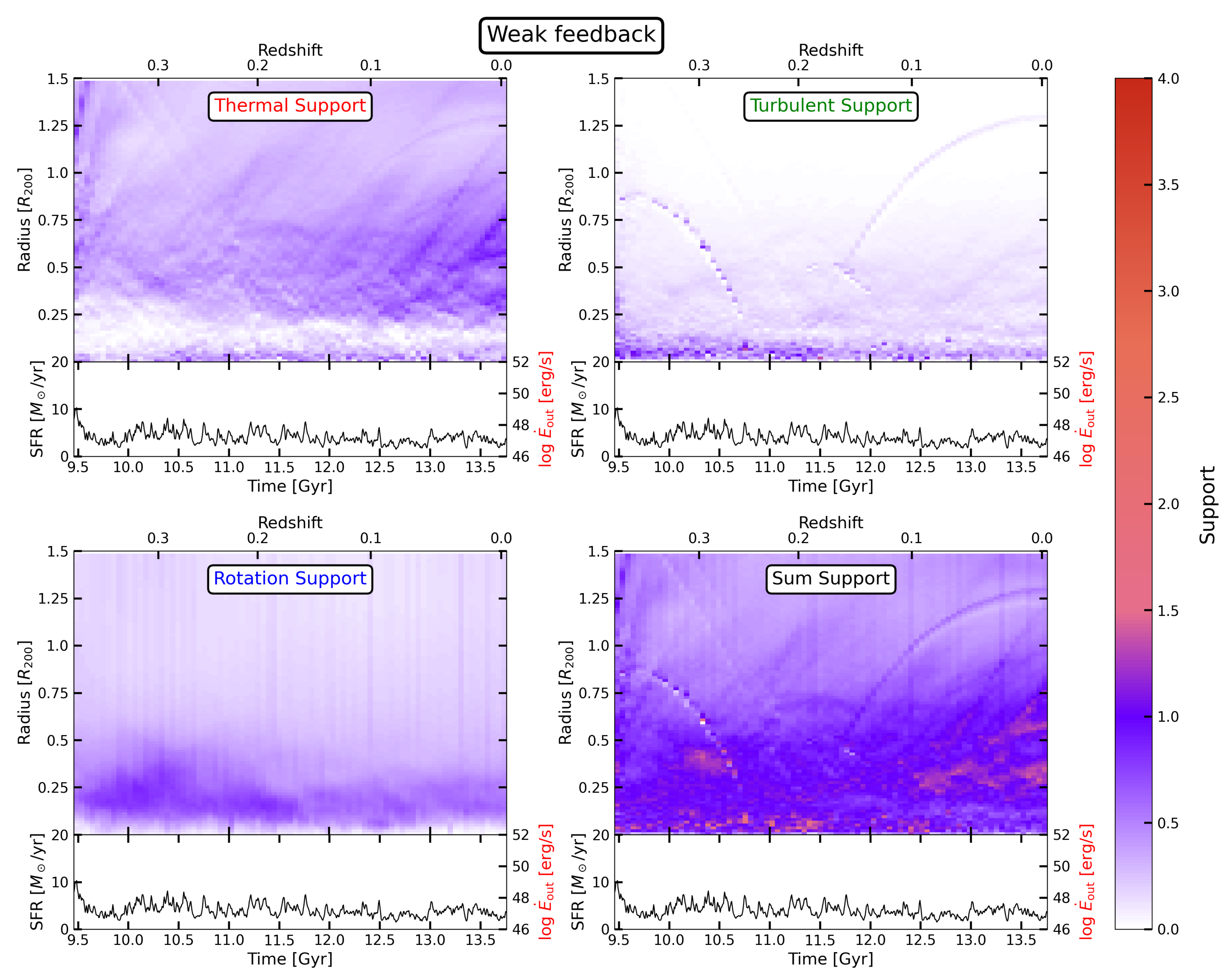}
    \caption{As in Figure~\ref{fig:support_vs_t_Tempest}, each 2D panel shows the support against gravity as functions of both time, from $z=0.4$ to $z=0$, and radius, in the weak feedback re-run of Tempest (see text). Each panel shows the support provided by a different type of force as in Equation~(\ref{eq:Fnet}): thermal (\textbf{\emph{top left}}), turbulent (\textbf{\emph{top right}}), and rotational (\textbf{\emph{bottom left}}) support and the sum support of all non-gravity forces (\textbf{\emph{bottom right}}). The color scale is chosen to emphasize fractional contributions to perfect support against gravity in the white-purple colors and over-support provided by outward forces stronger than gravity in the salmon-orange colors. Underneath each panel is the SFR of the central galaxy over time. The decreased feedback strength is present throughout the full time range shown.}
    \label{fig:support_feedback_weak}
\end{figure*}

Figures~\ref{fig:support_feedback_strong} and~\ref{fig:support_feedback_weak} show the support of CGM gas against gravity over time from $z=0.4$ to $z=0$ (bottom axis) and radius (left axis), for the strong burst run and the weak feedback run, respectively. These figures are analogous to Figure~\ref{fig:support_vs_t_Tempest}. When the feedback energy is increased for a short burst of time at $\sim10$ Gyr (Fig.~\ref{fig:support_feedback_strong}), the thermal and turbulent forces briefly become very strong, with the turbulent force dominating the inner CGM and the thermal force dominating the outer CGM following the burst. The increase in thermal support at large distances from the galaxy when the feedback strength is increased could be indicating that the turbulence within the outflows is dissipating into heat far from the galaxy, or could be an indication of direct heating by strong feedback. The sum support of all non-gravity forces remains elevated at values $>1$ everywhere within the CGM for $\sim500$ Myr following the strong burst. The elevated support level dissipates faster than the sound crossing time out to $R_{200}$, which is $\sim2$ Gyr.

In the weaker feedback re-start (Fig.~\ref{fig:support_feedback_weak}), the strong over-support in the inner parts of the CGM seen in the fiducial Tempest disappears, and instead the inner CGM is in nearly perfect force balance (sum of all non-gravity forces in bottom right panel has a value near one). The rotational support is stronger in the inner CGM to partially make up for the drastic decrease in turbulent support when the feedback strength is ``turned down," while the thermal support in the inner CGM is similar to the fiducial feedback run. Near the outskirts of the halo, the CGM is under-supported and thus likely dominated by inflows when the feedback is weaker. The outer halo's degree of turbulent and rotational support in the weak feedback run is similar to the fiducial case, but the thermal support in the outer halo is weaker, bringing the sum support of all non-gravity forces to values $<1$ indicating under-support. This seems to indicate that star formation feedback has two effects on the support of CGM gas: it creates over-support in the inner CGM by stirring up significant turbulence that can be elevated for $\sim500$ Myr after a burst of feedback, and it brings the outer CGM into force balance by providing thermal support at large scales far from the galaxy, persisting for at least the $\sim4$ Gyr over which the strong burst and weak feedback re-simulations were run. The importance of feedback strength in setting the force balance and support against gravity of CGM gas suggests that the physics of the CGM is intimately linked to the processes occurring in the central galaxy, rather than (only) being driven by the halo mass. This result is likely dependent on the implementation of feedback in FOGGIE, and different feedback schemes used by other simulations may change the relative contributions of each force to the support of the CGM against gravity. In particular, feedback from active galactic nuclei (AGN), which is not currently implemented in the FOGGIE simulations, may leave a drastically different signature on the force balance of CGM gas and is an interesting avenue for future study.

\section{Discussion}
\label{sec:discussion}

Sections~\ref{sec:local_results} and~\ref{sec:global_results} show that there are significant variations in the forces acting on the gas in the CGM, both spatially and temporally, such that any given parcel of gas is typically not in an equilibrium of inward and outward force balance at any given time. Despite this, summing the forces acting on gas over large scales in space and averaging over time recovers an inward-outward force balance in the outer CGM, but this balance still is not described by hydrostatic equilibrium: forces exerted by turbulence and rotation contribute to the support against gravity in addition to the expected thermal pressure force. The following subsections describe various implications of these results.

\subsection{Importance of simulation resolution}
\label{subsec:turb_res}

The strong, small-scale variation in the forces acting on CGM gas presented in Section~\ref{sec:local_results} suggests that the spatial resolution of the simulation may be important to capture this variation. In particular, Figure~\ref{fig:force_rays} indicates most forces switch from inward to outward on a typical scale of $\sim5$ kpc when the simulation spatial resolution is $\sim1$ kpc (at $z=0$). When the physical scale of the simulation resolution is improved by a factor of two at $z\approx1.5$, the scale of the variations becomes even smaller, by a factor of two. It is possible that variations on even smaller scales could be present, if those smaller scales could be resolved.

Even disregarding the small-scale variation to focus on the global force balance, high spatial resolution is required to accurately calculate the turbulent velocities to obtain the force exerted by the turbulent pressure. Because we define turbulence as a velocity dispersion on small spatial scales, being unable to resolve small-scale structure in the velocity field would reduce the importance of turbulence. This is especially important for the outflow CGM segment, which Figure~\ref{fig:support_segments_Tempest} shows has the strongest turbulent force.

\begin{figure}
    \centering
    \includegraphics[width=\linewidth]{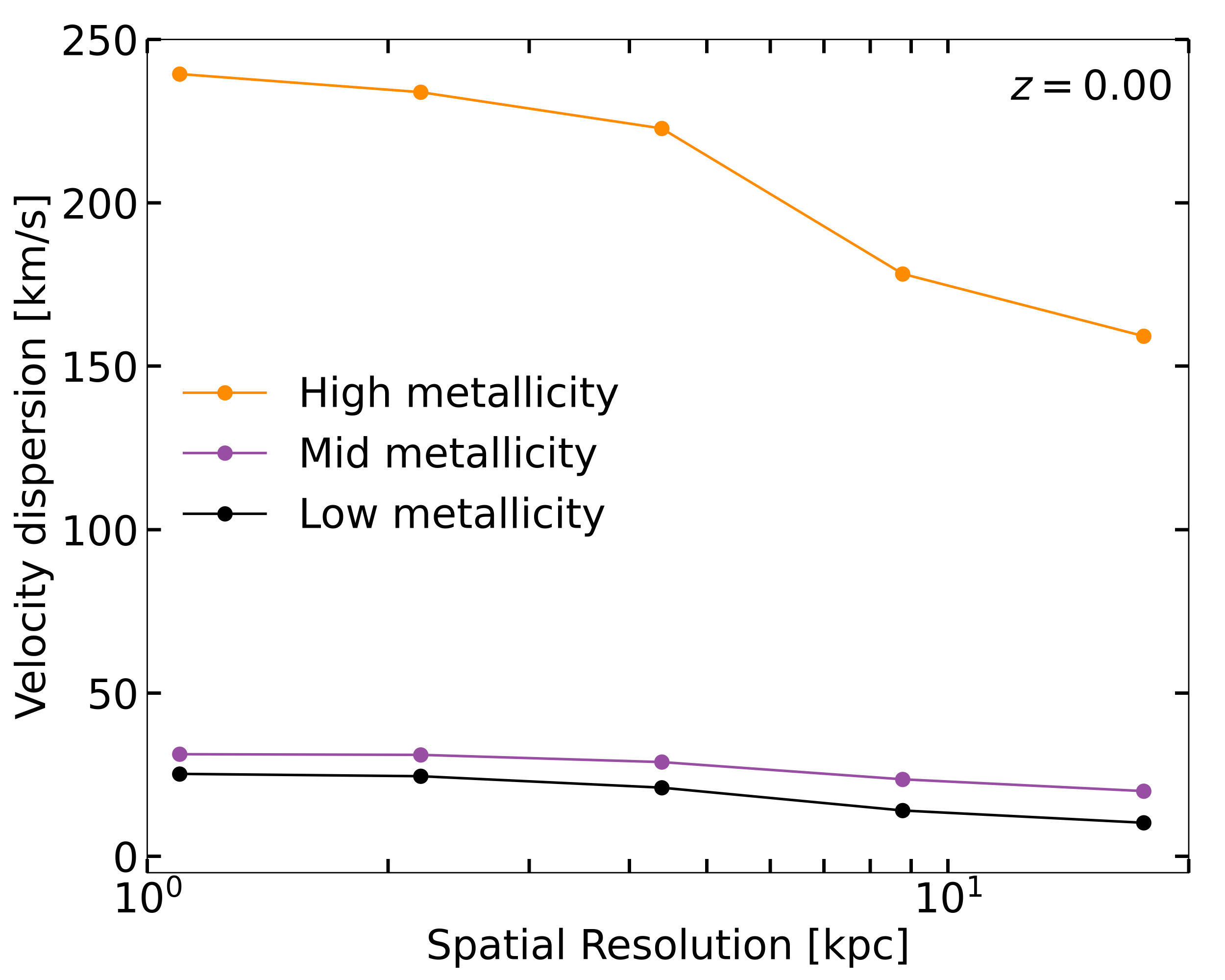}
    \caption{The velocity dispersion (equation~\ref{eq:vdisp}) as a function of simulation spatial resolution, computed as artificially degrading the resolution of the fiducial Tempest halo at $z=0$ by binning cells into larger and larger bins. Each curve shows the average velocity dispersion-resolution relation for a different CGM segment. The high metallicity segment traces galactic outflows, which have the largest average velocity dispersion and also the strongest dependence of velocity dispersion on spatial resolution.}
    \label{fig:vdisp_spatial_res}
\end{figure}

Figure~\ref{fig:vdisp_spatial_res} shows how the average velocity dispersion (equation~\ref{eq:vdisp}) within each CGM segment depends on the spatial resolution of the simulation, for the Tempest halo at $z=0$. We artificially degraded the resolution by binning cells in the fiducial simulation into larger and larger spatial bins and recalculating the velocity dispersion. Note that this method may not be equivalent to running a lower resolution simulation; any effects that the high resolution of the FOGGIE simulations has on the properties of the CGM or the galaxy are still present where they would not be in a simulation run at lower resolution. Figure~\ref{fig:vdisp_spatial_res} shows that the average velocity dispersion of CGM gas increases as the simulation cell size decreases, i.e., as the spatial resolution is improved. The increase is modest for the low and mid metallicity segments, but is drastic for the high metallicity (outflow) segment, where the velocity dispersion is highest. The velocity dispersion in the outflow segment continues to increase with improving resolution to the limit of the resolution in the simulation. This may indicate that even at the high resolution of the fiducial FOGGIE simulations, the turbulence in the outflow is under-resolved.

\begin{figure}
    \centering
    \includegraphics[width=\linewidth]{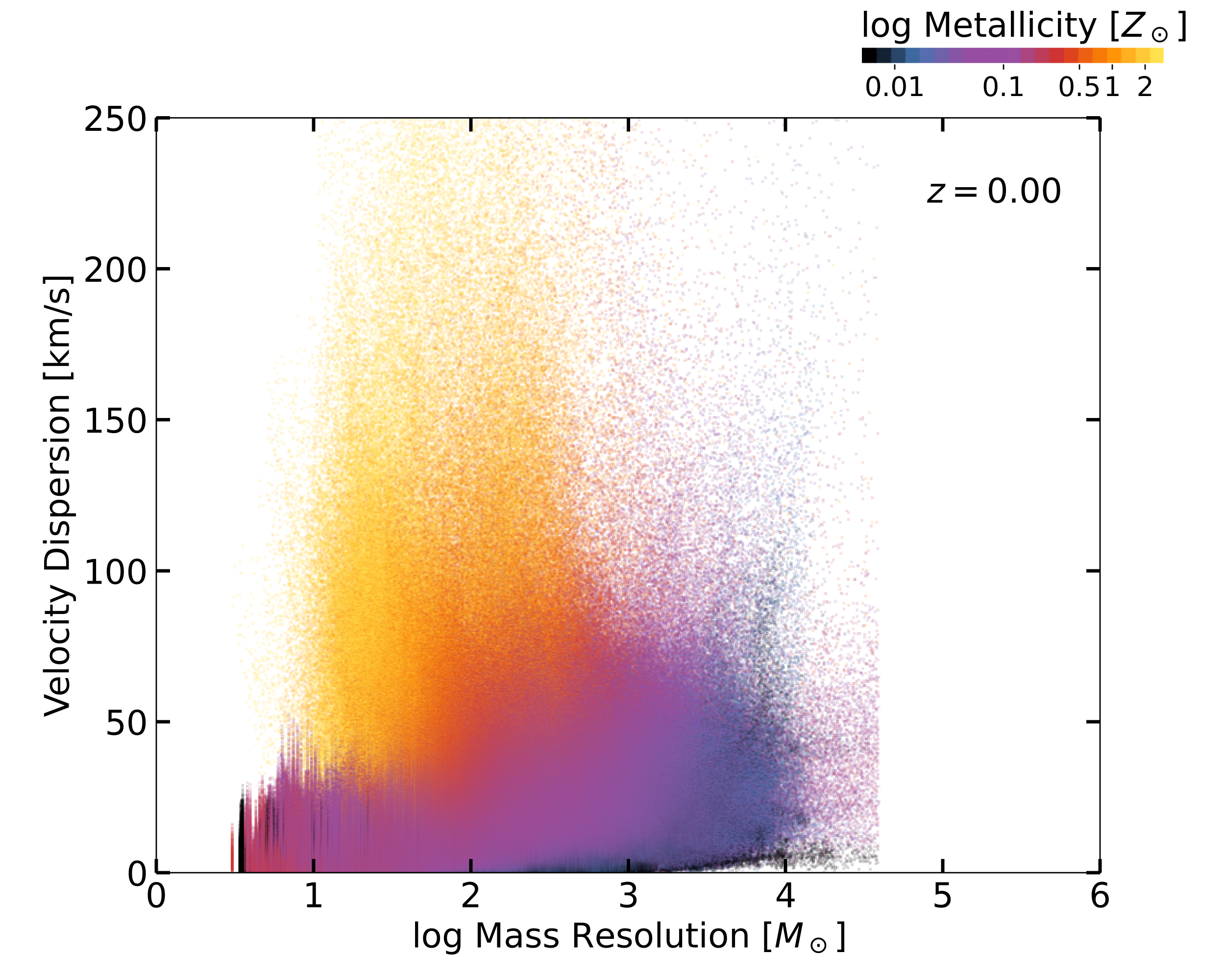}
    \caption{The velocity dispersion of each cell (equation~\ref{eq:vdisp}) in the fiducial Tempest halo's CGM at $z=0$ vs. the mass of gas in each cell, color-coded by the metallicity of the cell. The high metallicity cells have the largest velocity dispersion and some of the smallest cell masses, suggesting a need for very high mass resolution in order to capture the velocity dispersion in these outflow segments.}
    \label{fig:vdisp_mass_res}
\end{figure}

Many cosmological simulations are particle-based, rather than grid-based like FOGGIE \citep[e.g.,][]{Schaye2015,Hopkins2018,Pillepich2018}. This means that a more direct resolution to discuss for comparison to these simulations is the \emph{mass} resolution, rather than the spatial resolution. Figure~\ref{fig:vdisp_mass_res} shows the cell-centered velocity dispersion of the smoothing region around each cell (equation~\ref{eq:vdisp}) in the CGM of the Tempest halo at $z=0$ compared to the mass contained within that cell, color-coded by the metallicity of the cell. The high metallicity (outflow) segment is yellow-orange in color, the low metallicity (inflow) segment is black, and the intermediate metallicity segment is blue-purple-red. The velocity dispersion, and thus turbulent force, is clearly largest in the outflow segment. This segment also has cell masses of $10-300 M_\odot$, due to the gas in this segment being hot and diffuse. However, even some of the mid-metallicity segment (the no strong flow segment) has large velocity dispersions and small cell masses. These masses in each cell are smaller than the mass resolution of any current particle-based cosmological simulation (which typically have mass resolutions of $\sim10^4-10^5M_\odot$), suggesting that particle-based simulations are necessarily missing some of the high velocity dispersion material due to poor mass resolution, and thus also underestimating the effect of turbulence in outflow regions of the CGM.

Furthermore, simulations that resolve a larger dynamic range due to increased spatial resolution have higher effective Reynolds numbers, which means they have a lower numerical viscosity (which is the only type of viscosity at play in these calculations, as they do not include explicit viscous terms in the Euler equations).  This is a quantity that can be directly measured in idealized turbulence simulations (see, e.g., Table 1 and Figure 4 of \citeauthor{Grete2022} \citeyear{Grete2022}, as well as \citeauthor{Kritsuk2011} \citeyear{Kritsuk2011}).  Turbulence with larger Reynolds numbers takes longer to dissipate into thermal energy, and thus for comparable amounts of driving energy one would expect a better-resolved simulation to have proportionally more turbulent support. Similarly, simulations with higher-order numerical methods will also generally have effectively smaller numerical dissipation at comparable physical resolutions, and thus proportionally more turbulent support \citep[in particular, see Section 6 of][for a detailed discussion of this]{Kritsuk2011}.

To test the impact of resolution on the importance of turbulent support directly, we re-calculated the turbulent support as a function of radius for the Tempest halo at $z=0$ (as in Figure~\ref{fig:support_vs_r_Tempest}) at a lower resolution by down-sampling the simulation. We sampled the simulation at a resolution of $\sim4$ kpc, a factor of $\sim4$ worse than the fiducial simulation resolution. This lower resolution corresponds to an average mass per resolution element of $\sim10^4 M_\odot$, comparable to other cosmological zoom-in simulations. At this lower resolution, the amount of turbulent support dropped by a factor of 5 in the inner 50 kpc of the CGM. In the outer CGM, the turbulent support dropped from $\sim10\%$ of the total support against gravity to $\sim5\%$. This suggests that a lower-resolution simulation under-estimates the degree of turbulent support in the CGM, especially in the inner regions, but even the outer CGM's turbulent support would be under-estimated by a factor of two.

\subsection{Local force variation, absorption line observations, and small-scale turbulence simulations}
\label{subsec:obs}

One of the most common ways to observe the CGM is in absorption toward bright background light sources, such as a background quasar \citep[e.g.,][]{Wakker2009,Rudie2012,Werk2013,Stocke2013,Bordoloi2014,Lehner2015,Borthakur2016,Heckman2017,Keeney2017,Chen2018,Berg2018,Berg2019,Rudie2019,Chen2020,Lehner2020,Haislmaier2021}. In this method, the line of sight to the quasar is a 1D probe of the foreground galaxy's CGM, and the spatial location and galaxy association of any absorbers in the spectrum is estimated by assuming the velocity shift of the absorber must be similar to the systemic velocity of the nearby galaxy. Due to the point-like nature of a background quasar, the area probed transverse to the line of sight is very small. Coupled with generally small velocity widths for low ionization state absorbers \citep{Werk2014,Rudie2019}, the typical assumption is that the parcel of gas responsible for the observed absorption is small compared to the path length of the line of sight. In addition, photoionization modeling of low-ionization state absorbers suggests that such absorbers typically arise from small clouds on the order of $\lesssim1-2$ kpc in size \citep{Werk2016,Haislmaier2021}. Thus, this common method of observing the CGM is probing gas only on very small scales---scales where we have shown (Section~\ref{sec:local_results}) there is significant variation in the forces acting on the CGM gas.

In rare cases where a background quasar is gravitationally lensed into multiple images, measurements of absorption along the line of sight to each image can give an estimate of the size of absorbing clouds on scales of a few kpc. Studies using this method to probe the properties of CGM gas on small scales find a variety of results: the column density of cool clouds (traced by \ion{Mg}{2} absorption) either vary significantly on scales of $\lesssim3$ kpc, indicating small clouds \citep{Ellison2004}, or have coherence across large scales, indicating streams \citep{Rubin2018}, or both \citep{Chen2014,Augustin2021}. The column densities of higher ions that trace warmer gas tend to have longer coherence lengths, suggesting that the warmer phases are located in larger structures \citep{Lopez2007}. The high degree of column density variation on small scales found by some of these studies may be related to the large variations on small scales that we find in the forces acting on CGM gas in this study, if those forces are helping to compress gas into small, dense clouds. Even if the forces do not act for enough time to compress gas directly (and indeed we do find significant time variation as well in Section~\ref{sec:local_results}), the clumpy nature of the CGM revealed by lensed quasar observations provides more evidence against smooth equilibrium models.

While small cloud sizes are typically inferred for lower ions tracing cooler gas, the starting point of many multiphase CGM models is to assume equilibrium in total (thermal and non-thermal) pressure between the warm and cool gas \citep[e.g.,][]{Voit2019a,Voit2019b}. If such an equilibrium holds, the dynamics and support of the warm gas will have an important effect on the pressure of the cool gas. We find that the warm gas has strong variations in its pressure structure in both space and time, so the same may also be true of the cool gas that quasar absorption line surveys typically probe.

Because any given parcel of gas is not found to be in any kind of inward and outward force balance in the FOGGIE galaxy halos, we caution against using global equilibrium models to estimate the thermodynamic properties of observed absorbing gas. Hydrostatic equilibrium is clearly ruled out by the prevalence of turbulence and rotation, but even an equilibrium force balance model that includes these other non-thermal forces cannot describe the details of force variations on the small scales that absorption line observations seem to be probing. In addition, a line of sight through a galaxy's CGM may pass through multiple different CGM segments, such as galactic outflows or filamentary inflows, where the thermodynamic properties of the gas are significantly different. Just as we did in Section~\ref{sec:global_results}, we recommend using a statistical sample of many absorbers to determine the overall properties of the CGM for a given galaxy population, importantly including the velocity shifts and widths of the absorption lines to probe the non-thermal kinematic components of support in the CGM.

There is significant observational evidence of CGM turbulence, particularly in the warm gas of $L^*$ galaxies' CGM. \citet{Werk2016} analyzed the kinematic alignment and widths of both low- and high-ionization state absorbers observed in the CGM of $L^*$ galaxies and found very broad line widths for the O VI absorbers that are presumably tracing warm, $\sim10^{5.5}$ K gas. These line widths suggest some amount of turbulent broadening or radiatively cooling flows \citep{Heckman2002} that are prevented from accreting fully onto the central galaxy and instead are participating in a cycle of cool cloud formation and destruction through feedback heating \citep{McQuinn2018}. \citet{Werk2014} found that the low-ionization state absorbers, tracing cool clouds, in this sample had inferred densities and temperatures inconsistent with being in thermal pressure equilibrium with the presumed surrounding warm medium, thus requiring non-thermal pressure support. \citet{Rudie2019} found significant non-thermal contribution to line widths of a sample of absorbers in the CGM of $L^*$ galaxies at $z\sim2$, also indicating that turbulence and other non-thermal kinematics are playing an important role in the CGM gas. In the Milky Way halo, ratios of measured column densities of different ions require non-equilibrium and kinematic models to describe, such as turbulent mixing layers between hot and cool gas \citep{Savage2000,Fox2004,Wakker2012}.

Recent simulation and analytic work has explored the role of turbulence in producing multiphase gas, like is commonly observed in the CGM. These works focus on very small parcels of gas in idealized setups designed to be relevant to turbulent mixing layers between hot and cool gas \citep{Ji2019,Fielding2020a,Tan2020,Yang2022} or cool clouds embedded within a hot medium \citep{Buie2018,Gronke2018,Fielding2022,Gronke2022}. They all find that turbulence drives mixing in the interface layer between the gas phases, which allows the mixed, intermediate-temperature gas to radiatively cool and fuel cool gas production. Thus it appears that turbulence is not only necessary for helping to support CGM gas against gravity, but also important for cool gas production and survival to explain the abundance of kinematic, multiphase CGM observations.

\subsection{Global forces and galaxy evolution models}
\label{subsec:galaxy_evolution}

Section~\ref{sec:global_results} shows that despite significant small-scale variation, a global equilibrium of inward and outward forces emerges, where thermal pressure, turbulent pressure, and rotation all contribute to the support of CGM gas against inward gravity in different relative amounts in different parts of the CGM. The eventual equilibrium is unsurprising, as there are many indications that the overall population of galaxies can be described well by equilibrium models. Galaxy population scaling relations, such as the star forming main sequence or the mass-metallicity relation, are well-described by models that balance fresh gas inflow rates, star formation rates, and outflow rates (with some fraction of outflows recycling back onto the galaxy) in near-equilibrium to produce slow galaxy growth over time \citep{Finlator2008,Bouche2010,Dave2012,Lilly2013,Dekel2014,Peng2014,Mitra2015,Somerville2015a,Somerville2015b}. What is surprising is the fact that this equilibrium emerges \emph{despite} the significant variation of the inward and outward forces on small scales presented in Section~\ref{sec:local_results}, and without any form of equilibrium in the CGM being enforced \emph{a priori} in the simulation. Certain simulation parameters, particularly those controlling star formation feedback, were tuned to produce fairly realistic galaxies as compared to observations, but no CGM parameters were similarly tuned.

Another class of models uses analytic arguments and one-dimensional scaling relations to predict the observational properties of the CGM gas, and these models typically start from an assumption of hydrostatic equilibrium, which is an inherent assumption that the halo mass is the dominant factor in setting the properties of the CGM \citep[e.g.,][]{Faerman2017,Voit2019a,Voit2019b,Faerman2020}. These models are broadly successful at reproducing observed column densities of various low and high ions, although there is large scatter in the observational data. In this work, we have found that the strength of feedback, which varies over time with the star formation history of the central galaxy, plays a significant role in setting the balance of forces in the CGM so that a spatially-averaged radial force balance is only achieved at low redshift when the feedback becomes more constant. This suggests that any equilibrium model --- purely hydrostatic or additionally including non-thermal pressure support --- that is determined only by halo mass without allowing variation due to feedback, provides a close, but not complete, description of the CGM.

Many models of the gas surrounding galaxies were developed using more massive halos than the $L^*$ scale we explore in this work. In galaxy groups and clusters, observations of X-ray emission and the Sunyaev-Zel'dovich (SZ) effect, which trace predominantly hot gas, suggest hydrostatic equilibrium is an accurate and descriptive model, with $\lesssim15\%$ of the pressure support provided by non-thermal pressures \citep[see the recent review and references therein by][]{Donahue2022}. The deeper gravitational potential well of massive halos reduces the effect of feedback, which may act to damp turbulence. These tracers are sensitive to the integral of gas properties along the line of sight, which means they primarily probe the spatially largest segments of the CGM that contain hot gas, and are not sensitive to cooler segments, like inflow, or to physically smaller segments, like collimated outflows --- leaving only the no strong flow segment, which is closest to equilibrium, as the part of the CGM predominantly probed by X-ray or SZ. At $L^*$ halo masses and below, X-ray observations become significantly more difficult and generally require stacking the emission from many galaxies, other than in a handful of particularly nearby, or powerfully starbursting, galaxies \citep[e.g.,][]{Strickland2004,Das2020}. Instead most observations trace cooler gas in the UV or optical, in absorption against background light sources. At these lower mass scales, feedback can have a stronger effect on the halo gas, driving it further out of equilibrium. Indeed, feedback appears to be necessary to reproduce the observed baryon fractions below the cosmic value in low mass halos \citep[e.g.,][]{Eckert2021}, so it is no surprise that it may have other effects, like driving turbulence, as well.

One of the main results of this work is the different gas properties in the different CGM segments, from the cool, inflowing filaments with little internal structure (at our resolution scale) to the warm and mildly turbulent volume-filling medium and finally the hot, significantly turbulent outflow. Analytic models of the CGM aim to describe the small-scale physics from the starting point of global CGM properties, like hydrostatic equilibrium, and we suggest that a similar process can be applied from the starting point of the distinct CGM segments we identify here. Broadly, within $0.25R_{200}$, there is no equilibrium, and turbulence drives strong outward forces on the order of $2-3\times$ stronger than the force of gravity. In this innermost region, thermal pressure force is roughly $25\%$ of the force of gravity, and rotational support is an additional $\sim25\%$. Between $0.25R_{200}$ and $0.5R_{200}$, the CGM is in an overall equilibrium, with thermal pressure, turbulent pressure, and rotational support all contributing roughly equally to support the gas against gravity. In the outer CGM, beyond $0.5R_{200}$, thermal support provides roughly $50-60\%$ of the support against gravity, with rotational support providing another $20\%$ and turbulence becoming negligible. Moving to larger radii, the gas becomes under-supported against gravity, near $75\%$ of the force of gravity near $R_{200}$, perhaps indicating accretion onto the outer edges of the halo. These broad strokes can be used as the initial conditions and properties of gas in the CGM in analytic or idealized simulation works that aim to model the physics occurring on smaller scales than we can resolve in FOGGIE, such as thermal instability, phase mixing, and cooling.

Global emergent equilibrium arising from local and time-dependent non-equilibrium conditions may be a common property of the CGM or galaxy evolution, in areas other than forces acting on the CGM gas. For example, the long-term pseudo-equilibrium in gas flows into and out of galaxies used in many semi-analytic models of galaxy evolution may be an emergent phenomenon arising from small-scale or short-time non-equilibrium in gas flows. As new telescopes and instruments come online that allow us to observe smaller and smaller scales within and surrounding galaxies, it will become crucially important to understand how small-scale gas physics build up to produce the galaxy population relations that seem to imply global equilibrium in galaxy evolution. Exploring the connection between non-equilibrium local conditions and equilibrium global conditions in the galaxy-CGM connection will be a main topic of future work.

\subsection{Comparison to other simulation works}
\label{subsec:compare}

\citet{Oppenheimer2018} analyzed the forces responsible for supporting the CGM against gravity using the EAGLE simulations. They found that Milky Way-like galaxies, which we focus on here, can have significant support provided by non-thermal gas motions. They do not separate turbulence from rotation or ram pressure, but instead group these kinetic processes roughly into tangential and radial directions. The tangential direction includes both rotation and turbulence, and they find this component strongly dominates the inner CGM of Milky Way mass galaxies, just as we find here. They also find that the thermal support dominates in the outer CGM, and we are again in agreement with their results. Quantitatively, \citet{Oppenheimer2018} find a stronger contribution of thermal support in the outer CGM relative to the non-thermal forms of support than we find in any of our halos. This may be a mass-dependent effect, as the lowest halo mass bin explored in \citet{Oppenheimer2018} is slightly more massive than our four FOGGIE galaxies here, and \citet{Oppenheimer2018} found that the degree of non-thermal support increases as halo mass decreases. It could also be a resolution-dependent effect, as we find that decreasing resolution leads to a decrease in the amount of turbulence in the CGM, particularly in the outflow (\citeauthor{Oppenheimer2018} \citeyear{Oppenheimer2018} does not find any differences between two simulations run at different resolutions, but the resolution in FOGGIE is higher still than their high-resolution run). They also find that a simulation run without feedback results in an increased importance of thermal support throughout the CGM and decreased non-thermal forms of support, just as we find in this study. The EAGLE simulations have a different feedback model than the FOGGIE simulations, and the importance of feedback to the relative strengths of thermal and non-thermal supporting forces suggests that the differences between this work and \citet{Oppenheimer2018} could be an effect of the feedback model implemented as well.

\citet{Lochhaas2020} performed a similar analysis using the idealized, isolated, non-cosmological CGM simulations of \citet{Fielding2017}. They found that while non-thermal pressure support was required to bring a Milky Way mass halo into equilibrium, the thermal pressure support dominated everywhere within the halo and it was only the lower-mass halo that had significant support provided by non-thermal gas kinematics. However, these simulations were initialized with a hot halo in hydrostatic equilibrium, so it is unsurprising that this condition is present throughout the run time of the simulation. The FOGGIE simulations (and the EAGLE simulations analyzed in \citeauthor{Oppenheimer2018} \citeyear{Oppenheimer2018}) have the benefit of cosmological structure that allows galaxies and their halos to evolve ``naturally", without specifying hydrostatic equilibrium as an initial condition. Both \citet{Lochhaas2020} and \citet{Oppenheimer2018} used a small number of similar simulations with different feedback parameters (but without varying the feedback model implementation) to explore how the support of CGM gas depends on the strength of galactic winds. \citet{Lochhaas2020} used only the lower mass halo of \citet{Fielding2017} for this exploration, which was already not in equilibrium with the fiducial feedback, and found that weaker winds (with lower mass loading or slower launch speed) produced halos that were even further from equilibrium. On the contrary, \citet{Oppenheimer2018} used a simulation with no winds at all in Milky Way mass halos and found these halos were in equilibrium, with a much stronger contribution from the thermal component than the non-thermal components compared to the fiducial simulation including galactic winds. The tension between these two studies' results may be due to different implementations of feedback models, or could be a mass discrepancy, as the simulations with different feedback parameters in \citet{Lochhaas2020} had halo masses an order of magnitude smaller than those in \citet{Oppenheimer2018}. This could suggest that the thermal component of support is predominantly provided by the higher virial temperature of more massive halos rather than heating by galactic winds. In Section~\ref{subsec:support_feedback} of the present study, we found that weaker feedback leads to a larger thermal component, and accordingly smaller non-thermal component, to the support of the CGM, similar to the \citet{Oppenheimer2018} results.

The FOGGIE simulations have lower-mass CGM \citep{Zheng2020,Simons2020} than other cosmological simulation works, such as FIRE or EAGLE \citep{Davies2019,Hafen2019}, by nearly an order of magnitude. \new{The FOGGIE CGM is also lower in mass than observational estimates of the hot ($T\sim3\times10^6$ K) gas mass surrounding $L^\star$ galaxies \citep[e.g.,][]{Li2018,Bregman2022}, which suggests the feedback implementation in FOGGIE is not launching enough hot gas into the halo (\citealt{Lochhaas2021} shows the virial temperature for the FOGGIE halos is closer to $T\sim3\times10^5$ K and hotter gas tends to be associated with outflows).} Lower mass suggests a lower average gas density in the CGM of the FOGGIE simulations, which reduces the radiative cooling rate and increases the dissipation time of turbulence. A higher cooling rate in a denser CGM could mean less overall thermal support, requiring either more turbulent and rotation support to hold the CGM gas in balance, or leading to a net inflow if the CGM gas does not have enough support against gravity. A higher density CGM would cause faster turbulence dissipation, also leading to a lower degree of turbulent support, but would also increase heating from turbulent dissipation. However, since we find that turbulence appears to be strongly affected by star formation feedback, if a CGM is more massive because the feedback-driven outflows are more massive \new{(which would also be more in line with observational estimates of hot gas mass in the halo)}, it is conceivable that those outflows are also driving more significant turbulence. It is difficult to guess exactly how our results would change in a more massive or denser CGM like those found by other simulations, but a close examination of the sources of CGM support in a high-resolution simulation with a denser CGM is an interesting avenue for future study.

Finally, Figure~1 of \citet{Ji2020} shows radial gradients of different forms of pressure compared to the gradient of the gravitational field in the FIRE-2 simulations, which is a similar method of evaluating support, although by construction it cannot evaluate rotational support because there is no associated pressure with rotation. That study mainly focused on the impact of magnetic fields and cosmic rays in the CGM, neither of which are present in the FOGGIE simulations, but we can compare to the other forms of pressure support they explore. They find thermal support dominates, but there is also a significant amount of turbulent support, nearly as much as thermal support, across a range of halo masses $\sim10^{10}-10^{12}M_\odot$, with less turbulent support in the lower mass halos. They do not find this trend to vary with halocentric radius, unlike what we find in Section~\ref{sec:global_results} where turbulent support is stronger than thermal in the inner halos. Again, this slight discrepancy may be a resolution effect, as FIRE-2 has a mass resolution of $7000M_\odot$ per particle in the Milky Way mass galaxies \citep{Ji2020}, much larger than the mass resolution we find to be necessary to resolve the strongest turbulence (Figure~\ref{fig:vdisp_mass_res} and Section~\ref{subsec:turb_res}). Although we do not have cosmic rays or magnetic fields in the FOGGIE simulations, both \citet{Ji2020} and \citet{Butsky2020} found that cosmic ray pressure can also provide an important non-thermal supporting force for the CGM gas.

\section{Summary and Conclusions}
\label{sec:summary}

Using four roughly Milky Way mass galaxies from the Figuring Out Gas \& Galaxies In Enzo (FOGGIE) simulations, we have investigated the radially-directed forces acting on circumgalactic medium (CGM) gas on both small and large scales and as functions of time as each galaxy evolves. We calculate forces exerted by the thermal pressure gradient, the turbulent pressure gradient, the ram pressure gradient, and centrifugal rotation, and compare each to the force of gravity in these halos to determine which forces are responsible for supporting the CGM against the inward pull of gravity. We also evaluate whether there exists a balance between the inward and outward forces acting on the CGM, and if so, whether this balance fits with the common simplification of hydrostatic equilibrium. Our main results are:
\begin{enumerate}
    \item Casting the various pressures (thermal, turbulent, ram) as forces by taking their radial gradients allows us to compare the effect of these pressures to other processes that do not have an associated pressure, such as rotation. This framework naturally provides a way to evaluate the important forces cell-by-cell within the simulation, as well as in a large-scale summation over large swaths of the CGM (Section~\ref{sec:methods}).
    \item When calculating forces cell-by-cell for one of the FOGGIE galaxies at $z=0$, we find significant variation of all forces, but especially the thermal and turbulent forces, on small, resolution-dependent scales of $\sim5$ kpc, or $\sim5$ resolution elements in length (Figure~\ref{fig:force_slices} and Section~\ref{sec:local_results}). Not only do the forces vary in strength, but the thermal, turbulent, and ram pressure forces can oscillate between being directed inward and outward, indicating that these pressures do not necessarily act only to hold gas parcels aloft (Figure~\ref{fig:force_rays}). A given parcel of CGM gas is unlikely to be in perfect inward and outward force balance. Strong variation of forces on small scales suggests that CGM observations that probe such small scales, such as quasar absorption line observations, will not observe gas in any kind of equilibrium balance (Section~\ref{subsec:obs}).
    \item When summing the forces acting on CGM gas over large scales to smooth out the small-scale variation, we find that it is only the outer CGM that is in an average radial force balance. The inner CGM tends to have an over-abundance of outward-directed forces, primarily driven by strong feedback close to the galaxy (Section~\ref{sec:global_results} and Figure~\ref{fig:support_vs_r_Tempest}).
    \item The most relevant forces for supporting gas against gravity are thermal pressure forces, turbulent pressure forces, and centrifugal rotation (Section~\ref{sec:global_results}). Each of these forces are more dominant in different locations of the CGM: turbulence dominates in the inner CGM close to galaxies while thermal force dominates in the outer CGM, and rotation provides some supporting force throughout all halocentric radii, but is somewhat stronger at intermediate radii (Figures~\ref{fig:support_vs_r_Tempest} and~\ref{fig:support_t_avg_Tempest}).
    \item The contribution of each type of force to the support of the CGM against gravity is time-variant, in both the inner and outer CGM (Figures~\ref{fig:support_vs_t_Tempest} and~\ref{fig:support_t_avg_Tempest}). The turbulent force is the most time-dependent, with strong peaks following bursts of star formation (and thus periods of strong feedback) in the central galaxy. The thermal force is also affected by the time-variable feedback, but the rotational force is more steady. Turbulence strongly dominates the support of the inner CGM at high redshift $z\gtrsim0.5$ when the star formation rate of the FOGGIE galaxies is highly variable and bursty. At lower redshift, the star formation rate smooths out, and thermal force comes to dominate over the turbulent force, especially in the outer CGM.
    \item The CGM of the FOGGIE galaxies clearly separates into different ``segments": regions of outflowing galactic winds, inflowing accretion filaments, and material not participating in any strong radial flows (Section~\ref{subsec:segmenting} and Figure~\ref{fig:CGM_segments}). When evaluating the forces acting on CGM gas separately within each segment, we find that the inflow segment is under-supported with nearly equal contributions from thermal, turbulent, and rotational forces, the outflow segment is over-supported, highly time-variable, and supported primarily by turbulent forces, and the no strong flow segment is essentially in equilibrium, provided by a combination of thermal pressure, turbulent pressure, and rotational support in the inner CGM and primarily thermal force in the outer CGM (Section~\ref{subsec:support_segments} and Figure~\ref{fig:support_segments_Tempest}).
    \item We compared the fiducial simulation of one of the FOGGIE galaxies to two re-runs of this same galaxy from $z=0.4$ to $z=0$ with an order of magnitude smaller thermal feedback strength (``weak feedback") and with an enhanced burst of feedback (``strong burst"; after which the feedback strength was returned to fiducial). The strong burst case shows a drastic increase in turbulent support in the inner CGM and an increase in thermal support in the outer CGM following the burst, which takes $\sim1$ Gyr to dissipate back to the level it was before the burst. The weak feedback case is primarily supported by rotation in equilibrium in the inner CGM (as opposed to the fiducial run, which was over-supported in the inner CGM due to strong turbulence) and primarily supported by thermal force in the outer CGM, but not enough to bring it to equilibrium (Section~\ref{subsec:support_feedback} and Figures~\ref{fig:support_feedback_strong}-\ref{fig:support_feedback_weak}).
    \item Like other recent works, we find that some amount of non-thermal support is necessary to hold up the CGM against gravity (Section~\ref{subsec:compare}). However, we generally find a greater relative fraction of non-thermal support compared to thermal support in the FOGGIE galaxies. This is most likely a resolution effect, as high resolution is required to resolve the regions of strongest turbulence, which are generally in the warm, diffuse gas where cell masses are low (Section~\ref{subsec:turb_res} and Figures~\ref{fig:vdisp_spatial_res} and~\ref{fig:vdisp_mass_res}).
    \item We suggest that the results of this study, particularly within the different segments of the CGM, can be used as the initial conditions for further analytic or idealized simulation work that aims to describe the physics operating on scales below our resolution limit.
\end{enumerate}

One of the most surprising results of this study is the very strong variation of forces acting on halo gas on small scales. The small-scale variation smooths out to the expected equilibrium force balance on large scales, but it is important to note that such an equilibrium does not persist down to small scales. This suggests that global equilibrium models, such as hydrostatic equilibrium or even an equilibrium that includes non-hydrostatic forces, cannot specify the details of the forces acting on the gas on small scales. Because the common method of observing the CGM through absorption lines in the spectrum of background point sources probes only small scales, this result suggests that a global equilibrium model also cannot specify the conditions that observations probe. Instead, the apparent equilibrium in the force balance, and perhaps other forms of assumed equilibria in galaxy evolution theories, may be \emph{emergent} equilibria, where significant non-equilibrium conditions on small scales smooth out to the equilibrium on large scales. In this scenario, it makes the most sense to understand how small-scale gas processes produce equilibrium when scaled up to large scales, rather than using an assumption of equilibrium everywhere in the CGM to understand observations and the galaxy-CGM connection.

\acknowledgments{
We thank the anonymous referee for suggestions that improved the overall clarity of this work. This study was funded primarily by a {\it Hubble Space Telescope} Archival Research Theory Grant, HST AR \#16140 (PI C.\ Lochhaas).
CL and RA were additionally supported by NASA via an Astrophysics Theory Program grant 80NSSC18K1105. JKW acknowledges support from NSF-AST 1812521, NSF-CAREER 2044303 and an RCSA Cottrell Scholar grant, ID number 26842.  BWO acknowledges support from NSF grants \#1908109 and \#2106575 and NASA ATP grants NNX15AP39G and 80NSSC18K1105. JJ acknowledges support by the U.S. Department of Energy under Contract No. DE-AC02-09CH1146 via an LDRD grant. JT and ACW acknowledge support from the Roman Space Telescope Milky Way Science Investigation Team. AA is supported by NSF-AST 1910414. RA is additionally supported by HST AR \#15012. RCS appreciates support from a Giacconi Fellowship at the Space Telescope Science Institute, which is operated by the Association of Universities for Research in Astronomy, Inc., under NASA contract NAS 5-26555. For the purpose of open access, the author has applied a Creative Commons Attribution (CC BY) license to any Author Accepted Manuscript version arising from this submission.

Resources supporting this work were provided by the NASA High-End Computing (HEC) Program through the NASA Advanced Supercomputing (NAS) Division at Ames Research Center and were sponsored by NASA's Science Mission Directorate; we are grateful for the superb user-support provided by NAS. Resources were also provided by the Blue Waters sustained-petascale computing project, which is supported by the NSF (award number ACI-1238993 and ACI-1514580) and the state of Illinois. Blue Waters is a joint effort of the University of Illinois at Urbana-Champaign and its NCSA. 

\texttt{Enzo} \citep{Bryan2014,BrummelSmith2019} and \texttt{yt} \citep{Turk2011} are developed by a large number of independent researchers from numerous institutions around the world.  This research made use of Astropy (http://www.astropy.org), a community-developed core Python package for Astronomy \citep{Astropy1, Astropy2}. Their commitment to open science has helped make this work possible.

Pictures of the FOGGIE team's cats and dogs were responsible for improving the authors' mental health while carrying out this study.

\facilities{NASA Pleiades}

\software{astropy \citep{Astropy1,Astropy2},
          Cloudy \citep{Ferland2017}, 
          Enzo \citep{Bryan2014,BrummelSmith2019},
          grackle \citep{Smith2017},
          yt \citep{Turk2011}
          }
}

\appendix{
\section{Results for Squall, Maelstrom, and Blizzard}
\label{sec:other_halos}
Here we present all results presented in the main body of the paper, for the other three FOGGIE halos. Generally, the trends found and discussed in the main body of the paper for the Tempest halo also hold for the other three halos, except where noted.

\begin{figure}
    \centering
    \includegraphics[width=0.31\linewidth]{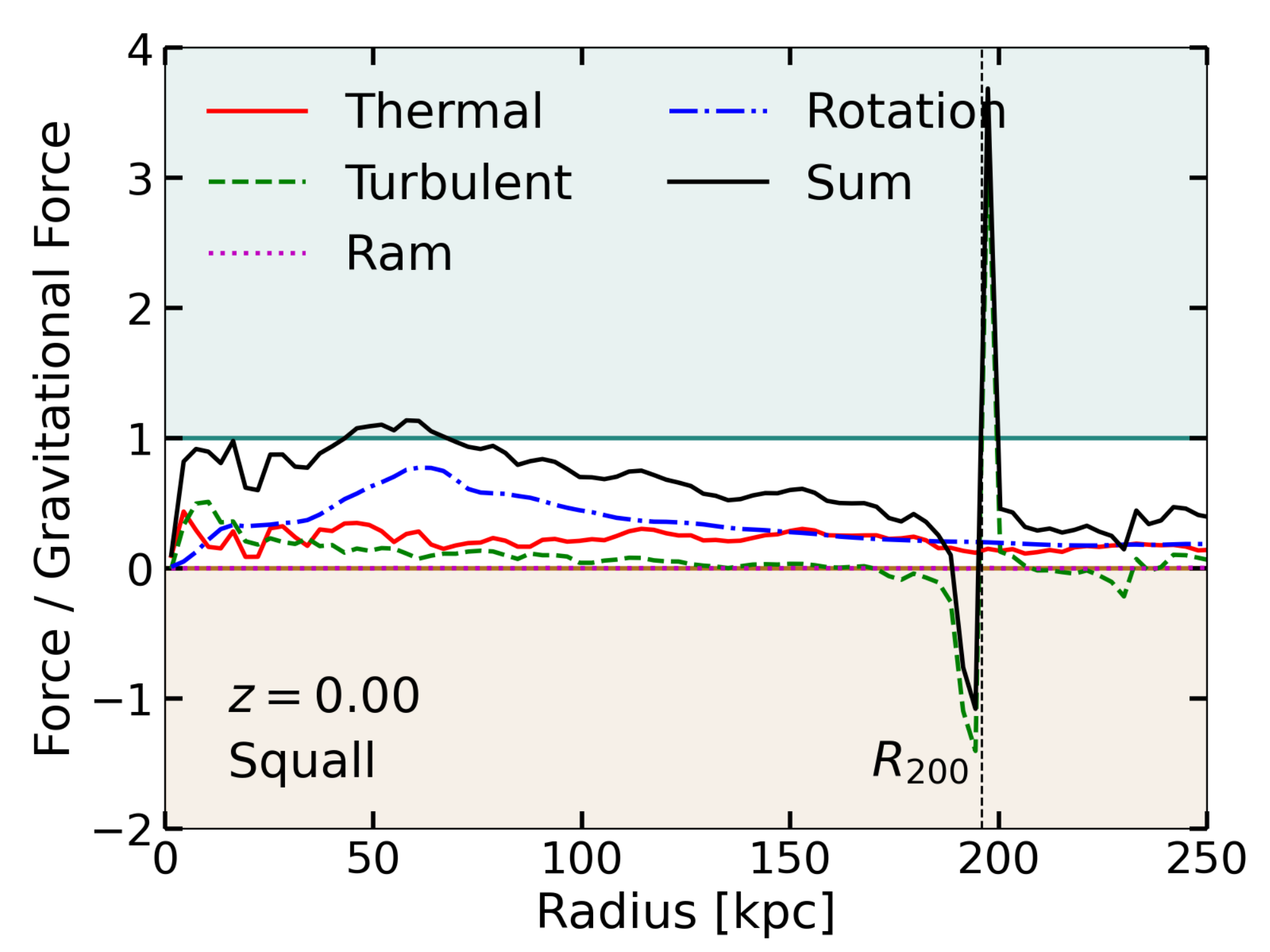}
    \includegraphics[width=0.31\linewidth]{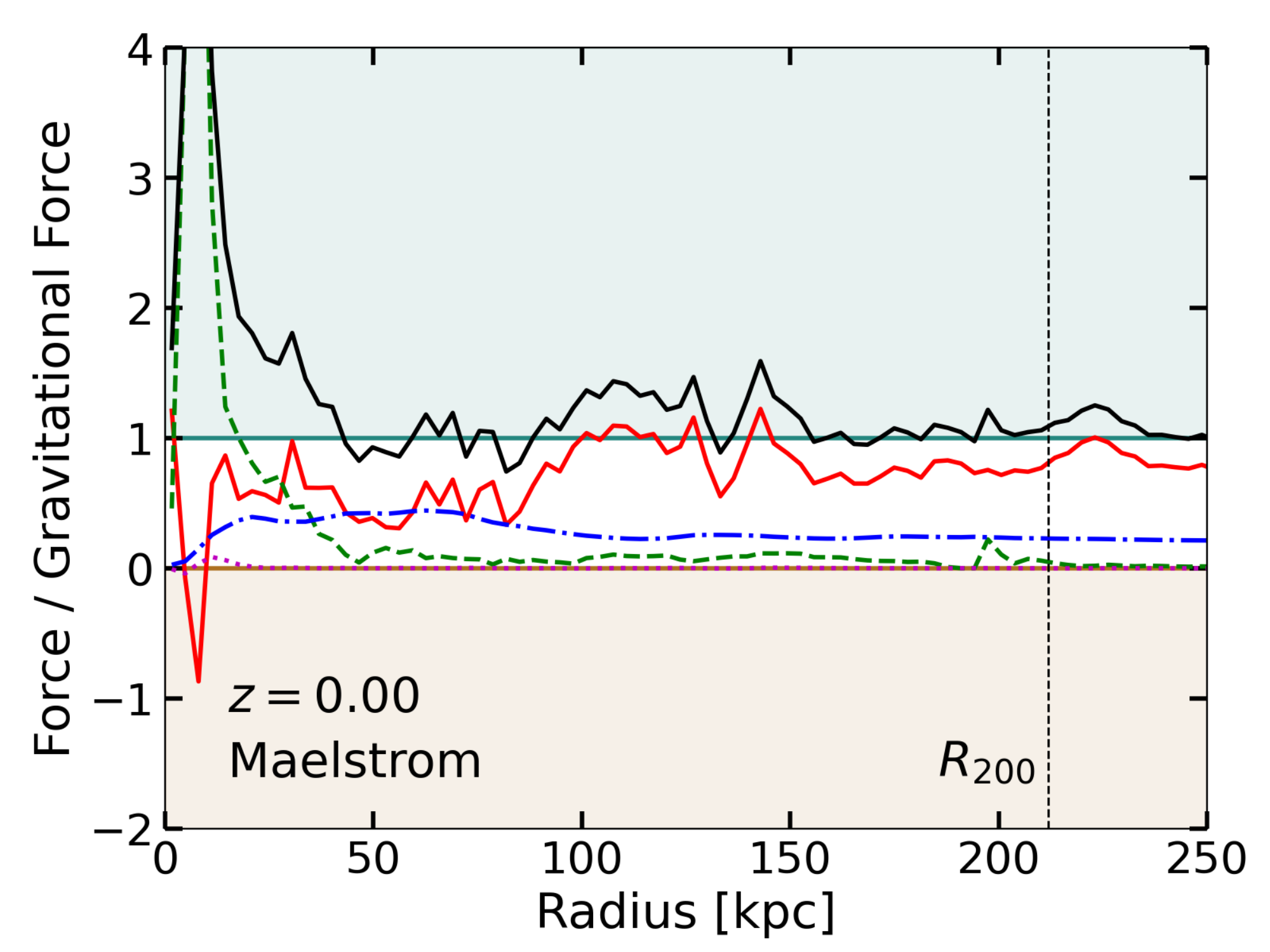}
    \includegraphics[width=0.31\linewidth]{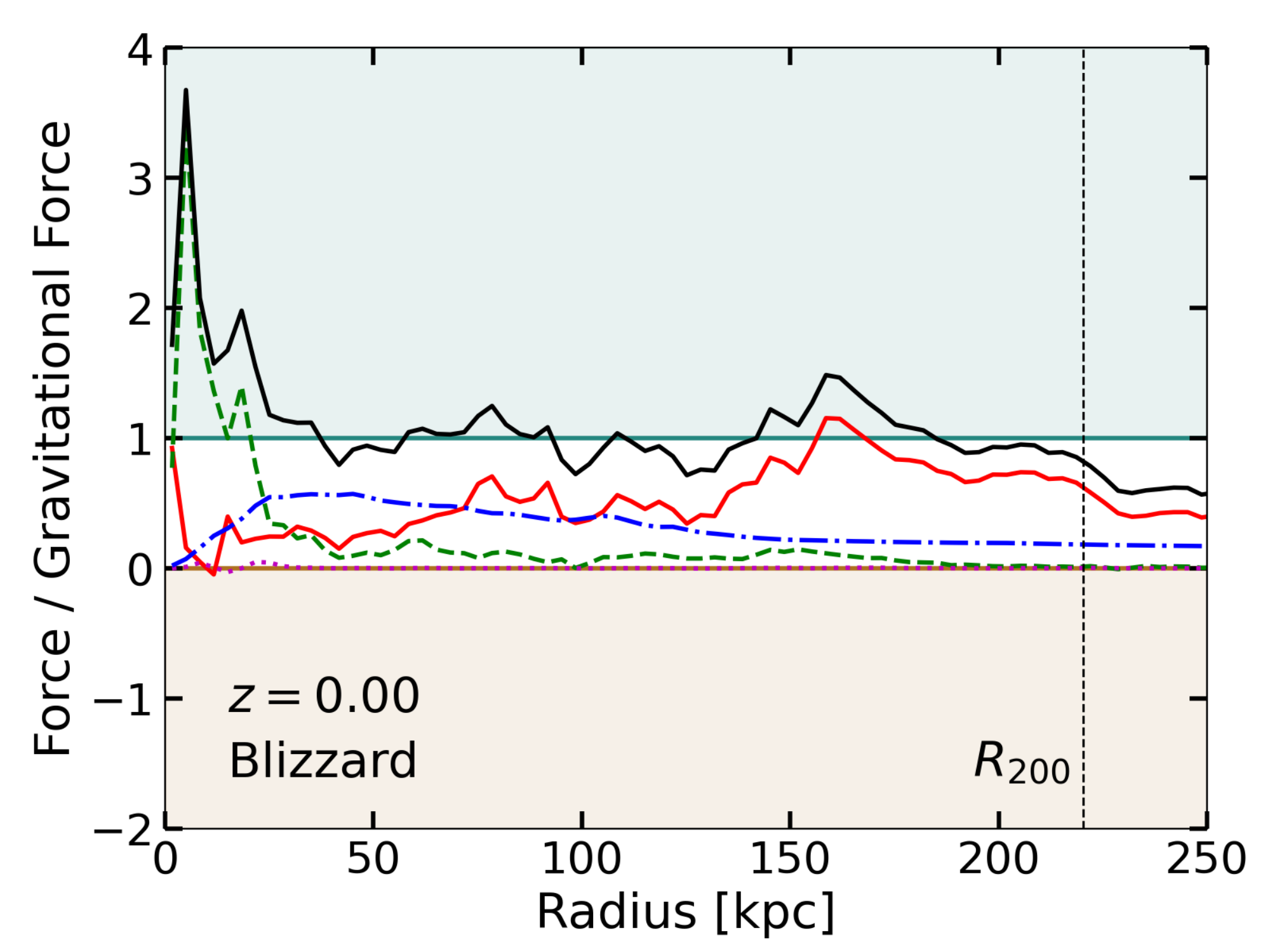}
    \caption{As in Figure~\ref{fig:support_vs_r_Tempest}, but for the other three FOGGIE halos at $z=0$.}
    \label{fig:support_vs_r}
\end{figure}

Figure~\ref{fig:support_vs_r} shows the contribution of each force type to the support of CGM gas against gravity, as in Figure~\ref{fig:support_vs_r_Tempest}, for the other three FOGGIE halos, at $z=0$. Squall is the only FOGGIE galaxy that does not follow the trends laid out by the other three galaxies. Near $z=0$, Squall appears to be in the middle of a strong accretion event and has very little feedback-driven outflow. Its inner CGM is roughly in force balance, with the thermal, turbulent, and rotation forces contributing roughly equally to support, while the outer CGM is under-supported and dominated primarily by rotational forces. \citet{Lochhaas2021} found that Squall is under-virialized near $z=0$, supporting the findings here that it is also under-supported.

\begin{figure*}
    \centering
    \includegraphics[width=\linewidth]{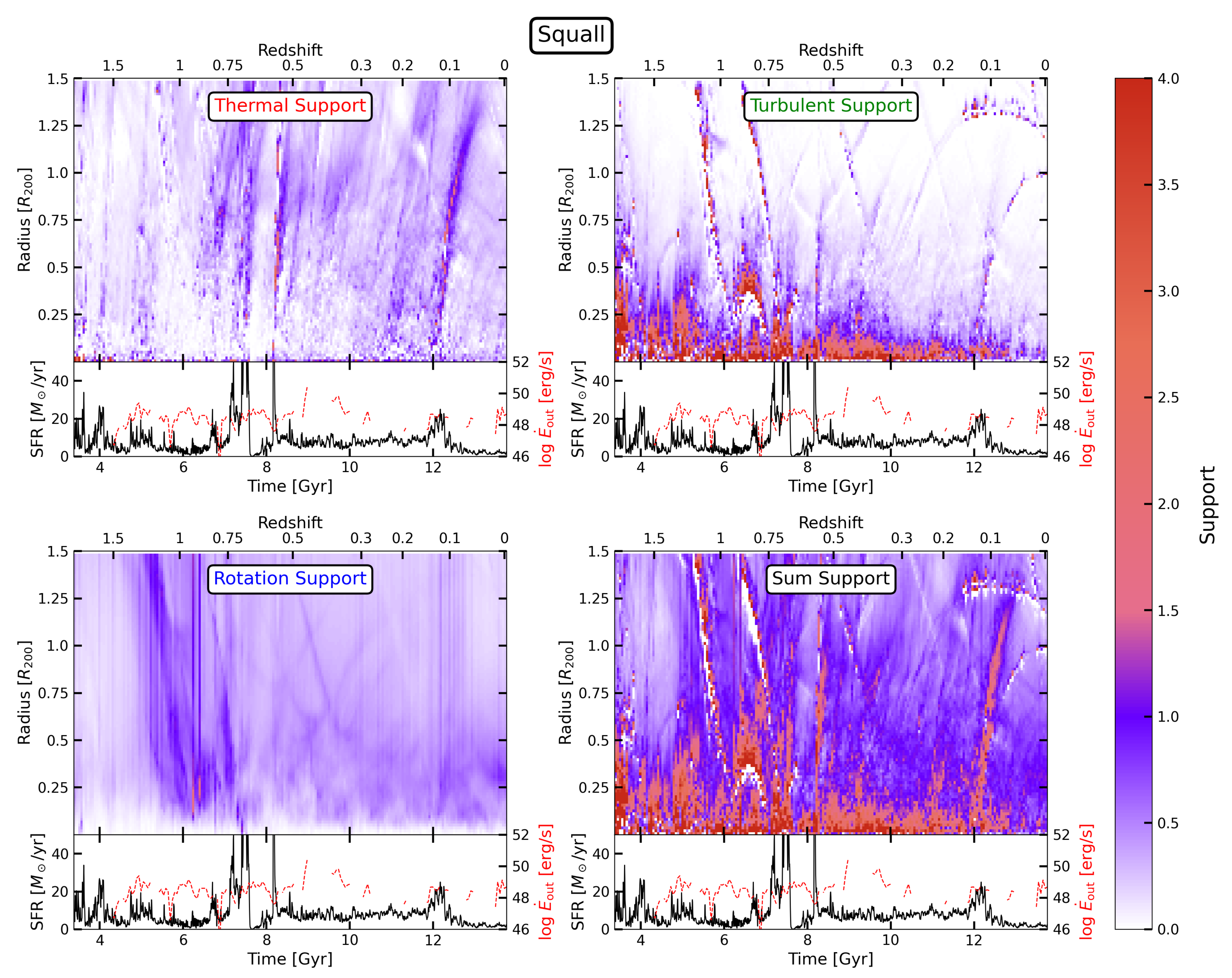}
    \caption{As in Figure~\ref{fig:support_vs_t_Tempest}, but for the Squall halo.}
    \label{fig:support_vs_t_Squall}
\end{figure*}

\begin{figure*}
    \centering
    \includegraphics[width=\linewidth]{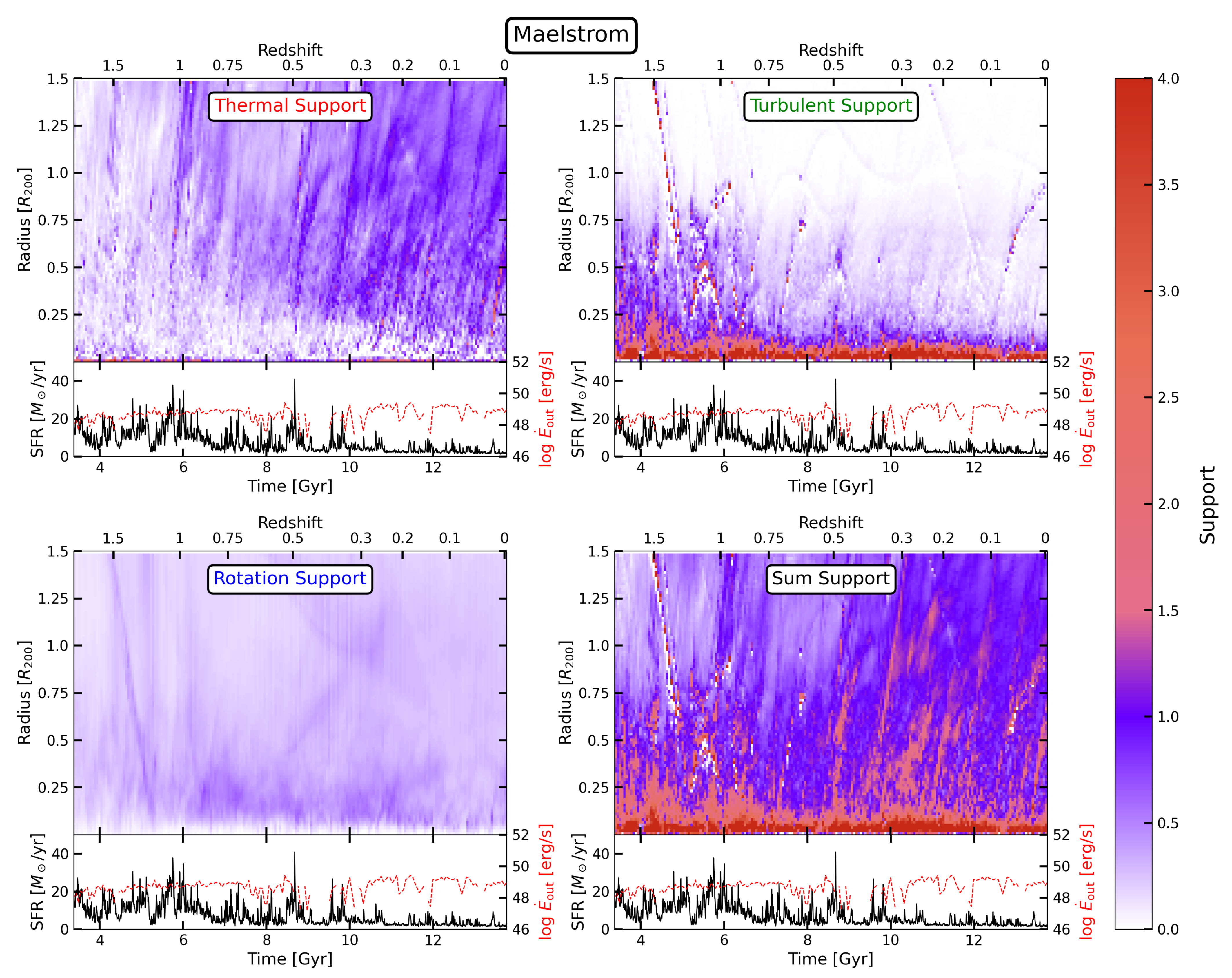}
    \caption{As in Figure~\ref{fig:support_vs_t_Tempest}, but for the Maelstrom halo.}
    \label{fig:support_vs_t_Maelstrom}
\end{figure*}

\begin{figure*}
    \centering
    \includegraphics[width=\linewidth]{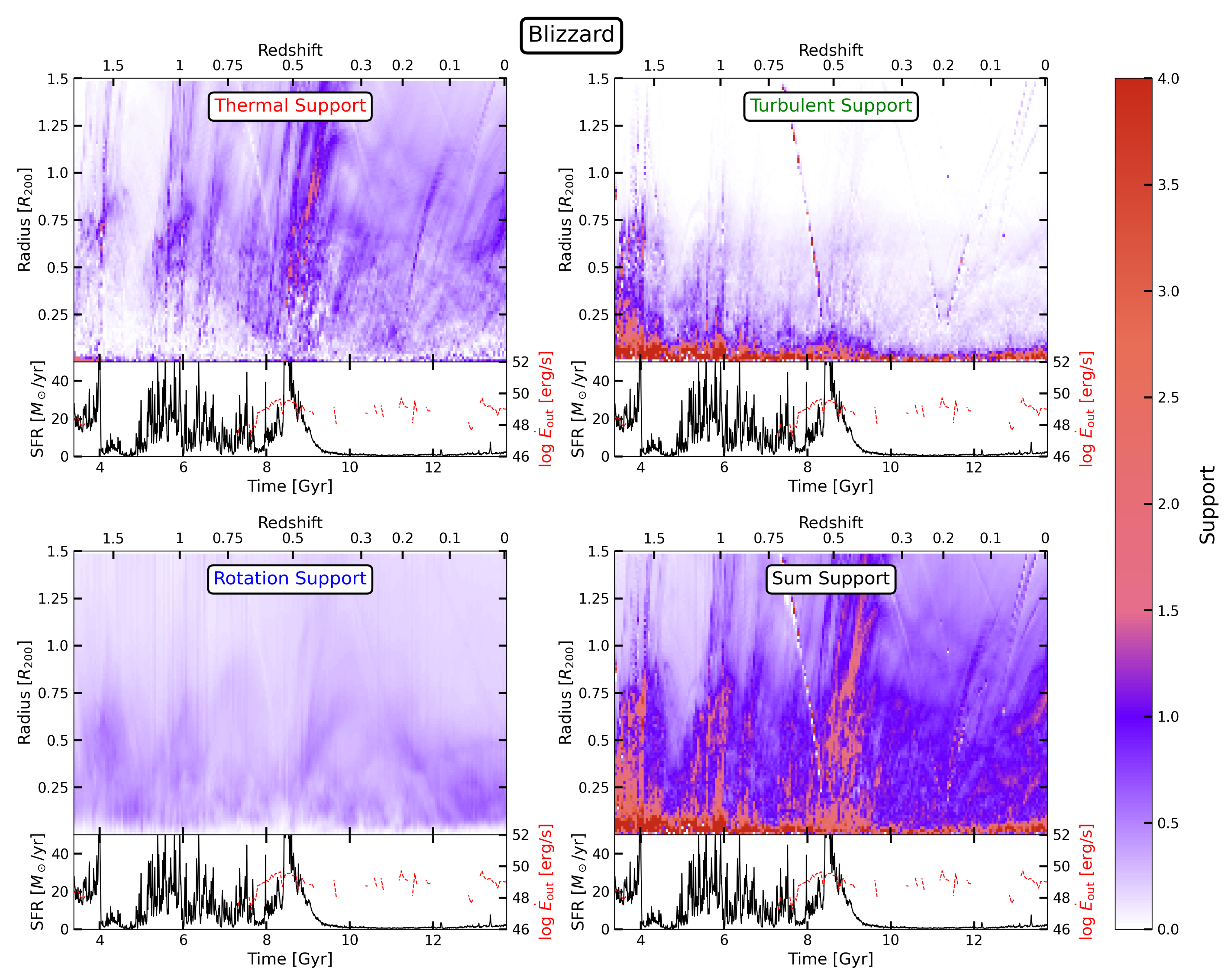}
    \caption{As in Figure~\ref{fig:support_vs_t_Tempest}, but for the Blizzard halo.}
    \label{fig:support_vs_t_Blizzard}
\end{figure*}

Figures~\ref{fig:support_vs_t_Squall}-\ref{fig:support_vs_t_Blizzard} show the contribution of each force type to the support of CGM gas against gravity in 2D plots, as functions of both time and radius, as in Figure~\ref{fig:support_vs_t_Tempest}, for the other three FOGGIE halos. Figure~\ref{fig:support_t_avg} shows the time-averaged contribution of each force type to the support of CGM gas against gravity as functions of radius, as in Figure~\ref{fig:support_t_avg_Tempest}, for the other three FOGGIE halos. Figure~\ref{fig:support_segments} shows the time-averaged contribution of each force type to the support of CGM gas against gravity as functions of radius, within the three identified CGM segments of inflow, outflow, and no strong flow, as in Figure~\ref{fig:support_segments_Tempest}, for the other three FOGGIE halos.

\begin{figure}
    \centering
    \includegraphics[width=0.31\linewidth]{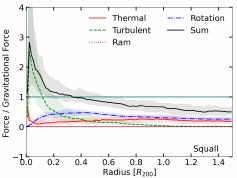}
    \includegraphics[width=0.31\linewidth]{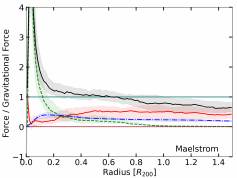}
    \includegraphics[width=0.31\linewidth]{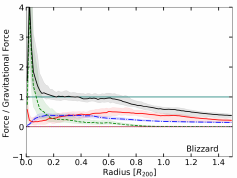}
    \caption{As in Figure~\ref{fig:support_t_avg_Tempest}, but collapsed into a single panel for each halo and without the teal and brown shaded regions, to eliminate shading confusion.}
    \label{fig:support_t_avg}
\end{figure}

\begin{figure*}
    \centering
    \includegraphics[width=0.32\linewidth]{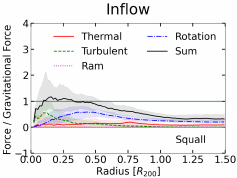}
    \includegraphics[width=0.32\linewidth]{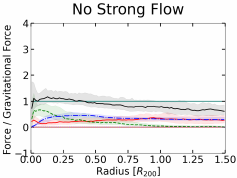}
    \includegraphics[width=0.32\linewidth]{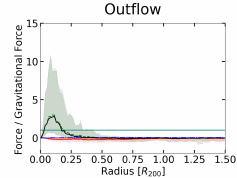}
    \includegraphics[width=0.32\linewidth]{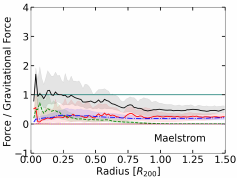}
    \includegraphics[width=0.32\linewidth]{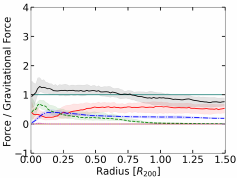}
    \includegraphics[width=0.32\linewidth]{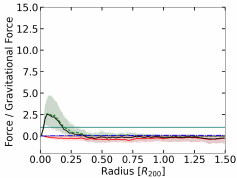}
    \includegraphics[width=0.32\linewidth]{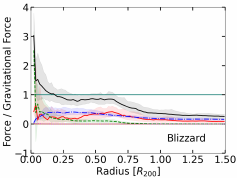}
    \includegraphics[width=0.32\linewidth]{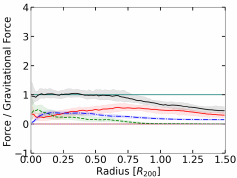}
    \includegraphics[width=0.32\linewidth]{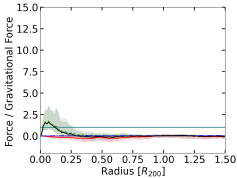}
    \caption{As in Figure~\ref{fig:support_segments_Tempest}, but for the other three FOGGIE halos.}
    \label{fig:support_segments}
\end{figure*}

}

\bibliography{bibliography}{}
\bibliographystyle{aasjournal}

\end{document}